The University of Nottingham

UNITED KINGDOM · CHINA · MALAYSIA

SCHOOL OF COMPUTER SCIENCE

# Twisted Cubes

## and their Applications in Type Theory

**Supervisor:**
Thorsten ALTENKIRCH

**Author:**
Gun PINYO

**ID:** 14274754

**Internal Examiner:**
Ulrik BUCHHOLTZ

**External Examiner:**
Benedikt AHRENS

May 2023

This page intentionally left blank.

(as an even page that follows the title page)

## Abstract

This thesis captures the ongoing development of *twisted cubes*, which is a modification of cubes (in a topological sense) where its *homotopy type theory* does not require paths or higher paths to be invertible. My original motivation to develop the twisted cubes was to resolve the incompatibility between *cubical type theory* and *directed type theory*.

The development of twisted cubes is still in the early stages and the intermediate goal, for now, is to define a *twisted cube category* and its *twisted cubical sets* that can be used to construct a potential definition of $(\infty, n)$-*categories*.

The intermediate goal above leads me to discover a novel framework that uses *graph theory* to transform *convex polytopes*, such as simplices and (standard) cubes, into base categories. Intuitively, an $n$-dimensional polytope is transformed into a *directed graph* consists 0-faces (extreme points) of the polytope as its nodes and 1-faces of the polytope as its edges. Then, we define the base category as the full subcategory of the graph category induced by the family of these graphs from all $n$-dimensional cases.

With this framework, the modification from cubes to twisted cubes can formally be done by reversing some edges of cube graphs. Equivalently, the twisted $n$-cube graph is the result of a certain endofunctor being applied $n$ times to the singleton graph; this endofunctor (called *twisted prism functor*) duplicates the input, reverses all edges in the first copy, and then pairwisely links nodes from the first copy to the second copy.

The core feature of a twisted graph is its *unique Hamiltonian path*, which is useful to prove many properties of twisted cubes. In particular, the reflexive transitive closure of a twisted graph is isomorphic to the simplex graph counterpart, which remarkably suggests that twisted cubes not only relate to (standard) cubes but also simplices.

## Acknowledgements

First of all, I would like to express my deep and sincere gratitude toward my supervisor, Thorsten Altenkirch, who believed in my passion for dependent type theory and accepted me as a PhD student under his supervision. He was also flexible and provided me with the freedom to follow my own ideas as well as providing necessary guidance.

Special thank goes to Nicolai Kraus, who I see as my own older brother. He is also the second author of my earliest paper, where most of the fundamental ideas in this thesis are based on; Without him, I would not finish the paper up to this quality. Besides that, during the time he was a research assistant in our group, he was always available to answer my questions or explore a problem together. His optimistic view of life helps me through many struggles and stressful events during my PhD journey.

I am particularly indebted to Paolo Capriotti, who introduced me to many invaluable mathematical concepts during the time he was a research assistant in our group. His teaching revolutionised the way I see mathematics as a whole including the type theory itself. We also surprisingly share many common interests and hobbies.

My two thesis examiners, Ulrik Buchholtz (internal examiner) and Benedikt Ahrens (external examiner) have both spent a substantial amount of effort with my thesis and I truly appreciate them. Their comments and feedbacks enable me to improve the readability and resolve many typographic mistakes in the thesis. In addition, their comments also contribute to the research itself. One particular example is Ulrik's comment on similarity between the Gray code and the unique Hamiltonian paths of twisted cubes.

The Functional Programming Laboratory of the University of Nottingham was a very pleasant research group to study in. It was relaxed, motivating, and joyful all at once. In addition to the people I have mentioned above, I would like to thank all members (and formal members) in our group, especially Jakob von Raumer, Ambrus Kaposi, and Christian Sattler, for many interesting discussions and joyful activities. I also would like to use this opportunity to thank Venanzio Capretta and Natasha Alechina for their helpful comments and feedbacks during the first and second annual reviews.

There are many more people who would deserve to be mentioned; Most of them were visitors to our research group and/or people who I met during many conferences. In particular, Andreas Nuyts, Jamie Vicary, Benedikt Ahrens, Fredrik Nordvall Forsberg, Jonathan Weinberger, Niels van der Weide, Alex Kavvos, Conor McBride, Andrew Pitts, Andreas Abel, Andrea Vezzosi, András Kovács, Anders Mörtberg, Simon Huber, Steve Awodey, Emily Riehl, Edwin Brady, and the many others whom I forgot to mention.

I am honoured to be a recipient of the DPST scholarship (The Development and Promotion of Science and Technology Talents Project), sponsored by the Thai government. This scholarship has been allowing me to study abroad since 2011, including bachelor, master, and doctorate degrees. I am grateful for many people who involve in this scholarship and make the possibility for my study in the United Kingdom.

During the thesis pending period, my career started at SiData+ (Siriraj Informatics and Data Innovation Center, Faculty of Medicine Siriraj Hospital, Mahidol University). Here, I would like to thank Prapat Suriyaphol (director of SiData+) and Watcharaporn Tanchotsrinon (data scientist at SiData+) for their voluntary and professional support, despite their having PhD degrees in different academic disciplines.

Finally, I would like to thank my parents and my girlfriend: Tawatchai Pinyo, Amnouy Pinyo, and Pawinee Chuawong, for their unconditional love, encouragement, and support throughout this journey of my PhD.

# Contents

Chapter

# 1

# Introduction

## § 1.1 Motivation and Overview

*Twisted cubes* was born from a motivation to fix the incompatibility issue between two of the most important sub-disciplines in *homotopy type theory*, which are *cubical type theory* and *directed type theory*:

- *Cubical type theory* interprets cubical sets (together with at least certain model structures) as $\infty$-groupoids; opening a new perspective of an equality type (a.k.a. path type) of some ambient type $A$ as a function from the interval type to $A$ (see section 2.6 for further detail).
- *Directed type theory* generalises the interpretation of types from $\infty$-*groupoids* to $\infty$-*categories*; in other words, it removes the constraint on the equality types to not necessarily be invertible types (see section 2.5 for further detail).

The incompatibility issue between these two sub-disciplines arises from the fact that cubical sets, when combined with the Kan filler condition, can produce an invertible counterpart for all paths. This "built-in" invertibility, although it is one intended interpretation for many areas of homotopy theory theory, is not desirable for directed type theory, e.g. in concurrency where a non-invertible entity such as time plays a vital role.

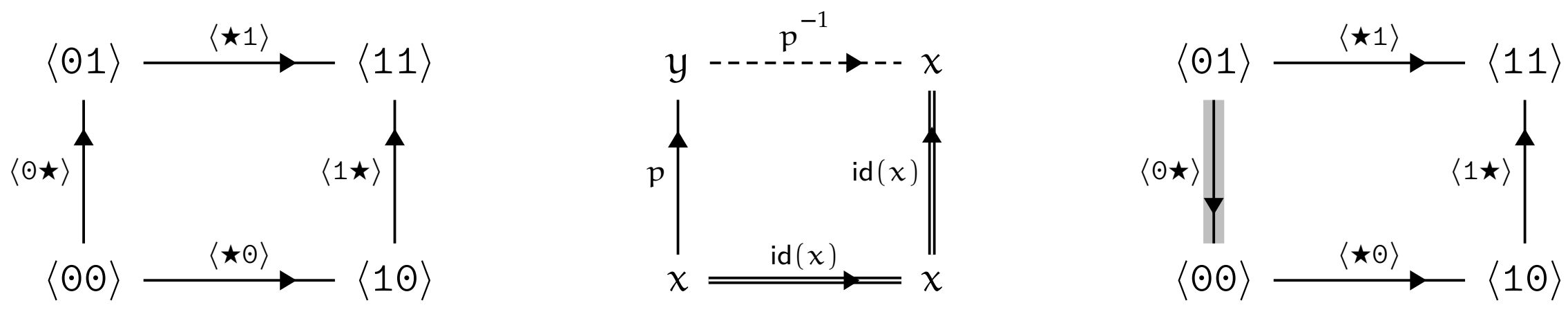

Figure 1.1: Standard 2-cube (left) has a built-in invertibility (middle). This can be removed by reversing the arrow ⟨0★⟩ and then become twisted 2-cube (right).

To really understand the problem, we need to know what *cubes* (in section 3.1) and *the Kan filler condition* (in subsection 3.1.7) really are; but for now, it is suffice to use the left square in figure 1.1 to illustrates a 2-dimensional cube; here, the Kan filler condition states that if we have a square with exactly one side missing, we can use the condition to find the missing one and fill the inside.

The middle square in figure 1.1 shows us that, given a path $p$ from point $x$ to point $y$, we can always find an inverse of $p$, denoted as $p^{-1}$, by assigning ⟨0★⟩, ⟨★0⟩, and ⟨1★⟩ with $p$, id($x$), and id($x$), respectively; then, the missing one is ⟨★1⟩, which we can use the Kan filler condition to find and assign it to be $p^{-1}$. Actually, $p^{-1}$ can be produced regardless of what square side is used as the missing one. Please see subsection 3.2.1 for a deeper explanation of this built-in invertibility.

To tackle this issue, we forcefully reverses ⟨0★⟩ from the right square in figure 1.1 upside down and restrict the Kan filling condition to accept only ⟨1★⟩ as the missing face. We call this modified square a *twisted* 2-*cube*, which invalidate the problematic method in the previous paragraph. In addition, the arrangement of ⟨0★⟩, ⟨★0⟩, and ⟨1★⟩ in the right square (i.e. twisted 2-cube) in figure 1.1 really resembles how composition in category theory works. Section 1.3 will later discuss about this.

This intuition of a twisted 2-cube, as a modification of a "standard" 2-cube, may seems artificial and specific to the case of 2 dimensions, but the number of dimensions can indeed be generalised to arbitrary natural number $n$. In other words, we can define a *twisted* $n$*-cube*, which is denoted as $\mathbb{I}^{n}_{\bowtie}$, as a modification of a *standard* $n$*-cube*, which is denoted as $\mathbb{I}^{n}_{\square}$.

We informally construct $\mathbb{I}^{n}_{\bowtie}$ by recursion on the natural number $n$. For the base case, we define $\mathbb{I}^{0}_{\bowtie}$ as $\mathbb{I}^{0}_{\square}$, which is just the unique point in 0-dimensional space.
For the recursive case, we use the process called *thickening-and-twisting* that transforms $\mathbb{I}^{n}_{\bowtie}$ into $\mathbb{I}^{(n+1)}_{\bowtie}$ as follows:

- The *thickening* phase — we prepend a new dimension as the first dimension and expand $\mathbb{I}^{n}_{\bowtie}$ along this new dimension to get $(\mathbb{I} \times \mathbb{I}^{n}_{\bowtie})$; this thickening phase is actually the same as constructing $\mathbb{I}^{(n+1)}_{\square}$ from $\mathbb{I}^{n}_{\bowtie}$, i.e. its *cylinder object*.
- The *twisting* phase — we reverse every direction in all other dimensions at the starting point of the new dimension.

Figure 1.2 shows how this process transforms $\mathbb{I}^{0}_{\bowtie}$ to $\mathbb{I}^{1}_{\bowtie}$ then $\mathbb{I}^{2}_{\bowtie}$ and then $\mathbb{I}^{3}_{\square}$, iteratively.

Every $n$-cube (either twisted or standard) will have $(n-1)$-cubes enclosed as its boundary that are called *facets* (see definition 3.5). We use a notation ${}^{b}\partial^{n}_{r}$ to represent the facet of an $n$-cube located at where the value at dimension $r$ is $b$ where $(0 \leqslant b \leqslant 1)$ and $(0 \leqslant r < n)$ . For example, figure 1.3 shows all possible facets with $n \leqslant 3$.

One important property of standard cubes which twisted cubes retain is that every facet of a twisted $n$-cube is a twisted $(n-1)$-cube (see subsection 3.2.4). An interesting example is the construction of $\mathbb{I}^{3}_{\bowtie}$ in figure 1.2 where the left and right facets of $\mathbb{I}^{3}_{\bowtie}$ are already $\mathbb{I}^{2}_{\bowtie}$ in thickening phase whereas the rest are $(\mathbb{I} \times \mathbb{I}^{1}_{\bowtie})$; the twisting phase doesn't affected the right facet but reversed the left facet entirely (yet it is still $\mathbb{I}^{2}_{\bowtie}$).

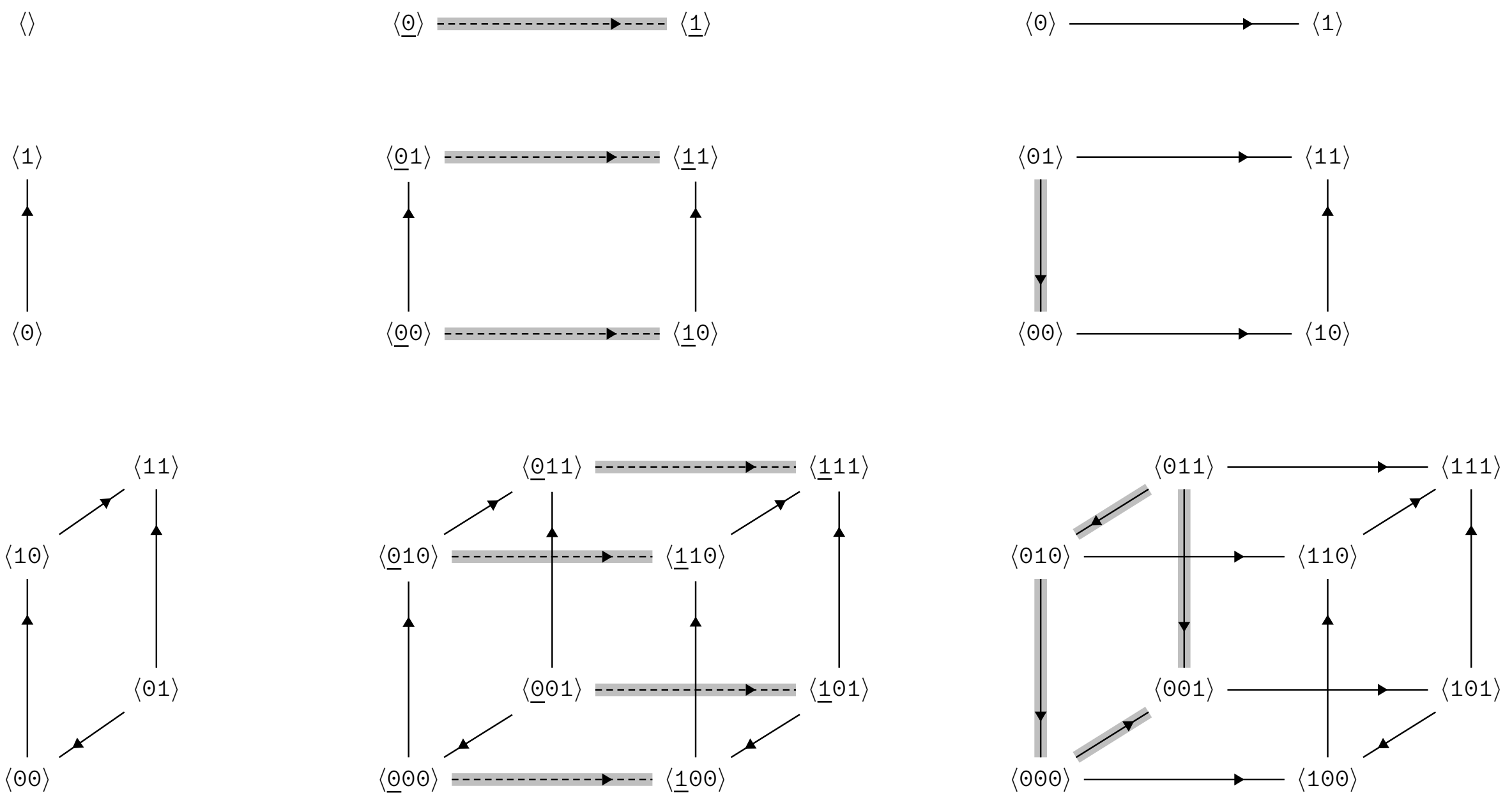

Figure 1.2: The *thickening-and-twisting* process for $(n \leqslant 2)$; where the first column ($\mathbb{I}^n_{\bowtie}$) is *thicken* into the next one ($\mathbb{I} \times \mathbb{I}^n_{\bowtie}$), which is then *twisted* into the last one ($\mathbb{I}^{(n+1)}_{\bowtie}$).

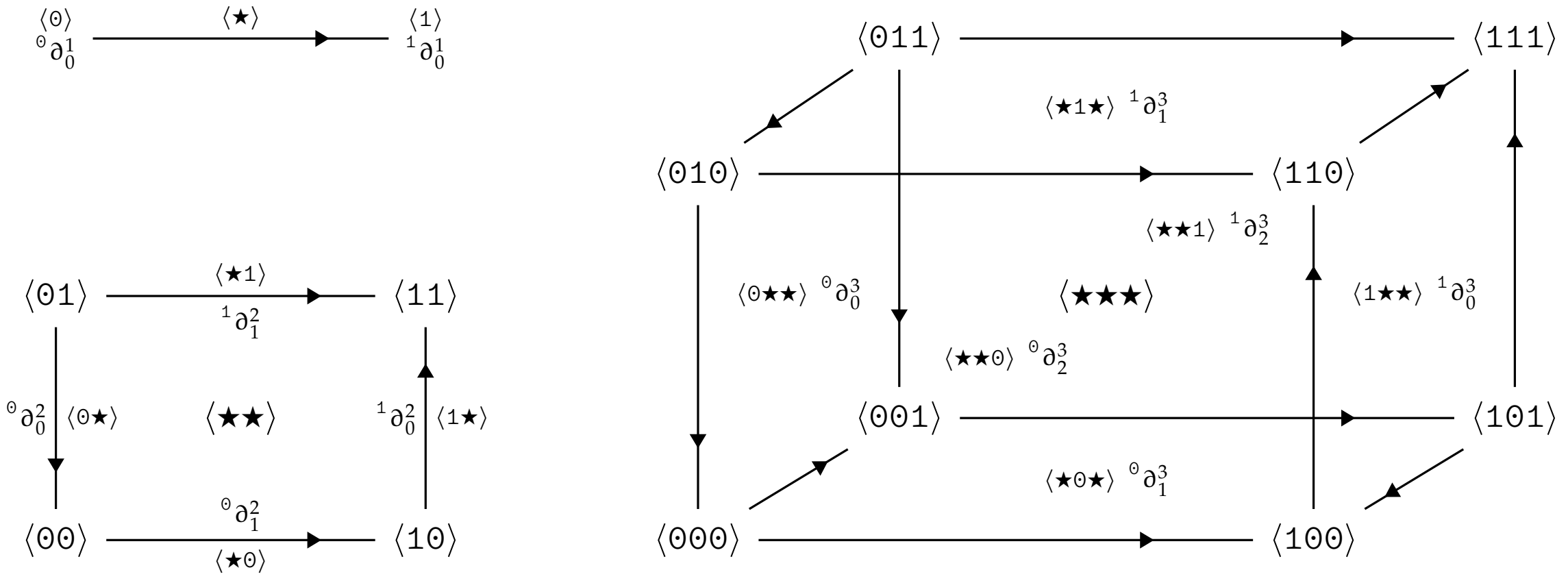

Figure 1.3: Illustration of $\mathbb{I}^1_{\bowtie}$, $\mathbb{I}^2_{\bowtie}$, and $\mathbb{I}^3_{\bowtie}$ annotated with ternary numbers and ${}^b\partial^n_r$.

## § 1.2 Related Work

Most of related references will be cited along chapter 2, which is about prerequisite disciplines; in particular, references in directed type theory (section 2.5) and cubical type theory (section 2.6). In order to avoid repeated citation, this section will only mention the related references that haven't been cited before (or will be cited along in chapter 2).

When I first came up with this idea of twisted cubes, I tried to search though the literature to assure that I had not coincidentally rediscovered someone else's idea. Although the idea itself did not exist before, the "twisted square" pattern does appear elsewhere in other usages, i.e. in the fundamental polygon of the Klein Bottle [Hat02, section 2.1, page 102] and the *twisted arrow categories* [Law70] (a.k.a. the *categories of factorisations* [BW85]). However, it is unclear how to generalise these ideas to more than squares; but that is not a problem because they have been developed to solve different problems; for example, there is a notion of *twisted systems* [Err99], that uses the twisted arrow categories to model categorical transition systems.

Prior to this thesis, the idea of twisted cubes has been published as a paper [PK20], with the title "From Cubes to Twisted Cubes via Graph Morphisms in Type Theory" (co-authored with Nicolai Kraus), in the post-proceedings of TYPES2019 (an international conference dedicated to type theory and its related fields). This paper is indeed used to form a backbone of this thesis. Beside the main paper, the idea of twisted cubes has been further published as two extended abstracts. The first extended abstract [PK19] was accepted to TYPES2019 under the title "Twisted Cubes" (co-authored with Nicolai Kraus), which is a sketch in preparation of the main paper above. The second extended abstract [Pin21] was accepted to TYPES2021 under the title "Interpreting Twisted Cubes as Partially Ordered Spaces". The main concept with the relevant perspectives of these three papers is properly utilized in order to improve the idea of twisted cubes proposed in this thesis.

Last but not least, I would like to reference two extended abstracts that I did during this PhD but prior to the discovery of the twisted cubes. This might not be directly related to twisted cubes but it was important for me to understand type theory anyway. The first extended abstract [AP17] was accepted to TYPES2017 under the title "Monadic Containers and Universes" (co-authored with Thorsten Altenkirch, my supervisor), which explores the relationship between monadic containers [Abo+03; AAG05] and type theoretic universes that are closed under sigma-types and unit-types. The second extended abstract [AP18] was accepted to TYPES2018 under the title "Integers as a Higher Inductive Type" (again, co-authored with Thorsten Altenkirch, my supervisor), which applies the feature of quotient inductive types [AK16] to modify the canonical type of natural numbers to be a type of integers by further stating that the constructor $\texttt{succ} : \mathbb{N} \to \mathbb{N}$ is an equivalence.

## § 1.3 Application to (Higher) Category Theory

Please note that, this section enforces convention 3.3 (first index as zero).

Twisted cubes do not only remove the discussed source of invertibility, but they also change the way we view the composition of morphisms. The filling of a standard 2-cube can be interpreted as the statement "the composition of two edges equals the composition of the other two edges" (see figure 3.8); if we want to see the lid as the composition of the three other edges, then one of them has to be inverted. In contrast, the lid of a twisted 2-cube can directly be seen as the single composition of the three other edges (see the right digram of figure 1.1).

**Embedding Categorial Composition:** In category theory, to compose any morphism $f$ to morphism $g$ in some category $\mathcal{C}$, it is necessary that the codomain of $f$ must be strictly equal to the domain of $g$. This equality is as strict as the associativity and unital laws; however, when generalising a category into a weak higher category, we never relax this equality into an equivalence in the same way that associativity and unital laws become associator and unitors. Sure, this relaxation hasn't had any practical motivations, but if someone want to try this, then the shape of composition will change from triangle to twisted 2-cube as figure 1.4 illustrates.

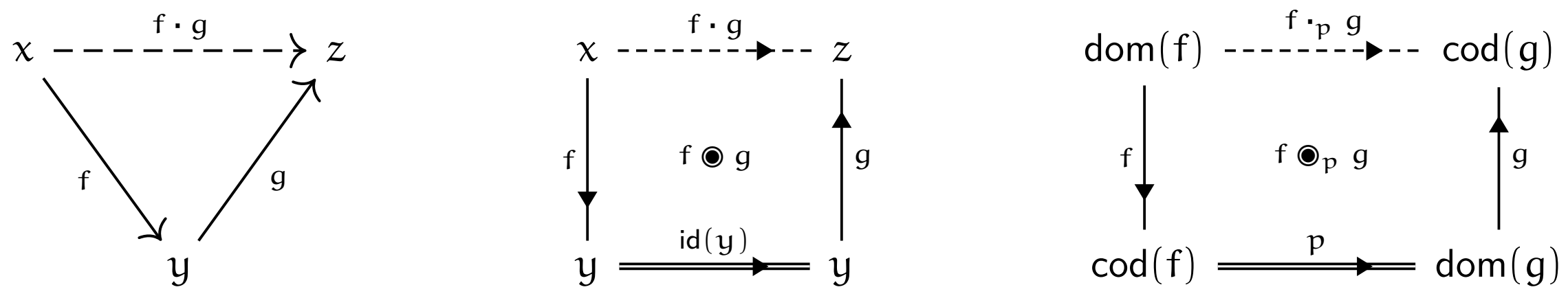

Figure 1.4: Generalising categorical composition to a twisted 2-cube.

This relaxation allows us to think that the composition is a (meta) function that takes any pair of morphisms $f$ and $g$ then returns another function that takes an equivalence between $\mathrm{cod}(f)$ and $\mathrm{dom}(g)$ then returns a morphism from $\mathrm{dom}(f)$ to $\mathrm{cod}(g)$.

$$(\boldsymbol{\lambda}(p) \;\Rightarrow\; f \cdot_p g) \;:\; \mathrm{equiv}(\,\mathrm{cod}(f)\,,\,\mathrm{dom}(g)\,) \;\rightarrow\; \mathrm{morph}(\,\mathrm{dom}(f)\,,\,\mathrm{cod}(g)\,)$$

We also use a notation $(f \circledcirc_p g)$ as the filler that has $(f \cdot_p g)$ as its lid. Please note that, unlike the traditional definition of composition, this (meta) function doesn't require an instantiation of the objects that are domains/codomains of the inputting morphisms, which is similar to how a path is encoded in existing models of cubical type theory.

**Embedding 2-Cell in Twisted 2-Cube:** Twisted 2-cube isn't only a target to generalise a composition but also a target to generalise 2-cell, i.e. morphism between morphisms; to construct a 2-cell $\alpha$ from morphism $f$ to morphism $g$, we need to ensure that the domain and codomain of $f$ are strictly equal to $g$ counterpart. These strictly equalities can be relaxed in same way that we do for the composition as illustrated in figure 1.5.

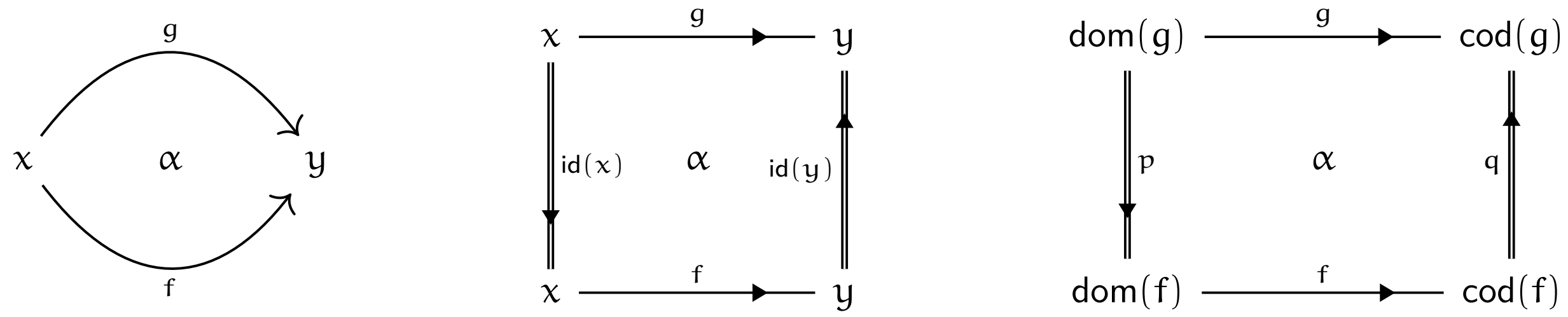

Figure 1.5: Generalising categorical 2-cell to a twisted 2-cube.

This relaxation allows a 2-cell from $f$ to $g$ to be seen as $\langle\, p \,,\, q \,,\, \alpha \,\rangle$ such that

- $p$ is an equivalence between $\mathrm{dom}(f)$ and $\mathrm{dom}(g)$,
- $q$ is an equivalence between $\mathrm{cod}(f)$ and $\mathrm{cod}(g)$, and
- $\alpha$ is a filler with the boundary $f$, $g$, $p$, and $q$.

Please note that, unlike the traditional definition of 2-cell, the "hom-set" from $f$ to $g$ doesn't require $f$ and $g$ to have the same domain and codomain, we just need to find equivalences $p$ and $q$ to prepend a 2-cell. This fits well with how cubical type theories construct a 2-path, i.e. they construct an arbitrary standard 2-cube first then assign its domain and codomain later.

<u>**Embedding n-Cell in Twisted n-Cube:**</u> The generalisation from a 2-cell to a twisted 2-cube can be lifted as the generalisation from an $n$-cell to a twisted $n$-cube; for example, figure 1.6 shows how a 3-cell can be generalised to a twisted 3-cube.

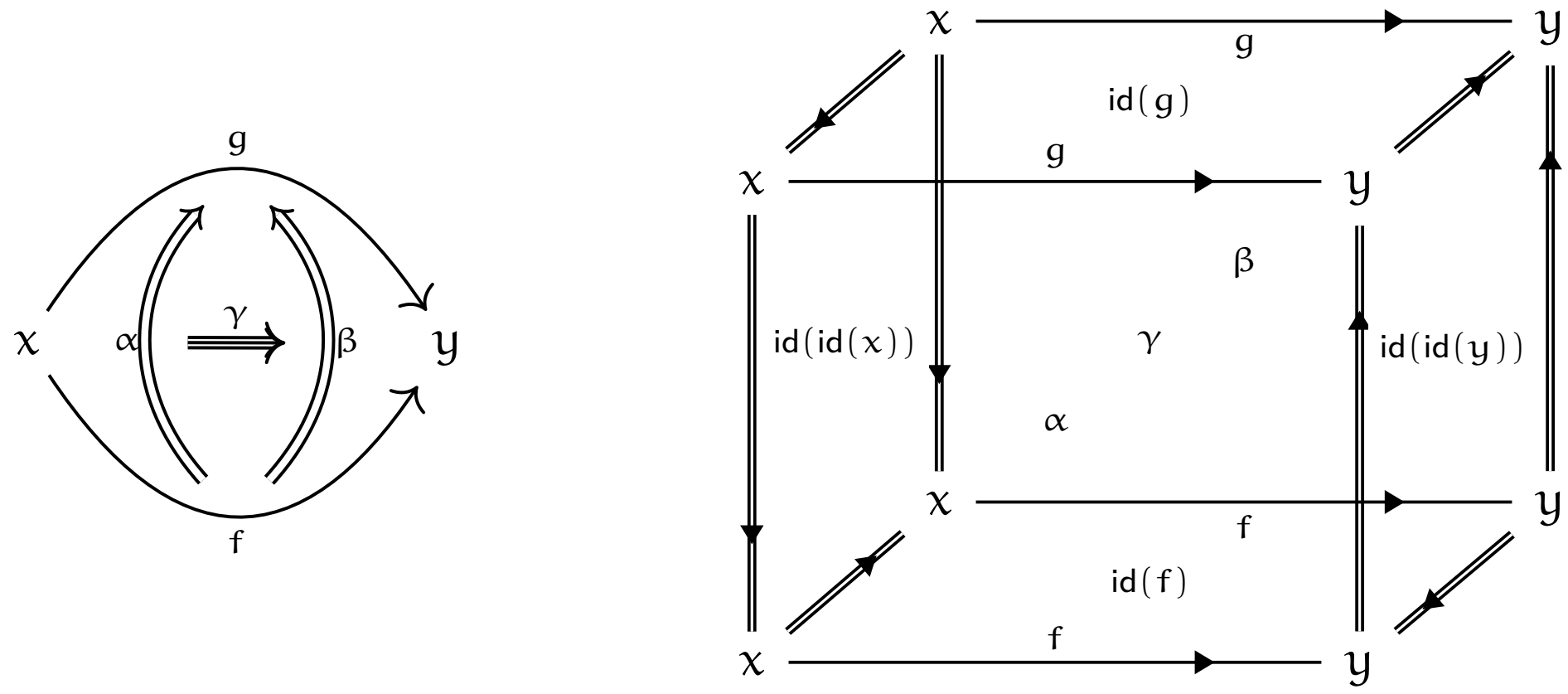

Figure 1.6: Generalising categorical 3-cell to a twisted 3-cube.

An interesting observation in figure 1.6 is that $\alpha$ and $\beta$ must use facets ${}^{1}\partial^{3}_{0}$ and ${}^{1}\partial^{3}_{1}$ because only at dimension 0 that both facets has the same orientation. This is also true when $n :\equiv 2$ in figure 1.5; in fact, for any twisted $n$-cube, the facet ${}^{\ominus}\partial^{n}_{r}$ will have the same orientation as the facet ${}^{1}\partial^{n}_{r}$ iff $(r = 0)$; the proof can be easily deduced from subsection 3.2.5. Please note that, this is not the case for standard $n$-cubes.

This leads me to the following conjecture.

◤ **Conjecture 1.1** ⬣ Let $f$ and $g$ be $(n-1)$ cells in some higher category and let $\alpha$ be an $n$-cell from $f$ to $g$; then, $\alpha$ can be embed as a twisted $n$-cube such that

- its facet ${}^{0}\partial_0^n$ is assigned to $f$,
- its facet ${}^{1}\partial_0^n$ is assigned to $g$, and
- the rest of facets are the canonical choices of degenerated $(n-1)$-cell determined by the shared boundaries with $f$ and $g$.

The embedding above can be further generalised as:

- if $f$ and $g$ don't share the same domains or codomains,
- there is a $(n-1)$-cell named $p$ as an equivalence from $\mathsf{dom}(f)$ to $\mathsf{dom}(g)$,
- there is a $(n-1)$-cell named $q$ as an equivalence from $\mathsf{cod}(f)$ to $\mathsf{cod}(g)$, and
- $\alpha$ is a generalised $n$-cell from $f$ to $g$ with $p$ and $q$ as boundaries (similar to the right diagram of figure 1.5);

then, $\alpha$ can also be embed as a twisted $n$-cube such that

- its facet ${}^{0}\partial_0^n$ is assigned to $f$,
- its facet ${}^{1}\partial_0^n$ is assigned to $g$,
- its facet ${}^{0}\partial_1^n$ is assigned to $p$,
- its facet ${}^{1}\partial_1^n$ is assigned to $q$, and
- the rest of facets are the canonical choices of degenerated $(n-1)$-cell determined by the shared boundaries with $f$, $g$, $p$, and $q$.

Moreover, if $p$ and $g$ don't share the same domains or codomains
but there are equivalences with their respective domains and codomains,
then those equivalences will be assigned as ${}^{0}\partial_2^n$ and ${}^{1}\partial_2^n$.
This iterative process can be repeated until the facets
at dimension $(n-1)$ are assigned. ◢

**Compositing Higher Twisted Cubes:** Recall that the composition for twisted 1-cubes can be seen as a construction of a twisted 2-cube by a Kan-filling condition that restrict the facet ${}^1\partial^2_1$ as the only possible lid. Similarly, the composition for twisted 2-cubes can be seen as a construction of a twisted 3-cube by a Kan-filling condition. However, there are 2 ways to compose twisted 2-cubes:

The first way uses the facet ⟨★1★⟩ as the lid; imagine that we look into $\mathbb{I}^3_{\bowtie}$ in figure 1.3 from the above through the facet ⟨★1★⟩, this results in the perspective projection shown as the left diagram in figure 1.7. The following composition algorithm transforms the path ⟨011⟩-⟨010⟩-⟨110⟩-⟨111⟩ to ⟨011⟩-⟨111⟩ using other five facets.

⟨011⟩-⟨010⟩-⟨110⟩-⟨111⟩
⇒ ⟨011⟩-⟨010⟩-⟨000⟩-⟨100⟩-⟨110⟩-⟨111⟩ [using ⟨★★0⟩ backwardly]
⇒ ⟨011⟩-⟨010⟩-⟨000⟩-⟨001⟩-⟨101⟩-⟨100⟩-⟨110⟩-⟨111⟩ [using ⟨★0★⟩ backwardly]
⇒ ⟨011⟩-⟨001⟩-⟨101⟩-⟨100⟩-⟨110⟩-⟨111⟩ [using ⟨0★★⟩ forwardly]
⇒ ⟨011⟩-⟨001⟩-⟨101⟩-⟨111⟩ [using ⟨1★★⟩ forwardly]
⇒ ⟨011⟩-⟨111⟩ [using ⟨★★1⟩ forwardly]

The second way uses the facet ⟨★★1⟩ as the lid; imagine that we look into $\mathbb{I}^3_{\bowtie}$ in figure 1.3 from the back through the facet ⟨★★1⟩, this results in the perspective projection shown as the right diagram in figure 1.7. The following composition algorithm transforms the path ⟨011⟩-⟨001⟩-⟨101⟩-⟨111⟩ to ⟨011⟩-⟨111⟩ using other five facets.

⟨011⟩-⟨001⟩-⟨101⟩-⟨111⟩
⇒ ⟨011⟩-⟨010⟩-⟨000⟩-⟨001⟩-⟨101⟩-⟨111⟩ [using ⟨0★★⟩ backwardly]
⇒ ⟨011⟩-⟨010⟩-⟨000⟩-⟨001⟩-⟨101⟩-⟨100⟩-⟨110⟩-⟨111⟩ [using ⟨1★★⟩ backwardly]
⇒ ⟨011⟩-⟨010⟩-⟨000⟩-⟨100⟩-⟨110⟩-⟨111⟩ [using ⟨★0★⟩ forwardly]
⇒ ⟨011⟩-⟨010⟩-⟨110⟩-⟨111⟩ [using ⟨★★0⟩ forwardly]
⇒ ⟨011⟩-⟨111⟩ [using ⟨★1★⟩ forwardly]

This pattern of composition leads me to the following conjecture.

◤ **Conjecture 1.2** ⬣ There should be $n$ ways to compose twisted $n$-cubes. Each way $r$, for all $(0 \leqslant r < n)$, will construct a twisted $(n+1)$-cube by a Kan-filling condition that uses the facet ${}^{1}\partial^{n+1}_{r+1}$ as its lid. ◢

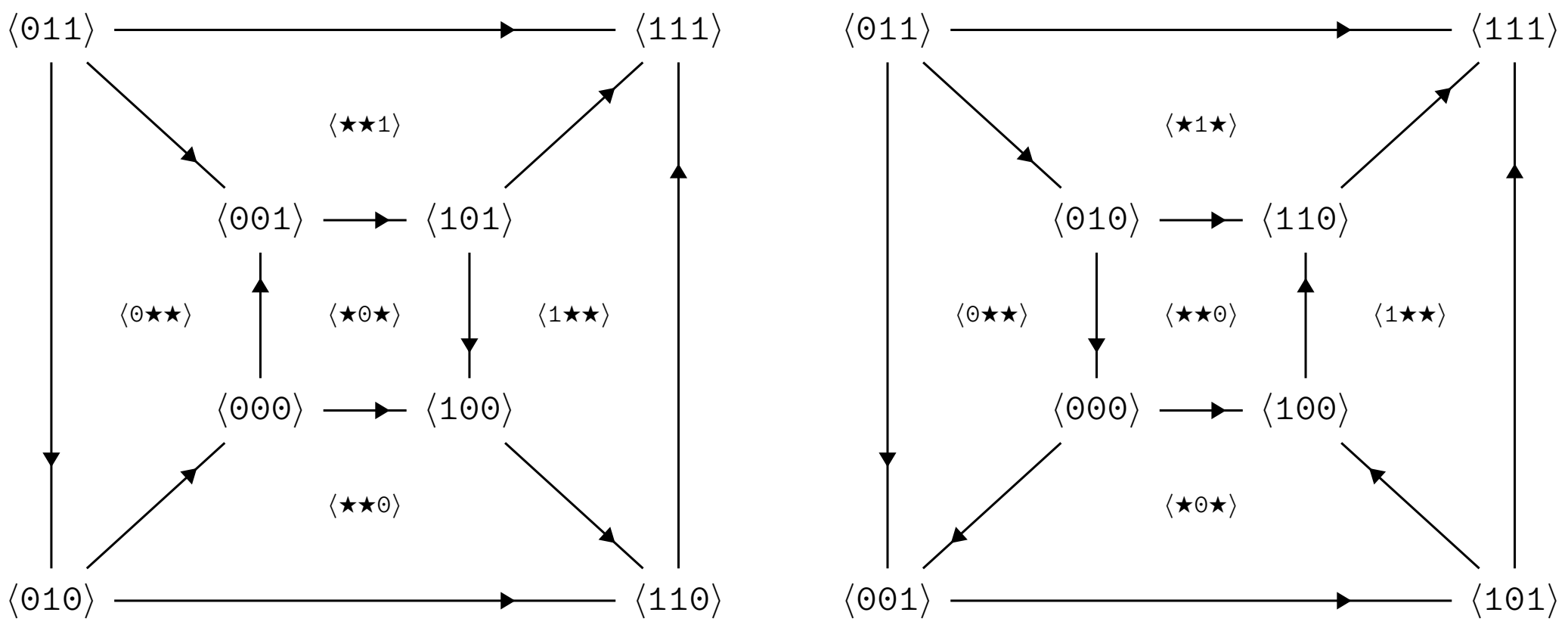

Figure 1.7: Looking $\mathbb{I}^3_{\bowtie}$ from $\langle \star 1 \star \rangle$ downward (left) and from $\langle \star\star 1 \rangle$ outward (right).

<u>**Embedding Higher Composition:**</u> Surprisingly, the first way to compose twisted 2-cubes generalises the horizontal composition between 2-cells whereas the second way generalises the vertical composition; please see figure 1.8 for their illustrations.
This observation lead me to conjecture 1.3.

◤ **Conjecture 1.3** ⬣ Each way $r$, for all $(0 \leqslant r < n)$, to compose twisted $n$-cubes discussed in conjecture 1.2 should generalises each corresponded way to compose $n$-cells in a higher category. For example:

- The first (and only) way to compose twisted 1-cubes is with the 1-composition.
- The first way to compose twisted 2-cubes is with the horizontal 2-composition.
- The second way to compose twisted 2-cubes is with the vertical 3-composition.

To be more precise, let $\alpha$ and $\beta$ be a pair of (way-r) composable n-cells then there is a twisted $(n+1)$ cube such that

- its facet ${}^{0}\partial_{r}^{n+1}$ is assigned to the twisted n-cube that embedded $\alpha$,
- its facet ${}^{1}\partial_{r}^{n+1}$ is assigned to the twisted n-cube that embedded $\beta$,
- its facet ${}^{1}\partial_{r+1}^{n+1}$ is assigned to the twisted n-cube that embedded the result of way-r composition from $\alpha$ to $\beta$, and
- the rest of facets are the canonical choices of degenerated $(n-1)$-cell determined by the shared boundaries with $\alpha$, $\beta$, and the result of composition. ◢

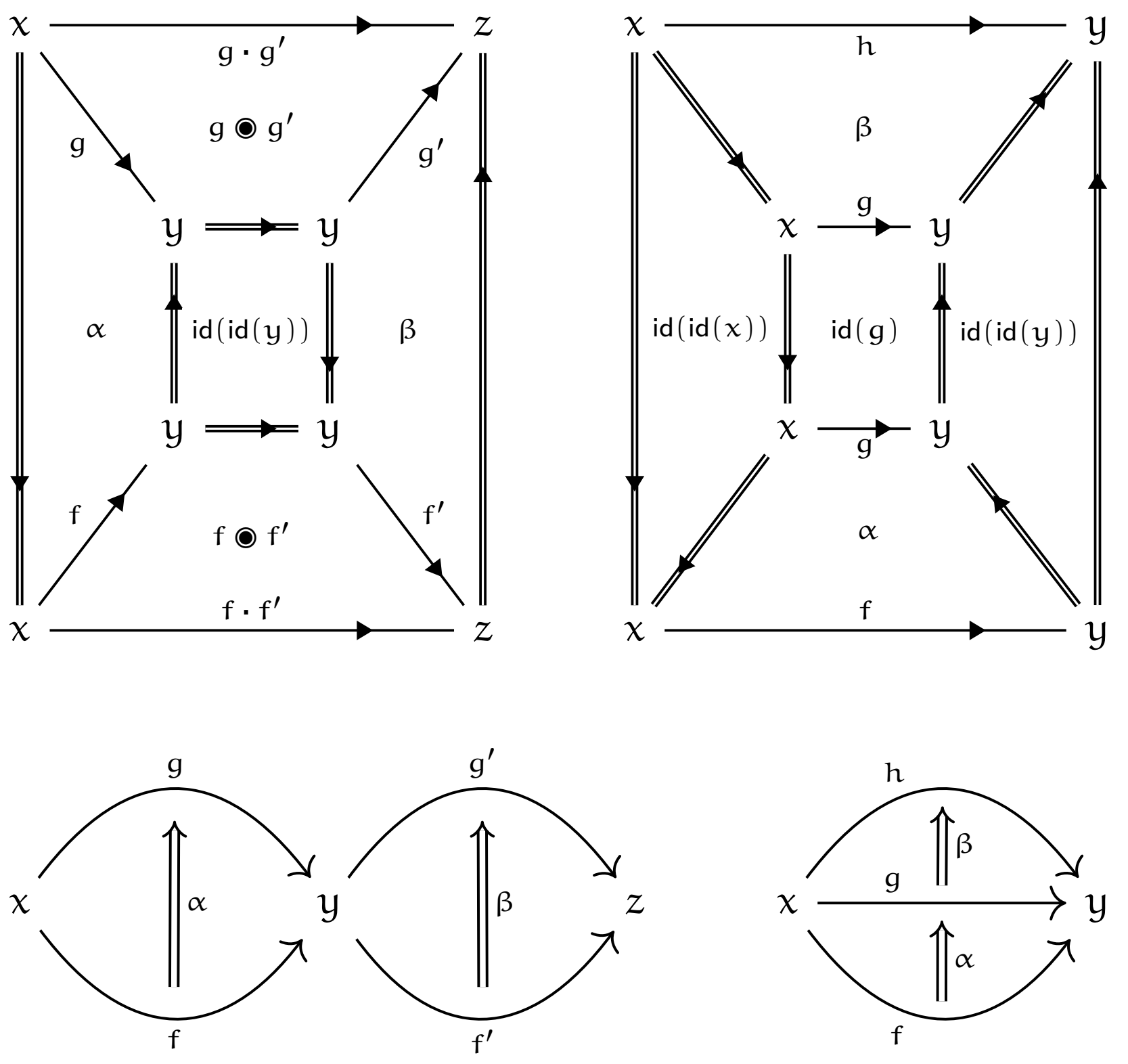

Figure 1.8: Generalising horizontal (left) and vertical (right) compositions.

Chapter

# 2

# Prerequisite Disciplines

In order to make the main ideas and concepts proposed in this thesis easier to understand and accessible to wider audience, this chapter introduces the important fundamental knowledge regarding twisted cubes.

Please note that, this chapter also take this opportunity to establish further non-standard conventions as well as review some more relevant literature.

## § 2.1 Dependent Type Theory

*Dependent type theory* (or just *type theory* for short) is a discipline that lies between pure mathematics and theoretical computer science.

### § 2.1.1 From a Mathematical Perspective

Type theory is the study of formal systems, which are also called *type theories*, that can be used as foundations of mathematics alternative to set theories. The notable example of type theories is the Martin-Löf intensional type theory (or MLTT for short) [Martin-Löf72; Martin-Löf73; Martin-Löf79; Martin-Löf84], which is the original dependent type theory proposed by Per Erik Rutger Martin-Löf.

The most important entities in any type theory are called *types*. A type will have zero or more associated entities called *terms*. We use the form of judgements

$$\mathtt{t} : A$$

to denote that a term $\mathtt{t}$ belongs to a type $A$. Intuitively, a type can be interpreted as a set where each of its terms will be interpreted into an element of the set correspond to the original type. In fact, many sets in *Zermelo-Fraenkel set theory*, which is the most commonly accepted standard foundation of mathematics, can be translated to a type theory. This allows most of the mathematical concepts that have been formalised in (axiomatised) set theories, to be naturally formalised in type theories; however, this process is non-trivial because sets and types are built by different concepts. Please see [HoTT, Section 1.1] for more detail.

In addition to the "set-interpretation" from the paragraph above, a type can also be interpreted as a proposition where any of its terms, if it exists, serves as "evidence" that the corresponding proposition is valid. In fact, many propositions in most logical systems, such as propositional or first-order logic, can be translated to a type theory[1]. One immediate and important benefit of this interpretation is a new technique to prove validity of a proposition. This technique encodes a proposition as a type, then try to find one of its term; if successful, then that term will serve as evidence that the original proposition is valid.

[1]Only if the propositional truncation is included as a type former.

The ability of a type theory to interpret a type as a set and a proposition simultaneously gives a type theory an advantage over a set theory. This is because a set theory needs another logical system, such as first-order logic, to reason about sets. In contrast, a type theory can use itself[2] to reason about types, which removes duplications in many places in formalisation and even allows us to reason about propositions themselves.

◤ **Convention 2.1** **(meta theory of this thesis)** ⬣

The entire thesis, including this chapter, will use the type theory[3] defined in [HoTT, Chapter 1] as the underlining foundation with modification explained in later conventions. Nevertheless, this meta theory is only for the completeness propose and this thesis is still aiming to be self-contained as much as possible; so readers who are unfamiliar with type theory should be able to follow the thesis anyway.

One particular notation that should be mention here anyway is the difference of between notations ( $:\equiv$ ) and $(=)$ where $x :\equiv t$ is a judgement stating that a variable $x$ is assigned to be a term $t$ whereas $(a = b)$ is a type of proofs that a term $a$ is propositionally equal to a term $b$.

The reader may intuitively use the traditional set-theoretic foundation for this thesis, but please keep in mind that the meaning of a set is redefined to be a type that contains no further information than being a collection of its elements. This is always the case until we introduce other type theories, e.g. homotopy type theory in section 2.4, which state that a type represents an $\infty$-groupoid, which is more general than a set. ◢

[2]Technically, a type theory can reason only a fraction of itself, otherwise, it will have a Russell-like paradox. To fully reason about the original type theory, we need another type theory that is large enough to encode the original type theory.

[3]Our type theory is not necessary HoTT because we use only chapter 1 where all HoTT concepts haven't been introduced.

### § 2.1.2 From a Programming Perspective

Type theory is also the study of certain (functional) programming languages, which allows programmers to reason about their program within the same language. The core idea behind this is called the *Curry-Howard correspondence* [How80] (a.k.a. *propositions as types*) stating that there is one-to-one correspondence between mathematical objects and computer programs. Therefore, a mathematical proof can be recursively translated into a computer program, and vice versa. For example:

- Logical conjunctions $(\wedge)$ correspond to Cartesian products $(\times)$:
  The statement "proposition $(\mathsf{P} \wedge \mathsf{Q})$ is valid if and only if both propositions $\mathsf{P}$ and $\mathsf{Q}$ are vaild" is similar to statement "a term of type $(\mathsf{P} \times \mathsf{Q})$ is constructed from terms of types $\mathsf{P}$ and $\mathsf{Q}$".
- Logical implications $(\rightarrow)$ correspond to functions $(\rightarrow)$:
  The statement "if proposition $(\mathsf{P} \rightarrow \mathsf{Q})$ is valid then the validity of $\mathsf{Q}$ can be derived from the validity of $\mathsf{P}$" is similar to statement "with a function $\mathsf{f}$ of type $(\mathsf{P} \rightarrow \mathsf{Q})$, a term of type $\mathsf{Q}$ can be constructed from a term of type $\mathsf{P}$ by applying it to $\mathsf{f}$". This is also the reason why we can overload the symbol $(\rightarrow)$ for both logical implication and function without any ambiguity.

Using the fact that every type theory is both a foundation of mathematics and a programming language, one can build a *proof assistant* (a.k.a. *interactive theorem prover*), which is a compiler of the specific programming language that allows mathematicians to *formalise* their mathematical objects by writing them as computer programs. If those programs are successfully compiled, then the original mathematical objects are well defined. By doing a formalisation, mathematicians can be confident that their theorems are consistent, at least up to the consistency of the underline type theory itself, and there are no human errors to be concerned with. The concept of mathematical formalisation is an active area of research and there are several well-known proof assistants such as: Coq [Coq], Agda [Agda; AgdaD], Idris [Idris], and Lean [Lean].

The paragraph above explains how mathematics benefits from computer science; conversely, computer science also benefits from mathematics because programmers can use a proof assistant to verify that their programs respect certain laws of its specification. The process is: First, programmers treat a proof assistant as an ordinary programming language and write their programs in it. Then, convert some laws (i.e. proposition) of its specification into types by the Curry-Howard correspondence. Finally, find a term for each of those propositions and compile them. If successful, those programs are guaranteed to respect those specifications (up to the consistency of the type theory behind each compiler).

Formal verification requires a high learning curve from programmers and takes an excessive amount of time just to prove some easy proposition. This is infeasible for most areas in software engineering where "absolute correctness" are not critical and testing is good enough. Nevertheless, it starts to have some application on some software that bugs are unacceptable, e.g. software in astronautics, military, finance, or medicine.

## § 2.2 Category Theory

*Category theory* is a discipline in pure mathematics that formalises mathematical structures using *categories*. It was originally developed by Samuel Eilenberg and Saunders Mac Lane [EM45] as a tool to study algebraic topology. Subsequently, mathematicians discovered strong connections from category theory to many other branches in mathematics and related disciplines such as computer science and physics. Nowadays, category theory is mainly used as a tool to abstract away and unify many concepts among different academic disciplines. Due to its accumulating complexity, category theory became an important discipline of its own right. There are several text books that introduce the subject including a book: *Categories for the Working Mathematician* [Mac98] by Saunders Mac Lane himself. There is even an encyclopedia [nLab] dedicated to category theory.

Category theory has a profound and substantial connection with type theory. In particular, the Curry-Howard correspondence can be extended into the *Curry-Howard-Lambek correspondence* which is a three-way one-to-one correspondence between mathematical objects, computer programs, and categorical entities. For instance, logical conjunctions and Cartesian products correspond to categorical entities called *products* while logical implications and functions correspond to categorical entities called *exponential objects*. In fact, one could map an entire structure of any type theory to a categorical entity called a *category with families* [CCD19]. Conversely, we can assign a categorical semantic of a type theory by defining its model using a category with families.

**Definition of Categories:** The main idea of category theory is about entities called *categories* that consist of arrows called *morphisms* that can be composed with one another together with associativity and unitality.

◤ **Definition 2.2** ⬣ A record $\mathcal{C}$ is called a *category*, if it contains these fields:

- $\mathcal{C}.\mathsf{obj}$ — a set of things called *objects*, which usually abbreviated as $\mathsf{ob}(\mathcal{C})$.
- $\mathcal{C}.\mathsf{hom}(X, Y)$ — a set of arrows called *morphisms* from an object $X : \mathsf{ob}(\mathcal{C})$ to another object $Y : \mathsf{ob}(\mathcal{C})$, usually abbreviated as $\mathcal{C}(X, Y)$, a.k.a. *hom-set.*
- $\mathcal{C}.\mathsf{comp}$ — a function called *composition* that takes two composable morphisms $f : \mathcal{C}(X, Y)$ and $g : \mathcal{C}(Y, Z)$ then combines them as another morphism

  $$\mathcal{C}.\mathsf{comp}(f, g) \quad : \quad \mathcal{C}(X, Z), \quad \text{which usually abbreviated as} \quad (f \cdot g),$$

  together with another field $\mathcal{C}.\mathsf{assoc}$ called the *associativity* law stating that

  $$\mathcal{C}.\mathsf{assoc}(f, g, h) \quad : \quad f \cdot (g \cdot h) \;=\; (f \cdot g) \cdot h,$$

  i.e. the order of composition doesn't matter, both sides can be written as $f \cdot g \cdot h$.
- $\mathcal{C}.\mathsf{ident}$ — a function that takes $X : \mathsf{ob}(\mathcal{C})$ then return an *identity* morphism

  $$\mathcal{C}.\mathsf{ident}(X) \quad : \quad \mathcal{C}(X, X), \quad \text{which usually abbreviated as} \quad \mathsf{id}(X),$$

  together with fields $\mathcal{C}.\mathsf{unitL}$ and $\mathcal{C}.\mathsf{unitR}$ called the *unitality* laws stating that

  $$\mathcal{C}.\mathsf{unitL}(f) \quad : \quad \mathsf{id}(X) \cdot f \;=\; f \qquad \text{and} \qquad \mathcal{C}.\mathsf{unitR}(f) \quad : \quad f \cdot \mathsf{id}(X) \;=\; f$$

  where $f : \mathcal{C}(X, Y)$, i.e. composing any morphism with an identity morphism (on either side) will result in the morphism itself. ◢

◤ **Convention 2.3** **(small and large sets)** ⬣

To avoid a boilerplate about sets verses proper classes, we redefine the meaning of sets to possibly include something bigger than (traditional) sets, e.g. a class of all sets is now a set. The traditional meaning of sets will be referred as *small sets* whereas proper classes or larger entities will be referred as *large sets*. These terminologies will also apply for categories and other mathematical entities that relate to sets as well. ◢

◤ **Convention 2.4** **(direction of composition operation)** ⬣
Traditionally, the composition operation in category theory is denoted by $(\circ)$ and the composite direction goes from right to left, i.e. the composition of $f : \mathcal{C}(X, Y)$ and $g : \mathcal{C}(Y, Z)$ is $(g \circ f) : \mathcal{C}(X, Z)$. This traditional convention is borrowed from the function composition, which is in the form of $(g \circ f)(x) :\equiv g(f(x))$.

This thesis, in contrast, composes morphisms from left to right because this is the direction that most other concepts operate. For example, transitivity in relations, concatenation in sequences, piping in functional programmings, and most importantly the direction of the arrow that notates function types. **To avoid the confusion, we change the composition notation from** $(\circ)$ **to** $(\cdot)$, and reserve the original notation $(\circ)$ for the situation where compositing from right to left is more appropriate. ◢

◤ **Example 2.5** ⬣ Prominent examples of categories including:

- the category **Set** where objects are small sets and morphisms are functions,
- the category **Rel** where objects are small sets and morphisms are relations,
- the category **Poset** where objects are partially ordered sets (a.k.a. posets) and morphisms are order-preserving functions (a.k.a. monotonic functions),
- the category **Prop** for propositions and logical implications,
- the category **Top** for topological spaces and continuous functions,
- the category **Vect** for vector spaces and affine maps. ◢

◤ **Convention 2.6** **(partial definitions of categorical entities)** ⬣
The example above omits composition operations and identity morphisms because they can be canonically derived from objects and morphisms. In addition, associativity and unitality are just mere propositions which can be easily proven. We will also partially define some of the upcoming categories and related categorical entities whenever they are obvious to do so. ◢

## § 2.2.1 Commutative Diagrams

One of outstanding features of category theory is the ability to visualise categorical components by drawing a diagram called a *commutative diagram*, which visualises a category (or a fraction of a category) by drawing nodes and arrows to represent objects and morphisms, respectively. For simplicity, identity morphisms and compositing morphisms are omitted in a commutative diagram; they must exist by definition of categories anyway.

Besides objects and morphisms, we need equalities between morphisms to define categories, e.g. to define associativity and unitality. Each equality can exist as a closed area in a diagram such that its boundary consists of two (possibly composing) morphism, we say that the area is commuting if and only if both arrows are equal.

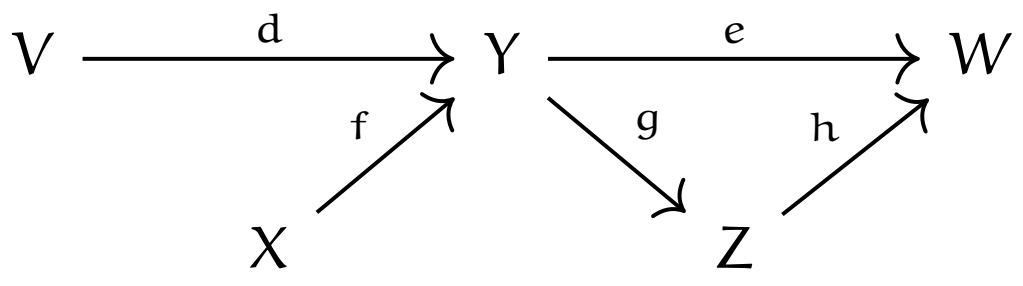

Figure 2.1: A diagram representing a category $\mathcal{C}$ as a toy example.

For example, the diagram in figure 2.1 shows a category $\mathcal{C}$ as a toy example that has $V$, $X$, $Y$, $Z$, and $W$ as its objects. This category $\mathcal{C}$ does not only have $d$, $e$, $f$, $g$, $h$ as its morphisms but also $\mathsf{id}(V)$, $\mathsf{id}(X)$, $\mathsf{id}(Y)$, $\mathsf{id}(Z)$, $\mathsf{id}(W)$, $d \cdot g$, $d \cdot e$, $f \cdot g$, $f \cdot e$, $g \cdot h$, $d \cdot g \cdot h$, and $f \cdot g \cdot h$ as morphisms as well. The diagram also contains one triangle that has boundary as $g \cdot h$ and $e$; therefore, $g \cdot h$ must equal to $e$.

Unless explicitly state otherwise, all of closed areas in any commutative diagram always commute. In addition, a commutative diagram usually represents only just a subset of objects and morphisms intending to visualise some information of the category.

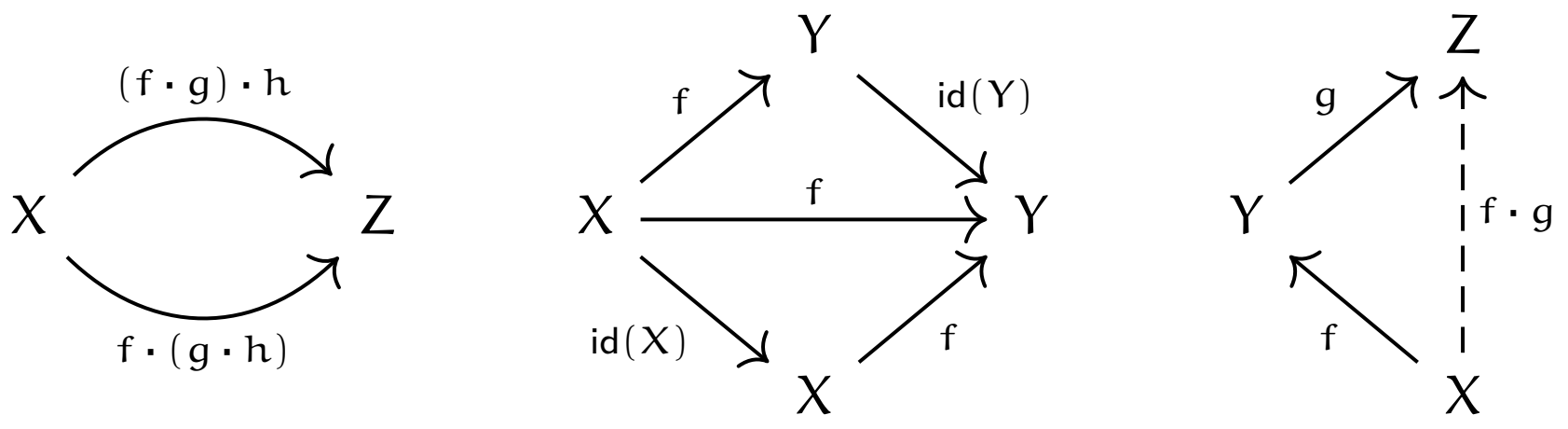

Figure 2.2: Diagrams for associativity, unitality, and composition operation.

In figure 2.2, the left diagram visualises the associativity. The middle diagram visualises the unitality, which doesn't only contain the lower and upper triangles that visualise $(\mathrm{id}(X) \cdot f = f)$ and $(f = f \cdot \mathrm{id}(Y))$, respectively, but also contains a square that visualises $(\mathrm{id}(X) \cdot f = f \cdot \mathrm{id}(Y))$. The right diagram visualises the composition operation; it is a common convention to use a dashed arrow for a newly defined morphism.

## § 2.2.2 Upgrading Sets to Categories

Recall that the main strength of category theory is an ability to transfer some concepts from one discipline into another. Since the category theory itself is a discipline; therefore, it can import some concepts from another discipline, such as set theory, which is the main objective of this subsection. In other words, this section will transfer some operations in set theory to category theory by upgrading them to operate on categories instead of sets.

◤ **Convention 2.7** **("upgrade" as a custom vocabulary)** ⬣

For simplicity, we will simply call the process that transfer some operations in set theory to category theory as *upgrading*. Please note that, this is *not* a standard convention in the literature and we will restrict its usage to be in this chapter exclusively. ◢

◤ **Example 2.8** ⬣ Many mathematical disciplines have an object of interest as a set (a.k.a. carrier set) equipped with the rest of components. For example:

- In topology, a topological space is a set (of its points) equipped with its topology.
- In order theory, a partially ordered set is a set equipped with a partial order.
- In logic, a proposition is a set (equipped with nothing) where truth and false are skeleton set and empty set, respectively.

Similar to how we can assemble those carrier sets to become the category **Set**; we can assemble topological spaces, posets, and propositions to be become the categories **Top**, **Poset**, and **Prop**, respectively. ◢

Similar to other objects in example 2.8, a category can be seen as a set of its objects equipped with the rest of components that we mention in definition 2.2. This allows us to upgrade any set to a category by adding a meaning behind that set. Since one set can have multiple meanings, so it can be upgraded to multiple categories depending our choices of additional morphisms. For instance, the set of all small sets can be upgraded as both categories **Set** and **Rel** by defining morphisms to be functions and relations, respectively.

Despite the non-uniqueness of the upgrading process, a set usually has a category that is the *canonically-upgraded* version; we can think about this canonically-upgraded category as the default choice of upgrading. For instance, e.g. **Set** is more sensible and has more applications than **Rel** or the rest of other choices; so, the canonical lifted version of the set of all small sets is **Set**.

◤ **Remark 2.9** ⬣ Our upgrading concept is closely related to *categorification* but they are not exactly the same. Categorification means finding a higher dimensional structure whose lower dimensional part recovers some existing set-level concept. For example, the category of finite sets is a categorification of the natural numbers. ◢

**Product Categories:** Now, let us start upgrading on Cartesian products. To upgrade the Cartesian product $(C \times D)$, we first to upgrade the sets C and D themselves as categories $\mathcal{C}$ and $\mathcal{D}$, then we upgrade $(C \times D)$ to the *product category* $(\mathcal{C} \times \mathcal{D})$, defined in definition 2.10. Please note that, the object of $(\mathcal{C} \times \mathcal{D})$ is indeed the Cartesian product $(C \times D)$.

◤ **Definition 2.10** ⬣ Let $\mathcal{C}$ and $\mathcal{D}$ be categories, we define the *product category* $(\mathcal{C} \times \mathcal{D})$ as another category that consists of:

- $\text{ob}(\mathcal{C} \times \mathcal{D})$ is the Cartesian product between $\text{ob}(\mathcal{C})$ and $\text{ob}(\mathcal{D})$.
- Let $X, X' : \text{ob}(\mathcal{C})$ and $Y, Y' : \text{ob}(\mathcal{D})$, a morphism from object $\langle X, Y \rangle$ to object $\langle X', Y' \rangle$ is a pair of morphism in $\mathcal{C}(X, X')$ and morphism $\mathcal{D}(Y, Y')$.
- The rest of the components are defined pairwisely. ◢

**Functors:** Another important entities to be upgraded are functions. To upgrade some function $f$ from a set C to a set D, we first upgrade sets C and D to categories $\mathcal{C}$ and $\mathcal{D}$, then we upgrade the function $f$ to a *functor* $F$, defined in definition 2.11, where the map on objects of $F$ is indeed the function $f$ itself.

If we look definition 2.11 from the perspective of example 2.8, then the functor $F$ is the function $f$ equipped with an additional map on morphisms and laws that preserve identity and composition.

◤ **Definition 2.11** ⬣ A record $\mathcal{F}$ is called a *functor* $\mathcal{F}$ from a category $\mathcal{C}$ to another category $\mathcal{D}$, if it contains these fields:

- $\mathcal{F}.\mathsf{mapObj}$ — a function that maps each object $X$ in $\mathsf{ob}(\mathcal{C})$ to another object in $\mathsf{ob}(\mathcal{D})$, Please note that, an image $\mathcal{F}.\mathsf{mapObj}(X)$ is usually abbreviated as $\mathcal{F}(X)$.
- $\mathcal{F}.\mathsf{mapHom}(X, Y)$ — a function that maps every morphism $f$ in $\mathcal{C}(X, Y)$ to another morphism in $\mathcal{D}(\mathcal{F}(X), \mathcal{F}(Y))$ for all $X, Y : \mathsf{ob}(\mathcal{C})$, Please note that, an image $\mathcal{F}.\mathsf{mapHom}(X, Y)(f)$ is usually abbreviated as $\mathcal{F}(f)$.
- $\mathcal{F}.\mathsf{presComp}$ — a proof ensures that the functor *preserves composition*, i.e.

  $$\mathcal{F}.\mathsf{presComp}(f, g) \;:\; \mathcal{F}(f \cdot g) = \mathcal{F}(f) \cdot \mathcal{F}(g).$$

  for each composable morphisms $f$ and $g$ in $\mathcal{C}$.
- $\mathcal{F}.\mathsf{presIden}$ — a proof ensures that the functor *preserves identity*, i.e.

  $$\mathcal{F}.\mathsf{presIden}(X : \mathsf{ob}(\mathcal{C})) \;:\; \mathcal{F}(\mathsf{id}(X)) = \mathsf{id}(\mathcal{F}(X)).$$ ◢

<u>**Natural Transformations:**</u> The set of functors from a category $\mathcal{C}$ to another category $\mathcal{D}$ can be canonically upgraded as the *functor category* from $\mathcal{C}$ to $\mathcal{D}$, denoted as $[\mathcal{C}, \mathcal{D}]$ where the morphisms are called *natural transformations* described below. Let $\mathcal{F}$ and $\mathcal{G}$ be functors from a category $\mathcal{C}$ to a category $\mathcal{D}$, then a natural transformation $\alpha$ from functor $\mathcal{F}$ to functor $\mathcal{G}$, which consists of a function that maps each object $X : \mathsf{ob}(\mathcal{C})$ to a morphism $\alpha(X) : \mathcal{D}(\mathcal{F}(X), \mathcal{G}(X))$ together with a law called a *naturality square*. Such a law illustrated as the second commutative diagram in figure 2.3, which is a visual way to say that

$$\alpha(X) \cdot \mathcal{G}(f) = \mathcal{F}(f) \cdot \alpha(Y) \qquad \text{for every} \quad f : \mathcal{C}(X, Y).$$

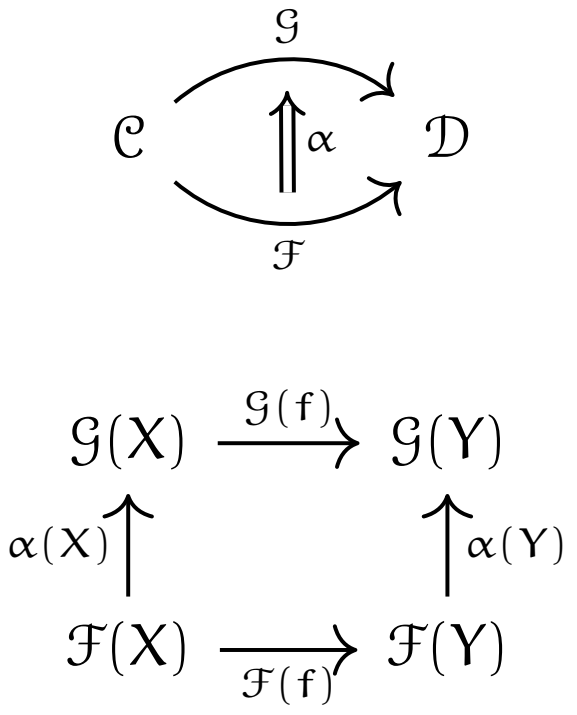

Figure 2.3

**Category of Categories:** Since sets and functions can be upgraded as categories and functors, so the category **Set** as a whole should be able to be upgraded as well. We may attempt to upgrade **Set** as another category called **Cat** where objects are small categories and morphisms are functors between them.

The hom-sets of **Set** are sets of functions, so their canonically-upgraded version must be functor categories. However the hom-sets of the category **Cat** must be sets by the definition of hom-sets. Therefore, we can't upgrade **Set** as another category. In fact, upgrading any category will always result in something more complex than the category itself because its hom-sets are sets and will be upgraded as *hom-categories*. This is where we need to enter the realm of *higher category theory* later in section 2.3.

**Isomorphisms:** The concept of upgrading even works on equalities, i.e. a proof of an equality between two objects in a category can be upgraded as *isomorphism*.

◤ **Definition 2.12** ⬣ Let $X$ and $Y$ be objects in some category $\mathcal{C}$, a morphism $f$ in $\mathcal{C}(X, Y)$ is called *isomorphism* if it is further equipped with another morphism

$$f^{-1} : \mathcal{C}(Y, X) \quad \text{such that} \quad f \cdot f^{-1} = \mathrm{id}(X) \quad \text{and} \quad f^{-1} \cdot f = \mathrm{id}(Y).$$

If there exists such an isomorphism $f$, we also say that "X is isomorphic to Y", denoted as $(X \cong Y)$, which can be seen as an upgraded version of the equality $(X = Y)$ itself. ◢

Isomorphisms are ubiquitous in category theory: Although the (strict) equalities itself still works on categorical entities, the uniqueness properties in category theory almost always hold only up to isomorphisms and the corresponding objects may not necessary (strictly) equal to one another. In other words, isomorphisms are the notion of sameness in categories. The role of isomorphism will become vital when we generalise strict higher categories to weak higher categories in section 2.3.

## § 2.3 Higher Category Theory

Higher category theory is an active discipline that generalises category theory. There are various ways to generalise categories to higher categories; each of them leads to a different definition and characteristics of higher categories.

### § 2.3.1 Strict Higher Categories

For pedagogical propose, this subsection still keep associativity and unital laws as equalities, which will be relaxed as isomorphisms, equivalences, even $\infty$-groupoids later in the section.

**2-Categories:** The first and easiest way to generalise a category is to upgrade its hom-sets to hom-categories, together with an appropriate replacement for the rest of the components. This results in a new kind of higher categories called 2-*categories*. To be more precise, a 2-category $\mathcal{B}$ is a higher category representing a upgraded version of some category $\mathcal{C}$ such that:

- The set objects $\mathsf{ob}(\mathcal{C})$ remains the same but relabelled as $\mathsf{ob}(\mathcal{B})$.
  Objects are also called 0-*cells*.
- Every hom-set $\mathcal{C}(X, Y)$ is upgraded to be a *hom-category* $\mathcal{B}(X, Y)$. Objects and morphisms in each hom-category are also called 1-*cells* and 2-*cells*, respectively. The composition operations here are called *vertical compositions*.
- A composition operation of $\mathcal{C}$ can be represented as a function from a Cartesian product of hom-sets $(\mathcal{C}(X, Y) \times \mathcal{C}(Y, Z))$ to the hom-set $\mathcal{C}(X, Z)$. This com-position function is now upgraded as a functor from the product category of hom-categories $(\mathcal{B}(X, Y) \times \mathcal{B}(Y, Z))$ to the hom-category $\mathcal{B}(X, Z)$. This kind of composition of functors is called *horizontal compositions* to disambiguate them from the vertical compositions; the notation $(\cdot)$ is also replaced with $(\blacklozenge)$. The associativity is also upgraded accordingly.

- An *identity* morphism $\mathsf{id}(X)$, which is equivalently a function from a singleton set (i.e. a set with a single element) to the hom-set $\mathcal{C}(X, X)$, can be upgraded as a functor from a terminal category (i.e. a category with a single object and morphism) to the hom-category $\mathcal{B}(X, X)$. The unitality is also upgraded accordingly.

The archetypical example of 2-categories is 1**Cat** which is a 2-category that expands the category **Cat** (that has been defined in subsection 2.2.2) by replacing each set of functors with the functor category counterpart. The identity morphisms can be easily upgraded as *identity functors*. However, upgrading the composition operations is more involved and will be omitted here for reasons of brevity.

**$\omega$-Categories:** After a set is upgraded into a category, which, in turn, upgraded into a 2-category; we can further upgrade a 2-category into a new categorical entity called a 3-*category* which is a category where its hom-sets are replaced with hom-2-categories and so on. The archetypical example of 3-categories is 2**Cat** where objects are 2-categories and morphisms are categorical entities called 2-*functors*, the canonically-upgraded version of functors. The iteration process can continue to infinity called *$\omega$-category*.

### § 2.3.2 Weak Higher Categories

There are many interesting mathematical entities that resemble $\omega$-categories (or their special cases) but can't really fit in them. The problem is that associativity and unitality may no longer hold in the form of equality but still hold in some weaker sense of sameness, e.g. isomorphisms. Such mathematical entities will be called *weak higher categories* to disambiguate them from the definitions of higher categories that we described before, which will now also be called *strict higher categories*. Please note that, a weak higher category is not necessary an actual category, in the sense that we can't undo the replacement of hom-sets, e.g. hom-categories, back to hom-sets; this is because the definition of a category require associativity and unitality as equalities.

**Bi-Categories:** We will start by generalising 2-categories to their weaker version called *bi-categories*. The first thing that must be done is to replace each of associativity and unitality from equalities to isomorphisms, which now are called *associators* and *unitors*. Unlike equalities, two isomorphisms that have the same domain and codomain may not be equal to each other; therefore, we need ensure that they are indeed equal to one other by asserting additional equalities, a.k.a coherence laws, between associators and unitors, i.e. Mac Lane's pentagon and triangles. Please see [Lei98] as a reference of definitions of a bi-category and its coherence theorem; see also [Lac09] for the motivation and further explanation of bi-category.

**Complexity of Weak $n$-Categories:** Next, we generalise 3-categories to its weaker version called *tri-categories*, which is similar to the generalisation from 2-categories to bi-categories but the associators and unitors are now also upgraded by replacing their equalities by isomorphisms. To make things worse, the coherence laws above are also upgraded into isomorphisms, which, in turn, need other coherence laws between them. This makes the complete definition of a tri-category become very long and complicated. To appreciate the complexity of tri-categories, please see the PhD thesis [Gur06] of Michael Gurski and the article [DH16] by Douglas and Henriques. Despite the complexity of tri-categories, Todd Trimble managed to go beyond and write down the definition of a *tetra-category*, i.e. a weak 4-category, the result has later been polished [Hof13] by Alexander E. Hoffnung.

**Potential Definitions of Weak $n$-Categories:** Due to their complexity, we cannot continue this trend of manually defining a weak higher category. Mathematicians in this area have been trying to systematically define a weak $n$-category, for all natural number $n$ (possibly also including the case where $n :\equiv \infty$); and there are already many proposed definitions for a weak $n$-category, each of them has many interesting properties, especially some of them even managed to cover the case when $n :\equiv \infty$, i.e. weak $\omega$-categories. For further details of the potential definitions of a weak $n$-category, please see this survey [Lei01] by Tom Leinster and this guidebook [CL04] by Eugenia Cheng and Aaron Lauda.

Despite having many potential definitions, mathematicians are not confident to use any of them as "the" definition of a weak $n$-category. The problem is about how to compare the strength among these candidates; in fact, we haven't found any suitable way to universally define a map from one candidate to another because doing so would require the definition of a weak $n$-category in the first place; and this is where the problem self-referencing begins. Nevertheless, the is a result that all of the main candidates of the potential definition of $(\infty, n)$ agree up to homotopy (except the case of $n :\equiv \infty$). For further reference, please see [INCat; Ber10; BS11].

### § 2.3.3 Commutative Diagrams for Higher Categories

Commutative diagrams are also upgraded for higher categories. Nodes and arrows still represent 0-cells and 1-cells, respectively, but the closed areas now represent 2-cells annotated by a double-line arrow pointing from one boundary arrow to another; for example, $\alpha$ in the first commutative diagram in figure 2.3 is annotated by double-line arrow pointing from $\mathcal{F}$ to $\mathcal{G}$.

In general, an $n$-category has $(n+2)$ kinds of data, which are 0-cells, 1-cells, ..., $n$-cells, and equalities between $n$-cells. Commutative diagrams can encode these additional kinds of data by introducing $n$-*arrows*, each of them will intuitively be an $n$-dimensional ball that has boundary consisting of two (possible composed) $(n-1)$-arrows annotated by $n$-lines arrow from one boundary arrow to another.

### § 2.3.4 Infinity-Groupoids

Although the ultimate definition of a weak $n$-category is still an open research problem, there are special cases of them that are well-defined and well-understood. For instance, an $\infty$-*groupoid*, which is a weak $\infty$-categories that all morphisms in every dimension are invertible, has many several definitions but all of them are equivalent to one another.

The $\infty$-groupoids, despite being just a special case of weak $\infty$-categories, are considered ubiquitous in algebraic topology (in particular, a sub-discipline called *homotopy theory*) because every topological space has a *fundamental* $\infty$-*groupoid* that captures many characteristics of such the topological space. In particular, two topological spaces with different fundamental $\infty$-groupoids can't be homeomorphic to each other.

## § 2.4 Homotopy Type Theory

*Homotopy type theory* (or just HoTT for short) is a sub-discipline in type theory that studies variations of MLTT that the main interpretation of a type is enriched from a set to an $\infty$-groupoid. This is useful for algebraic topology because a type can also be interpreted as a topological space due to the *Grothendieck's homotopy hypothesis*, which states that $\infty$-groupoids are equivalent to topological spaces.

### § 2.4.1 History of Homotopy Type Theory

HoTT is relatively a young discipline starting around 2006 from independent works by Awodey and Warren [AW09] and Voevodsky [Voe06], but it was inspired by Hofmann and Streicher's earlier groupoid interpretation [HS98].

Later, there is an important event of HoTT in 2012–2013; a special year on univalent foundations of mathematics was held at the Institute for Advanced Study (IAS); due to the accumulative results, the participants decided to write a book [HoTT] dedicated to HoTT, which I highly recommend a newcomer to read as the starting point to study this discipline.

## § 2.4.2 Weakening the Equalities to Infinity Groupoids

The motivation of upgrading MLTT to HoTT is because equalities in MLTT are sometime too strict to formalise many of the important mathematical structures, in particular, structures that arise in algebraic topology. This problem is similar to why those mathematical structures are incompatible with strict higher categories. To remedy this, we need to “weaken” an equality in a type theory down to an *equivalence*, which is the correct notion of sameness in an $\infty$-groupoid.

The “weakening” modification above requires a slight adjustment to the original MLTT by upgrading the equality types from being just mere propositions to arbitrary types which can contain multiple elements. In other words, let each of $x$ and $y$ be an element of a type $A$, the equality types $(x = y)$ can contain multiple proofs that are not necessary equal to each other. Moreover, let $p$ and $q$ be proofs of $(x = y)$ then there is the equality type $(p = q)$ that can contain multiple proofs that are not necessary equal to each other. Then, let $\alpha$ and $\beta$ be proofs of $(p = q)$ then there is the equality type $(\alpha = \beta)$ that can contain multiple proofs that are not necessary equal to each other. This process can repeats up to infinity.

To mimic the topological spaces interpretation, we may call an element of $(x = y)$ a *path* from a point $x$ to a point $y$. As a consequence, equality types can also be interpreted as $\infty$-groupoids themselves; therefore, most (if not all) components of an $\infty$-groupoid can be represented in the syntax of HoTT: Let $A$ be a type and $\mathcal{A}$ be the $\infty$-groupoid that represents $A$. An object $X$ in $\mathcal{A}$ is represented by the element $x$ in $A$, a hom-$\infty$-groupoid $\mathcal{A}(X, Y)$ is represented by the type $(x = y)$, and so on.

### § 2.4.3 The Univalence Axiom

Recall that the outstanding property of HoTT is an ability to construct equalities using a weaker sense of sameness. To do this, we introduce a new type called *equivalence*, denoted as $(A \simeq B)$, where its elements are interpreted as equivalences between the $\infty$-groupoids that interpret types A and B (see [HoTT, Chapter 4] for more detail). Then we axiomatise the *univalence axiom* by postulating that

$$\mathsf{idToEquiv}_{A,B} : (A = B) \to (A \simeq B) \qquad \text{is an equivalence,}$$

which allows us to construct an equality between types by first constructing an equivalence between them and apply $\mathsf{idToEquiv}_{A,B}$ backwards; thus, making it possible to formalise mathematical concepts that involve some degree of sameness but might not as strict as equality.

Moreover, the univalence axiom implies other important axioms such as *function extensionality* (which states that two functions are equal if they return the same output for every given input) and *propositional extensionality* (which states that two propositions are equal if they imply each other); thus, the univalence axiom simplifies a type theory considerably.

## § 2.5 Directed Type Theory

*Directed type theory* is a sub-discipline in type theory that generalises HoTT where the main interpretation of a type is generalised from an $\infty$-groupoid to an $\infty$-category. To put in simply, directed type theory is HoTT but equalities (or whatever takes place of equalities) may not be invertible. There are many proposals for directed type theories including Peter Lumsdaine [Lum10]; Licata and Harper [LH11], Nuyts [Nuy15]; Riehl and Shulman [RS17]; North [Nor19]; Benjamin, Finster, and Mimram [BFM21].

**Choices of the Definition of $\infty$-Categories:** Although section 2.3 says that the definition of an actual $\infty$-category is still an open research problem, we can pick any potential definition as our definition of an $\infty$-category. Alternatively, there are other special cases of $\infty$-categories that are more complex than $\infty$-groupoids from which we can choose. The notable examples are $(\infty, n)$-categories, which are $\infty$-categories that all $m$-cells (a.k.a. $m$-morphisms) for all $(m > n)$ are invertible. The archetypical example of $(\infty, 1)$-categories is the (weak higher) category of small $\infty$-groupoids called $\infty\mathbf{Gpd}$, which can also be seen as the interpretation of a universe in type theory, i.e. a type that has types as elements, in HoTT.

**Connection to Directed Spaces:** Analogous to an $\infty$-groupoid that represents a topological space, an $\infty$-category represents a *directed space*, which is a topological space equipped with some notion of direction. This connects directed type theory to a discipline outside type theory, such as *concurrency*, because concurrent processes can be represented by *directed spaces* (please see [FRG06]).

We also need a notion of directed spaces to geometrically realise the twisted cubes in chapter 7. Rather than using one of the most-developed theories by Marco Grandis [Gra09] or Sanjeevi Krishnan [Kri08], we utilise only the main concept and use simpler notion called *partially-ordered spaces*, or *pospaces* for short, which is less expressive than those theories above but covers all of the directed spaces that we target.

## § 2.6 Cubical Type Theory

*Cubical type theory* is a sub-discipline in HoTT that implements $\infty$-groupoids using *cubical sets*, i.e. presheaves[4] on a cube category. Please note that, the meaning of a cube in abstract topology (and this thesis) is actually a hypercube in general context.

Cubical type theory has recently gained significant attention from the HoTT community because many of its models allow their type theories to have desired properties, such as *constructivity* (which means a type theory doesn't use certain kind of axioms, such as law of excluded middle or axiom of choice) and *canonicity* (which means the evaluation of every closed term of natural numbers will lead to a term only consisting of zero and the successor function).

### § 2.6.1 History of Cubical Type Theory

Originally, *simplicial sets*, i.e. presheaves on the simplex category (which will be later defined in section 4.2), were considered to be the best candidate to interpret $\infty$-groupoids in HoTT because simplicial sets are the most ubiquitous shapes in algebraic topology [Hat02] together with many interesting properties when being used to implement topological spaces (please see section 4.2 for further detail).

The first model of HoTT, given by Voevodsky [Voe10] (see also the presentation by Kapulkin and Lumsdaine [KL12]), uses simplicial sets to implement $\infty$-groupoids. How-ever, it is still an open question how simplicial sets can be used to build a constructive model of type theory with univalent universes [GS17].

[4] A presheaf on a category $\mathcal{C}$ is a functor from $\mathcal{C}^{\mathrm{op}}$ to **Set**, where $\mathcal{C}^{\mathrm{op}}$ is the category $\mathcal{C}$ where the domain and codomain of each morphism are swapped.

This is where cubical sets start to shine. The first constructive model for cubical sets has been discovered by Bezem, Coquand, and Huber [BCH]. Soon after that, the first model that has the canonicity property has been discovered by Cohen, Coquand, Huber, and Mörtberg [CCHM]; this is a big discovery because the canonicity implies that the univalence axiom can be defined as a theorem rather than only an axiom. Consequently, cubes have gathered a lot of attention in the type theory community, leading to various *cubical type theories*, e.g. [BCH; CCHM; CHM; AFH; Ang+; Sat17; Awo18; OP18; Uem19; CMS20].

### § 2.6.2 Diversity of Cube Categories and their Common Pattern

Please note that, the cubical type theories above are different to one another mainly because they use different definitions of the cube category [BM17], which leads to different syntactic construction of the *interval type* explained later in subsection 3.1.1.

Nevertheless, these categories share equivalent configuration of objects and face maps in the sense that if we filter out all of morphisms that are not face maps on each of these categories, then the resulting categories will become equivalent to one another, which we will call them the *categories of semi-cubes* and generally denoted in this thesis as $\square_{\mathsf{semi}}$. This equivalent configuration allows us to generally discuss about the main cubical concept that applies to all of those cubical type theories in section 3.1.

Chapter

# 3

# Twisted Cubes as a Modification of Standard Cubes

This chapter gives a another overview of twisted cubes, which can be seen as a detailed expansion of chapter 1. We also change the narrative to tackle the twisted cubes from explaining the twisted cubes as-is to comparatively explaining the twisted cubes from standard cubes perspective.

## § 3.1 Common Framework to Reason about Standard Cubes

Recall that twisted cubes is a modification of standard cubes so this section provides a common framework[1] for standard cubes in such a comfort way that the framework can be modified to accommodate the motivation and intuition of twisted cubes.

◤ **Convention 3.1** **(standard cubes vs twisted cubes)** ⬣

Unlike other variations of cubes mentioned in section 2.6, the face-maps configuration of twisted cubes is *not* equivalent to those variations counterparts in subsection 2.6.2. To disambiguate twisted cubes from those variations, we now refer them as *standard cubes* and redefine the definition of cubes to be any cubes including twisted cubes. ◢

◤ **Convention 3.2** **(abbreviation for an n-dimensional shape)** ⬣

A word "$n$-dimensional shape" will be abbreviated to "$n$-shape". In particular, an $n$-dimensional simplex, an $n$-dimensional standard cube, and an $n$-dimensional twisted cube will be abbreviated into an $n$-simplex, a standard $n$-cube, and a twisted $n$-cube, respectively. ◢

◤ **Convention 3.3** **(first index as zero)** ⬣

Everything that looks like sequence will have the first index start at 0, unless explicitly stated otherwise. In other words, the element at index $i$ is indeed the $(i+1)^{th}$. For example the index 0 is the first index, the index 1 is the second index, and so on. ◢

[1]Not to be confused with the graph-theoretic framework defined in chapter 5.

## § 3.1.1 Paths as Functions from the Interval Type

There is an outstanding feature that all cubical type theories enjoy, which is an ability to treat a path as a function from the *interval type* to the type of its endpoints.

The *interval type*, denoted as $\mathbb{I}$, is interpreted as the closed interval $\{\ x : \mathbb{R} \mid 0 \leqslant x \leqslant 1\ \}$. Syntactically, we only have 2 elements[2] in $\mathbb{I}$, denoted as $\mathtt{0}$ and $\mathtt{1}$ that represent the real numbers 0 and 1, respectively. For instance, let $x$ and $y$ be elements of a type $A$, and let $p$ be a path between $x$ and $y$ (i.e. $p : x = y$) then $p$ can be represented as a function

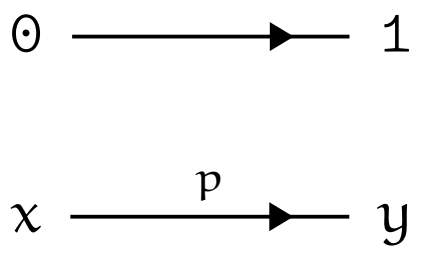

Figure 3.1

$$p : \mathbb{I} \to A \quad \text{where} \quad p(\mathtt{0}) :\equiv x \quad \text{and} \quad p(\mathtt{1}) :\equiv y.$$

This technique can easily construct some properties of equalities, such as the reflexivity of $x$ and the symmetry of $p$, which can be defined as

$$\mathsf{id}(x)(i) :\equiv x \quad \text{and} \quad p^{-1}(i) :\equiv p(1 - i), \quad \text{respectively.}$$

The constant path $\mathsf{id}(x)$, for any element $x : A$, is derivable from its endpoints (both of them must be $x$ anyway) so we may remove the label $\mathsf{id}(x)$ on the diagram and use double line to emphasize them instead.

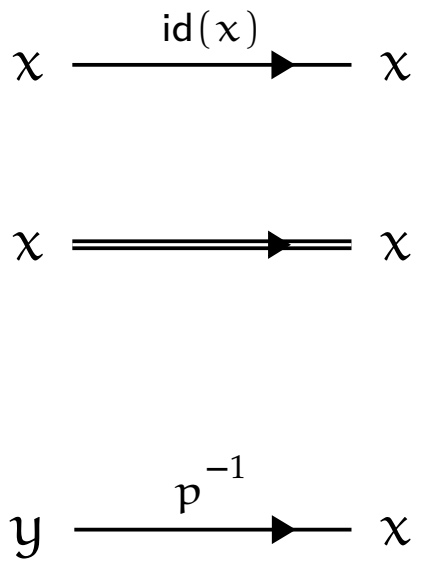

Figure 3.2

These constant paths are semantically interpreted as identity morphisms in the $\infty$-groupoid that interprets $A$. Intuitively, we may treat a constant path as a point that acts like a path, which is now called a *degenerated* path; more on this later in subsection 3.1.6.

[2] Each cubical type theory may have more components but these two elements are required.

**Partial Definitions as Types:** An element of type $(\mathbb{I} \to A)$ will semantically be interpreted as a morphism in an $\infty$-groupoid $\mathcal{A}$ (that interprets the type $A$) but without information about its domain or codomain. To assert such information we need a new type former to construct a subtype of $(\mathbb{I} \to A)$ such that some components are partially defined. Therefore, we introduce a new type former

$$\{\ t : T \mid t_0 :\equiv u_0 \ ;\ t_1 :\equiv u_1 \ ;\ t_2 :\equiv u_2 \ ;\ \dots\},$$

which is a subtype of a type $T$ such that an element $t$ must agree that the evaluation of a term $t_i$ (that may depend on $t$) is equal to $u_i$. Now we can define a type of functions from $\mathbb{I}$ to $A$ such that each function $p$ of this type must agree that $p(0)$ and $p(1)$ are equal to $x$ and $y$, respectively, as

$$\{\ p : \mathbb{I} \to A \mid p(0) :\equiv x \ ;\ p(1) :\equiv y\ \}.$$

Each element of this type will still semantically be interpreted as a morphism in $\mathcal{A}$ but now equipped with the additional information that its domain and codomain interpret $x$ and $y$, respectively.

### § 3.1.2 Generalising the Interval to the Standard $n$-Cube

To access higher components of $\infty$-groupoids, we generalise the interval type to the *template*[3] *standard $n$-cube type*, denoted as $\mathbb{I}^n_\square$, for all $n : \mathbb{N}$, which is defined to be the type $\mathbb{I}$ multiplied by itself $n$ times. This allows any $n$-cell of the $\infty$-groupoid $\mathcal{A}$ to be represented as a function of type $(\mathbb{I}^n_\square \to A)$, possibly equipped with partial definitions explained in subsection 3.1.5, which has now become the definition of a *standard $n$-cube*.

[3]The word "template" is added to the name of $\mathbb{I}^n_\square$ to disambiguate elements of $\mathbb{I}^n_\square$ from the actual standard $n$-cubes which are functions.

<u>**Using Ternary Numbers to Encode Faces:**</u> Combinatorially, the type $\mathbb{I}$ does not only consist of points $\mathtt{0}$ and $\mathtt{1}$ but also a single path that connects both points, now denoted as $\star$; this allows us to think that $\mathbb{I}$ has 3 components. Recall that $\mathbb{I}^n_\square$ is the type $\mathbb{I}$ multiplied by itself $n$ times, so it should abstractly have $3^n$ components corresponding to points (0-faces), lines (1-faces), squares (2-faces), and so on. We denote each component as $\langle t_0 t_1 \dots t_{n-1} \rangle$ where each $t_i$ can be either $\mathtt{0}$, $\mathtt{1}$, or $\star$ (as shown in figure 3.3). To be precise, an $m$-face of $\mathbb{I}^n_\square$ is a ternary number of length $n$ that has $m$ occurrences of $\star$.

◤ **Remark 3.4** ⬣ Binary and ternary numbers will be explicitly defined later in definitions 4.16 and 4.17, respectively. Regarding the order of digit, the first index means the *most significant digit*, which is also at the left-most position. ◢

◤ **Definition 3.5** **(facets)** ⬣

Due to the frequent usage of $(n-1)$-faces, we will simply call them as *facets*. Each of these facets can be uniquely represented by a ternary number of length $n$ that every index contains $\star$ except at index $r$ which contains $b$. We denote such the facet as ${}^b\partial^n_r$ (see figure 3.3 for ${}^b\partial^n_r$ examples) where $b : \mathbb{2}_{\mathsf{fin}}$, $n : \mathbb{N}$, and $r : \mathtt{fin}(n)$ (see definition 4.12 for $\mathtt{fin}(n)$). ◢

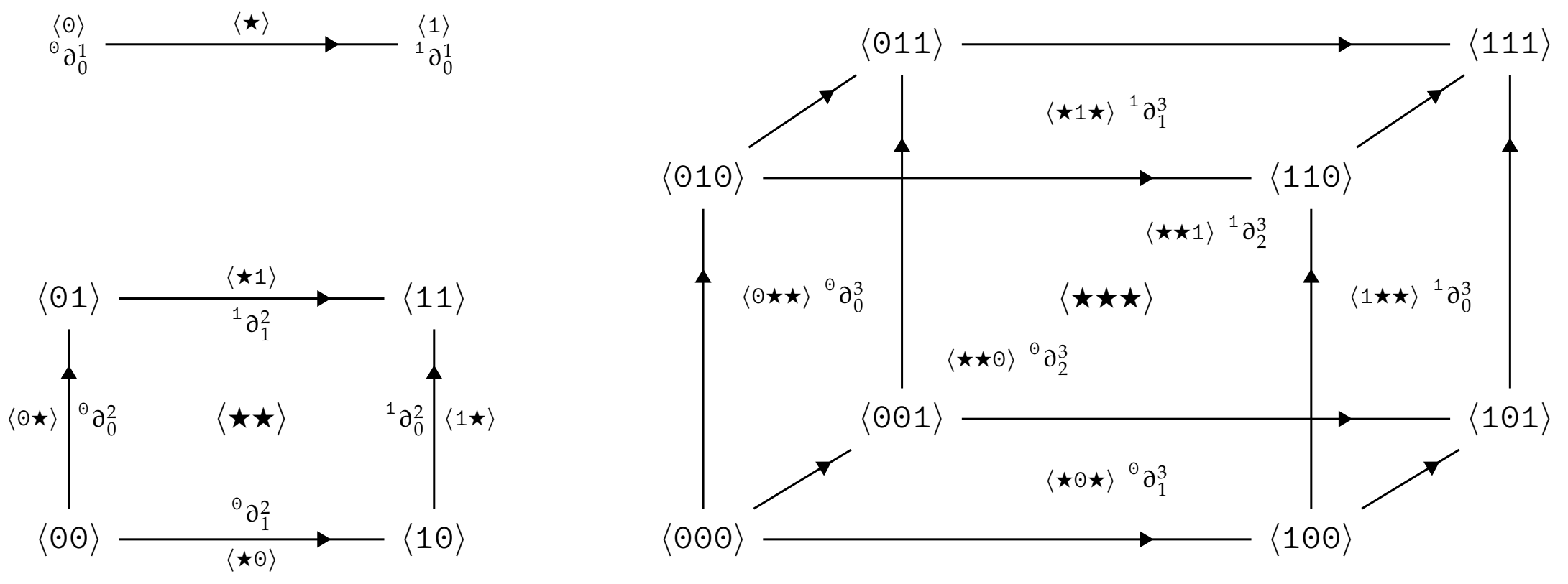

Figure 3.3: Illustration of $\mathbb{I}^1_\square$, $\mathbb{I}^2_\square$, and $\mathbb{I}^3_\square$ annotated with ternary numbers and ${}^b\partial^n_r$.

### § 3.1.3 Ternary Numbers as Face Maps

A ternary number of length $n$ may not necessarily be an element of $\mathbb{I}^n_\square$. For instance, every ternary number that contains $\star$ can't be expressed an element of $\mathbb{I}^n_\square$; this is because an element of $\mathbb{I}^n_\square$ is an $n$-tuple containing elements of $\mathbb{I}$, of which $\star$ is not an element.

Instead, a ternary number $\langle t_0 t_1 \dots t_{n-1} \rangle$ that represents an $m$-face of $\mathbb{I}^n_\square$ should be seen as an injective function from $\mathbb{I}^m_\square$ to $\mathbb{I}^n_\square$ such that, when given an input $\langle x_0 , x_1 , \dots , x_{m-1} \rangle : \mathbb{I}^m_\square$, the function returns $\langle t_0 , t_1 , \dots , t_{n-1} \rangle : \mathbb{I}^n_\square$ but substituting any $t_j$ that is the occurrent $i$ of $\star$ with $x_i$.

By seeing an $m$-face of $\mathbb{I}^n_\square$ as a function of type $(\mathbb{I}^m_\square \to \mathbb{I}^n_\square)$, we can post-compose it with a standard $n$-cube of type $(\mathbb{I}^n_\square \to A)$ to become a standard $m$-cube of type $(\mathbb{I}^m_\square \to A)$.

**Standard Semi-Cubes Category:** Instead of formal definition mentioned in definition 4.34, we can for now simply define a *category of standard semi-cubes*, denoted as $\square^{\mathsf{comb}}_{\mathsf{semi}}$, where objects are natural numbers, morphisms in $\square^{\mathsf{comb}}_{\mathsf{semi}}(m, n)$ are $m$-faces of $\mathbb{I}^n_\square$, and the rest of components behave exactly the same as in **Set**.

**Face Maps:** All morphisms in $\square_{\mathsf{semi}}(m, n)$ are morphisms that represent $m$-faces of a standard $n$-cubes, so we will call them *face maps*. This phenomenon also applies to simplex categories and twisted cube categories.

### § 3.1.4 Cubical Diagrams

Unlike earlier diagrams, figures 3.1 to 3.3 use different style of arrows to indicate that they are not commutative diagrams but rather a new kind of diagrams called *cubical diagrams* (please note that, this is not a standard terminology), which is a affine transformation of[4] an $n$-cube from an $n$-dimensional vector space into a 2-dimensional page in this thesis.

Unlike commutative diagrams that use double-line arrows to represent 2-cells, cubical diagrams use double-line arrows to represent degenerated paths. Also, moving the position of points around will not alter the underlining information in commutative diagrams, but it is indeed the case for commutative diagrams.

◤ **Notation 3.6** ⬣ We refer the *direction of dimension* $i$ to be the image of the affine transformation from the basis at dimension $i$ of the original vector space. ◢

◤ **Convention 3.7** **(order of cubical dimension)** ⬣
Given a cubical diagram of a standard $n$-cube, although it is easy to identify all of $n$ directions using the positions of points and directions of arrows, but it is impossible to determine the order among these dimensions without extra indicators, since we want to refer the components as ternary numbers.

[4]Technically, each diagram in figure 3.3 illustrates a *template* standard $n$-cube, i.e. $\mathbb{I}^n_\square$, rather a standard $n$-cube, i.e. $(\mathbb{I}^n_\square \to A)$ for some type $A$. We implicitly assume that there is a function $f : (\mathbb{I}^n_\square \to A)$ that each cubical diagram in figure 3.3 projected where each label $\vec{t}$ represents $(\vec{t} \cdot f)$.

To avoid the redundant specification on the order of dimensions on every cubical diagram, we use the order in figure 3.3 as the default reference. In other words, unless state otherwise, the directions of dimensions are ordered as: from left to right, from bottom to up, and from bottom-left to top-right. ◢

### § 3.1.5 Partially Defining Standard $n$-Cubes

Like the case of interval, an element of type $(\mathbb{I}^n_\square \to A)$ is semantically interpreted as an $n$-cell in an $\infty$-groupoid $\mathcal{A}$ but without any further information such as domain and codomain. To regain those detail, we assign all of facets using the mechanism of subtypes defined above, but with a restriction that every $(n-2)$-face, i.e. facet of facets, must be equal to component-wisely.

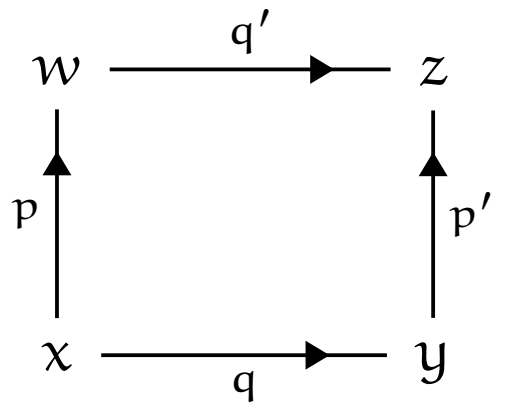

Figure 3.4

Unlike the case of the interval, we don't partially define the standard $n$-cube by function application; instead, we partially define it by assigning a standard $(n-1)$-cube to the composition between each facet ${}^{b}\partial^n_r$ and the original standard $n$-cube. For example, the type 2-cells illustrated in figure 3.4 can be defined as

$$
\begin{array}{llllll}
\{\ h : \mathbb{I}^2_\square \to A \mid \langle 0\star\rangle \cdot h :\equiv p & \langle 1\rangle \cdot p & = & w & = & \langle 0\rangle \cdot q' \\
\quad ;\ \langle 1\star\rangle \cdot h :\equiv p' & \langle 0\rangle \cdot p & = & x & = & \langle 0\rangle \cdot q \\
\quad ;\ \langle \star 0\rangle \cdot h :\equiv q & \langle 0\rangle \cdot p' & = & y & = & \langle 1\rangle \cdot q \\
\quad ;\ \langle \star 1\rangle \cdot h :\equiv q'\ \}, & \langle 1\rangle \cdot p' & = & z & = & \langle 1\rangle \cdot q'
\end{array}
$$

which has already satisfied the conditions for each of 4 points on the right.

## § 3.1.6 Defining $n$-Paths from Standard $n$-Cubes

An $n$-path is a standard $n$-cube such that the facets ${}^{0}\partial^{n}_{i}$ and ${}^{1}\partial^{n}_{i}$, for some dimension $i$, are assigned to be source and target, respectively, whereas other facets are degenerated. For example in case of 2-path, let $x$ and $y$ be elements in a type $A$, and let $p$ and $q$ be paths from $x$ to $y$, then a type of 2-paths from $p$ to $q$ can be constructed vertically as

$$
\begin{aligned}
\{\ h : \mathbb{I}^2_{\square} \to A \ \mid\ & \langle \star 0 \rangle \cdot h :\equiv p \ ;\ \langle 0 \star \rangle \cdot h :\equiv \mathsf{id}(x) \\
& ;\ \langle \star 1 \rangle \cdot h :\equiv q \ ;\ \langle 1 \star \rangle \cdot h :\equiv \mathsf{id}(y)\ \}.
\end{aligned}
$$

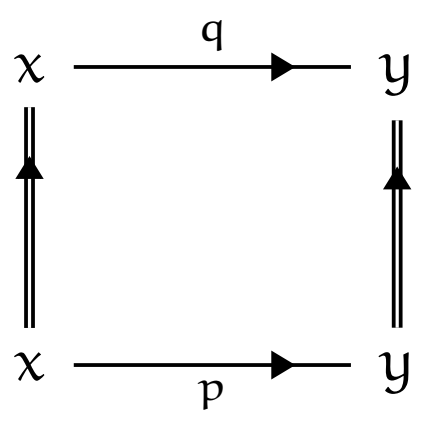

Figure 3.5

Alternatively, another type of 2-paths from $p$ to $q$ can be constructed horizontally by using the first dimension for endpoints instead of the second dimension, i.e. using facets $\langle 0 \star \rangle$ and $\langle 1 \star \rangle$ instead of facets $\langle \star 0 \rangle$ and $\langle \star 1 \rangle$ for domain and codomain, respectively. In general, there will be $n$ ways to construct $n$-paths depending on what dimension that we choose to assign endpoints.

### § 3.1.7 Kan Filler Condition

The last important component of cubical type theories is the *Kan filler condition*, which states that there exist a *lid* together with its *filler* for every $n$-*horn* where; an $n$-*horn* is syntactically[5] a type of standard $n$-cubes that partially defines every facet except one of them; a *lid* of an $n$-horn is a standard $(n-1)$-cube that has type as the missing facet of the horn; and a *filler* of the lid above is a standard $n$-cube that has type as the horn of its lid that also include its lid in its partial definition.

We can use the lids of these horns to construct the transitivity property of paths: Let $x$, $y$, and $z$ be elements of type $A$, and let $p$ be a path from $x$ to $y$ and $q$ be a path from $y$ to $z$. Then we define $p \cdot q$ to be the lid of the following 2-horn.

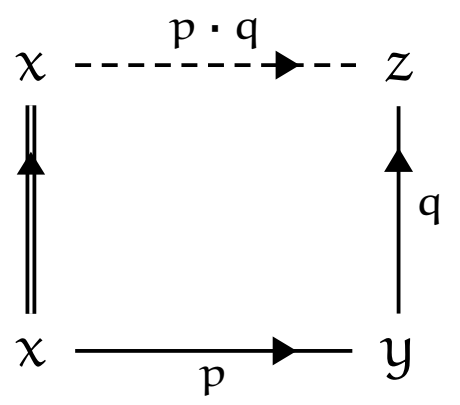

Figure 3.6

$$\{\ h : \mathbb{I}_{\square}^{2} \to A \ \mid \ \langle 0\star\rangle \cdot h :\equiv \mathsf{id}(x) \ ;\ \langle \star 0\rangle \cdot h :\equiv p \ ;\ \langle 1\star\rangle \cdot h :\equiv q \ \}$$

The Kan filler condition is important for every cubical type theory because it enables cubical sets to be interpreted as $\infty$-groupoids: Cubical sets alone only provide data structures to store $n$-cells and their identity $(n+1)$-morphisms. Then, the Kan filler allows these cells to be composed and generate their associators and unitors, and handle the rest of coherent structures. For example, composable morphisms can be seen together as a horn, then a lid of this horn acts as a composing morphism while the filler of this lid acts as an equivalence between the horn and the lid.

---

[5] To avoid technical detail, we simplified the actual definition of a horn, which is originally defined categorically.

## § 3.2 Motivation and Intuition of Twisted Cubes

Twisted cubes originated from my attempt to modify cubical type theory in such the way that it is compatible with directed type theory. In other words, I want to make a *directed cubical type theory*, which is a cubical type theory that generalises the interpretation of types from $\infty$-groupoids to $\infty$-categories.

### § 3.2.1 Invertibility of Standard Cubes

As the first step of designing a directed cubical type theory, we may try to artificially modify (i.e. hack) the syntax of some existing cubical type theory to forbid the construction of $p^{-1}$ for any path $p$ in the context of figure 3.1.

However, this is not as easy as removing the constructor $(1 - i)$ because the inverse of path $p$ can also be constructed by the Kan filling condition using the 2-horn in figure 3.7.

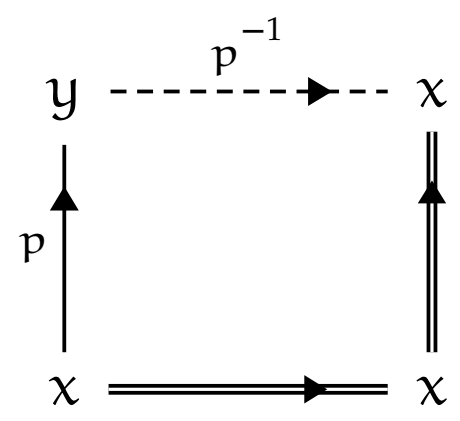

Figure 3.7

$$\{\ h : \mathbb{I}^2_\square \to A \mid \langle 0\star \rangle \cdot h :\equiv p \ ;\ \langle \star 0 \rangle \cdot h :\equiv \mathsf{id}(x) \ ;\ \langle 1\star \rangle \cdot h :\equiv \mathsf{id}(x) \ \}$$

Obviously, we can't simply remove all lids and fillers because they define the composition operation and the rest of components. In fact, we can't even partially forbid some of them because every another lid can construct $p^{-1}$ in similar fashion anyway. The cause of this invertibility is the orientation of the standard cubes itself; to illustrate this, let us transform the cubical diagram of $\mathbb{I}^2_\square$ in figure 3.3 to be the first commutative diagram in figure 3.8, and assign $[\![\star\star]\!]$ to be a 2-cell from $[\![\star 0]\!] \cdot [\![1\star]\!]$ to $[\![0\star]\!] \cdot [\![\star 1]\!]$.

The Kan filler condition says that if we miss an arrow then we can compose the rest and get the missing one (as a lid) together with a square that connects them (as a filler). Now, if we miss one of those four arrows, then the existing arrow on same side of boundary of the missing one must act inversely compose with the other two arrows.

### § 3.2.2 The Twisted 2-Cube as a Modification of the Standard 2-Cube

We change our approach from hacking around the syntax to modifying the orientation of the cubes themselves. Recall from figure 3.7 that if we want to fill a 2-horn such that $\langle \star\mathtt{1} \rangle$ is the missing facet, then the path assigned to the facet $\langle \mathtt{0}\star \rangle$ must act inversely. This suggests us to *forcefully reverse* the direction of the facet $\langle \mathtt{0}\star \rangle$ in $\mathbb{I}^2_{\square}$. This action transforms the first commutative diagram in figure 3.8 to the second one.

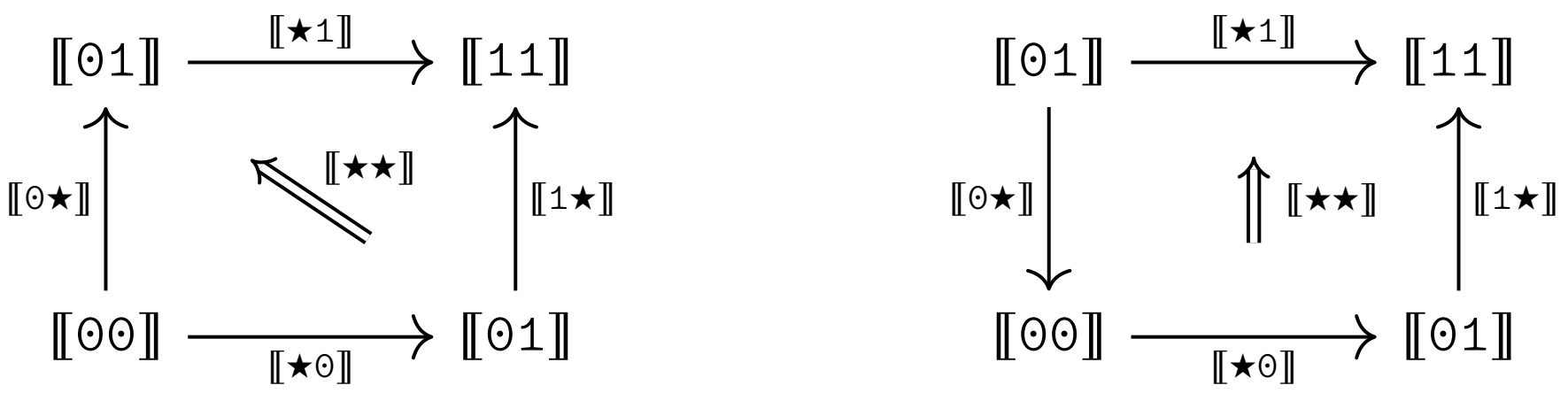

Figure 3.8: Transforming $\mathbb{I}^2_{\square}$ into $\mathbb{I}^2_{\bowtie}$ as commutative diagrams

The 2-cell $[\![\star\star]\!]$ in the second diagram is now pointing from $[\![\mathtt{0}\star]\!] \cdot [\![\star\mathtt{0}]\!] \cdot [\![\mathtt{1}\star]\!]$ to $[\![\star\mathtt{1}]\!]$; so, we can safely fill a horn whose missing facet is $\langle \star\mathtt{1} \rangle$ without worrying that it will cause the invertibility. Then, we relax the Kan filler condition to fill only the horns whose missing facet is $\langle \star\mathtt{1} \rangle$. Now, figure 3.7 is no longer valid. Syntactically, we call the type of squares that forcefully reverse the facet $\langle \mathtt{0}\star \rangle$ as the *template twisted* 2*-cube type*, denoted as $\mathbb{I}^2_{\bowtie}$, which has components as a 2-digits ternary number similar to $\mathbb{I}^2_{\square}$ except that the compositions $\langle \mathtt{0} \rangle \cdot \langle \mathtt{0}\star \rangle$ and $\langle \mathtt{1} \rangle \cdot \langle \mathtt{0}\star \rangle$ don't evaluate to $\langle \mathtt{00} \rangle$ and $\langle \mathtt{01} \rangle$ but rather evaluate to $\langle \mathtt{01} \rangle$ and $\langle \mathtt{00} \rangle$, respectively.

### § 3.2.3 Twisted $n$-Cube via Thickening-and-Twisting Process

The modification in subsection 3.2.2 may seem artificial and specific to the case of 2 dimensions, but in fact it can be generalised to be the *template twisted $n$-cube type*, denoted as $\mathbb{I}^n_{\bowtie}$, by recursion on the natural number $n$. For the base case, we define $\mathbb{I}^0_{\bowtie}$ as $\mathbb{I}^0_{\square}$, which is a type containing a 0-tuple. For the recursive case, we use the process called *thickening-and-twisting* that constructs $\mathbb{I}^{(n+1)}_{\bowtie}$ from $\mathbb{I}^n_{\bowtie}$ as follows.

First, the *thickening* phase, we prepend a new dimension as the first dimension and expand $\mathbb{I}^n_{\bowtie}$ along this new dimension to get $(\mathbb{I} \times \mathbb{I}^n_{\bowtie})$; this thickening phase is actually the same as constructing $\mathbb{I}^{(n+1)}_{\square}$ from $\mathbb{I}^n_{\square}$, i.e. its *cylinder object*. Then, the *twisting* phase, we reverse all other dimensions at the starting point of the new dimension.

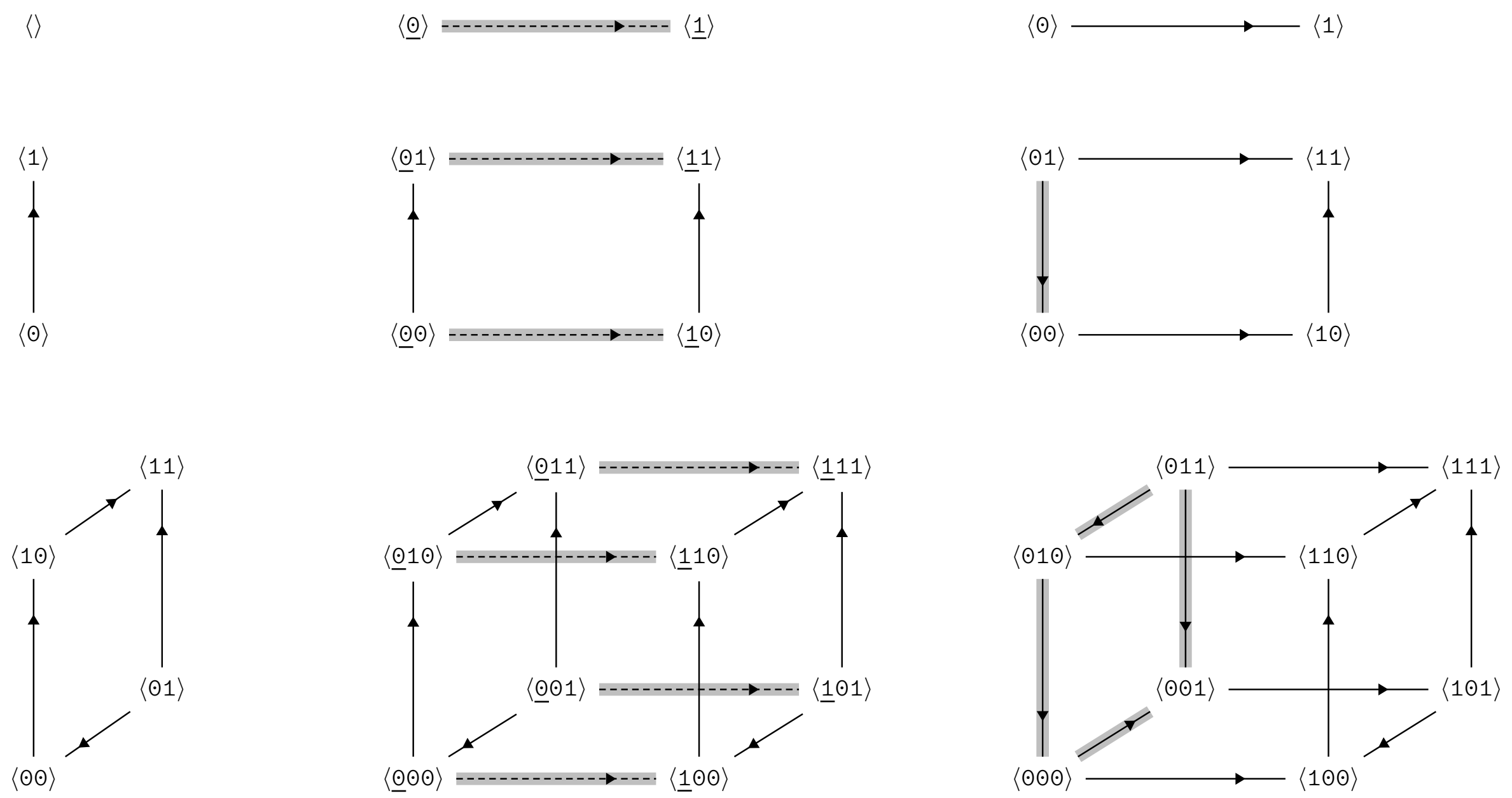

Figure 3.9: The *thickening-and-twisting* process for $(n \leqslant 2)$; where the first column $(\mathbb{I}^n_{\bowtie})$ is *thicken* into the next one $(\mathbb{I} \times \mathbb{I}^n_{\bowtie})$, which is then *twisted* into the last one $(\mathbb{I}^{(n+1)}_{\bowtie})$.

### § 3.2.4 All Faces of a Twisted Cube are Instances of Twisted Cubes

Twisted cubes preserve an important property from standard cubes; that is, every $m$-face of $\mathbb{I}^n_{\bowtie}$ is an instance of $\mathbb{I}^m_{\bowtie}$. As a sanity check, we can exhaustively verify this property for all $(m \leqslant n \leqslant 3)$ by inspecting figure 3.9. In addition, theorem 3.9 shows that we can actually prove this.

◤ **Lemma 3.8** ⬣ Every facet of $\mathbb{I}^{(n+1)}_{\bowtie}$ is an instance of $\mathbb{I}^n_{\bowtie}$, for all $n : \mathbb{N}$. ◢

*Proof.* Let's imagine that $\mathbb{I}^n_{\bowtie}$ is going though the thickening-and-twisting process.

- First, in the thickening phase, $\mathbb{I}^n_{\bowtie}$ is thickened into $(\mathbb{I} \times \mathbb{I}^n_{\bowtie})$.
  - The facets $\langle 0\star\star \dots \star\rangle$ and $\langle 1\star\star \dots \star\rangle$, which are respectively the left and right facets in the middle column of figure 3.9, are still instances of $\mathbb{I}^n_{\bowtie}$.
  - The rest of facets become instances of $(\mathbb{I} \times \mathbb{I}^{(n-1)}_{\bowtie})$.
- Then, in the twisting phase, $(\mathbb{I} \times \mathbb{I}^n_{\bowtie})$ is twisted into $\mathbb{I}^{(n+1)}_{\bowtie}$:
  - The right facet $\langle 1\star\star \dots \star\rangle$ is unaffected.
  - The left facet $\langle 0\star\star \dots \star\rangle$ is reversed in all dimensions, but still keep itself as an instance of $\mathbb{I}^n_{\bowtie}$ (as the completely upside down version).
  - The rest of facets are recursively twisted and change from being instances of $(\mathbb{I} \times \mathbb{I}^{(n-1)}_{\bowtie})$ to being instances of $\mathbb{I}^n_{\bowtie}$. QED

◤ **Theorem 3.9** ⬣ A face $\langle t_0 t_1 \dots t_{n-1}\rangle$ of $\mathbb{I}^n_{\bowtie}$ is also a twisted cube. ◢

*Proof.* We will iteratively prove through the smaller cases for $\langle\rangle$, $\langle t_{n-1}\rangle$, $\langle t_{n-2} t_{n-1}\rangle$, ... all the way until the original face $\langle t_0 t_1 \dots t_{n-1}\rangle$. To be more precise, we want to prove that, $\langle t_i t_{i+1} \dots t_{n-1}\rangle$ is an instance of $\mathbb{I}^k_{\bowtie}$ where $k$ is the number of $\star$ among $t_i t_{i+1} \dots t_{n-1}$ for all $(0 \leqslant i < n)$.

We will prove this by induction on $(n-1-i)$, i.e. $i$ but in backward order. The base case is when $i :\equiv (n-1)$, which is automatically true by the definition of $\mathbb{I}_{\bowtie}^{0}$. Regarding the inductive step, we have the induction hypothesis witnessing that $\langle t_{i+1} t_{i+2} \dots t_{n-1} \rangle$ is an instance of $\mathbb{I}_{\bowtie}^{k}$, this only leaves us to prove that $\langle t_i t_{i+1} \dots t_{n-1} \rangle$ is an instance of $\mathbb{I}_{\bowtie}^{k'}$ where $k' :\equiv (k+1)$ if $(t_i = \star)$ otherwise $k' :\equiv k$.

Before proving the remaining goal, we provide an intermediate result stating that the face $\langle \star t_{i+1} t_{i+2} \dots t_{n-1} \rangle$ is an instance of $\mathbb{I}_{\bowtie}^{(k+1)}$. Recall that the face $\langle t_{i+1} t_{i+2} \dots t_{n-1} \rangle$ is an k-face of $\mathbb{I}_{\bowtie}^{(n-1-i)}$ and imagine that $\mathbb{I}_{\bowtie}^{(n-1-i)}$ is going though the thickening-and-twisting process.

- First, during the thickening phase that $\mathbb{I}_{\bowtie}^{(n-1-i)}$ becomes $(\mathbb{I} \times \mathbb{I}_{\bowtie}^{(n-1-i)})$, the face $\langle t_{i+1} t_{i+2} \dots t_{n-1} \rangle$, which is an instance of $\mathbb{I}_{\bowtie}^{k}$, also becomes the face $\langle \star t_{i+1} t_{i+2} \dots t_{n-1} \rangle$, which is an instance of $(\mathbb{I} \times \mathbb{I}_{\bowtie}^{k})$.
- Then, during the twisting phase that $(\mathbb{I} \times \mathbb{I}_{\bowtie}^{(n-1-i)})$ becomes $\mathbb{I}_{\bowtie}^{(n-i)}$, the face $\langle \star t_{i+1} t_{i+2} \dots t_{n-1} \rangle$ changes from being an instance of $(\mathbb{I} \times \mathbb{I}_{\bowtie}^{k})$ to being an instance of $\mathbb{I}_{\bowtie}^{(k+1)}$ because of the same reason in the special case of facets.

Now, let's prove the remaining goal by splitting the scenario into three cases depending on the value of $t_i$:

- If $t_i :\equiv \star$, then the face $\langle t_i t_{i+1} \dots t_{n-1} \rangle$ is the face $\langle \star t_{i+1} t_{i+2} \dots t_{n-1} \rangle$, so we can apply lemma 3.8 to prove the goal.
- If $t_i :\equiv 1$, then the face $\langle t_i t_{i+1} \dots t_{n-1} \rangle$ is the face $\langle 1 t_{i+1} t_{i+2} \dots t_{n-1} \rangle$, which in turn, is a facet of $\langle \star t_{i+1} t_{i+2} \dots t_{n-1} \rangle$, which in turn again, is an instance of $\mathbb{I}_{\bowtie}^{(k+1)}$ by lemma 3.8. Therefore, $\langle t_i t_{i+1} \dots t_{n-1} \rangle$ is an instance of $\mathbb{I}_{\bowtie}^{k}$ due to the special case of facets.
- If $t_i :\equiv 0$, then we use the same reasoning as the case when $t_i :\equiv 1$. Please note that, this case does reverse the face $\langle t_i t_{i+1} \dots t_{n-1} \rangle$ in addition to the case when $t_i :\equiv 0$, but this reasoning above doesn't take the reversal into account so it is fine. Nevertheless, the fact that reversal happens in this case will matter when subsection 3.2.5 refers back here to count the number of reversals. QED

### § 3.2.5 Direction of an Arrow inside Twisted $n$-Cube

Recall from subsections 3.1.3 and 3.1.4 that an *arrow* in a cubical diagram is a 1-face of some $n$-dimensional cube, which can be represented as a ternary number of length $n$ that has exactly one occurrence of $\star$. To be more precise, an *arrow at dimension* $i$ is defined to a ternary number of length $n$ such that the digit at index $i$ is $\star$ and the rest of digits are booleans.

In the case of standard cubes, all arrows at dimension $i$ will have the same *direction* as dimension $i$ in the cubical diagram (i.e. the projection of the basis vector of axis $i$ in the $n$-dimensional vector space into the diagram). This is because both endpoints of an arrow share the same values in all dimensions except at dimension $i$ where the value of source (which is $\mathtt{0}$) is less than the target counterpart (which is $\mathtt{1}$).

In the case of twisted cubes, the direction of an arrow at dimension $i$ can either be in the same direction or in the *opposite* direction (relative to dimension $i$) depending on how many times that arrow is reversed. Since reversing an arrow two times is the same as not reverse it at all; therefore, the arrow is in the opposite direction iff it is reversed with an *odd* number of times.

**The Parity Function:** To precisely capture this concept of direction, we introduce a function $\mathtt{parity}_{\mathsf{arr}}$ that takes an arrow $\vec{t}$ at dimension $i$ and returns a boolean representing the truth value of the proposition "the direction of the arrow $\vec{t}$ is in the opposite direction relative to dimension $i$". We also introduce an auxiliary function $\mathtt{numRev}_{\mathsf{arr}}$ that takes an arrow $\vec{t}$ and returns the number of reversals of the arrow $\vec{t}$. With this auxiliary function, we can define $\mathtt{parity}_{\mathsf{arr}}(\vec{t})$ to be $\mathtt{numRev}_{\mathsf{arr}}(\vec{t})$ modulo 2. However, the actual definition of $\mathtt{numRev}_{\mathsf{arr}}$ requires more context and we will come back to define it later at the end of this subsection.

**The First Application of the Parity Function:** One immediate application of $\texttt{parity}_{\mathsf{arr}}$ is to *syntactically* determine the source and target of an arrow. For pedagogical purposes, let's first briefly review the case of standard cubes. Recall from subsection 3.1.3 that the position of each endpoint $b$ of an arrow is the ternary representation of that arrow itself except that now the occurrence of $\star$ is substituted by $b$ where $b :\equiv \texttt{0}$ if this endpoint is the source, otherwise, $b :\equiv \texttt{1}$ if this endpoint is the target.

Now in the case of twisted cubes, we will repeat the same process as in the case of standard cubes counterpart except that we change the substitution of $\star$ from $b$ to $(b \oplus b')$ where $b'$ is the parity of the arrow. The boolean operator $\oplus$, a.k.a. exclusive or, is used here because if $b' :\equiv \texttt{0}$, i.e. the arrow is in the same direction, then $(b \oplus \texttt{0})$ becomes $b$, which is the same as the case of standard cubes; otherwise, if $b' :\equiv \texttt{1}$, i.e. the arrow is in the opposite direction, then $(b \oplus \texttt{1})$ becomes $(1 - b)$, which is essentially the same as swapping the value between source and target.

◤ **Theorem 3.10** **(counting the number of reversals for an arrow)** ⬣

Let $n,, i : \mathbb{N}$ such that $(i < n)$ and let $\langle t_0 t_1 \ldots t_{i-1} \star t_{i+1} t_{i+2} \ldots t_{n-1} \rangle$ be an arrow, then $\texttt{numRev}_{\mathsf{arr}}(\vec{t})$ is equal to the number of occurrences of $\texttt{0}$ among $\langle t_0 t_1 \ldots t_{i-1} \rangle$. ◢

*Proof.* As a sanity check, we can exhaustively verify this proposition for all $(0 \leqslant i < 3)$ by inspecting figure 3.9. In the general case, we will backwardly iterate through $\vec{t}$ in the same way as we did for $\langle t_0 t_1 \ldots t_{n-1} \rangle$ in theorem 3.9 but now we rename variable $i$ in there as $j$ and detect any potential reversal of $\langle t_j t_{j+1} \ldots t_{n-1} \rangle$ from each iteration $(n - 1 - j)$. There are a total of $n$ iterations of the thickening-and-twisting process, which will be split as follows:

- Each iteration $(n - 1 - j)$ where $(j > i)$ can't produce any reversals. This is because the iteration concludes that $\langle t_j t_{j+1} \ldots t_{n-1} \rangle$ is an instance of $\mathbb{I}^0_{\bowtie}$ due to the fact that there are no occurrences of $\star$ among $t_j t_{j+1} \ldots t_{n-1}$; therefore, any reversal happens here will only apply to the $\langle\rangle$ of $\mathbb{I}^0_{\bowtie}$ and change nothing.

- Each iteration $(n-1-j)$ where $(j \leqslant i)$ will produce at most one reversal because the iteration concludes that $\langle t_j t_{j+1} \ldots t_{n-1} \rangle$ is an instance of $\mathbb{I}^1_{\bowtie}$ due to the fact that there is one occurrence of ★ at $t_i$ so we can reverse the arrow here; however, the reverse will only happen if $t_j :\equiv \mathtt{0}$. QED

**Explicit Definitions of $\mathtt{numRev}_{\mathsf{arr}}$ and $\mathtt{parity}_{\mathsf{arr}}$:** Since the value of $\mathtt{numRev}_{\mathsf{arr}}(\vec{t})$ only depends on $t_0 t_1 \ldots t_{i-1}$ where all digits there must be boolean; therefore, occurrences of $\mathtt{0}$ must be $i$ subtracted by the occurrences of $\mathtt{1}$. This allow us to explicitly define the functions $\mathtt{numRev}_{\mathsf{arr}}$ and $\mathtt{parity}_{\mathsf{arr}}$ as definition 3.11. Please note that, we define the function $\mathtt{parity}$ like that because $\mathtt{parity}_{\mathsf{arr}}$ is the modulo-2 arithmetic version of $\mathtt{numRev}_{\mathsf{arr}}$ and modulo-2 arithmetic has the property

$$(b + b') \;=\; (b \oplus b') \;=\; (b - b').$$

**Definition 3.11**

$$\begin{aligned}
\mathtt{numRev}(\langle b_0 b_1 \ldots b_{i-1} \rangle) &:\equiv i - (b_0 + b_1 + \ldots + b_{i-1}) \\
\mathtt{parity}(\langle b_0 b_1 \ldots b_{i-1} \rangle) &:\equiv (i \bmod 2) \oplus b_0 \oplus b_1 \oplus \ldots \oplus b_{i-1} \\
&:\equiv (1 - b_0) \oplus (1 - b_1) \oplus \ldots \oplus (1 - b_{i-1})
\end{aligned}$$

$$\begin{aligned}
\mathtt{numRev}_{\mathsf{arr}}(\langle t_0 t_1 \ldots t_{i-1} \bigstar t_{i+1} t_{i+2} \ldots t_{n-1} \rangle) &:\equiv \mathtt{numRev}(\langle t_0 t_1 \ldots t_{i-1} \rangle) \\
\mathtt{parity}_{\mathsf{arr}}(\langle t_0 t_1 \ldots t_{i-1} \bigstar t_{i+1} t_{i+2} \ldots t_{n-1} \rangle) &:\equiv \mathtt{parity}(\langle t_0 t_1 \ldots t_{i-1} \rangle)
\end{aligned}$$

## § 3.2.6 The Order of Dimensions in Twisted Cubes

When we draw a cubical diagram for a standard cube, it is impossible to recover the original order of dimensions unless convention 3.7 is enforced. Interestingly, this is not the case for twisted cubes.

To recover the order of dimensions in a cubical diagram of a twisted $(n+1)$-cube, we first separate all of $(n+1) \cdot 2^n$ arrows into buckets where two arrows will be in the same bucket if their direction is either the same or completely in the opposite direction, otherwise those arrows must be in different buckets.

Recall that all directions of dimensions (see notation 3.6) are linearly independent to one another. It is obvious to see that there will be exactly $(n+1)$ buckets where each bucket contains $2^n$ arrows and represents exactly one dimension (of the main $(n+1)$-cube).

So far, this separation method also works in the case of a standard $(n+1)$-cube with an additional property that each bucket having all arrows in the same direction. However, in the case of a twisted $(n+1)$-cube, there will be exactly one bucket having all arrows in the same direction whereas other buckets will have half of arrows pointing in the same direction whereas other half pointing in the opposite direction. This interesting property of arrows in a twisted $(n+1)$-cube comes from the fact that:

- All arrows in dimension 0 has the parity as `0`. This is because, each arrow $\vec{t}$ here must has the occurrence of $\star$ at the first index and it is impossible to have any occurrences of `0` that come before the first index.
- Half of arrows in dimension $(i+1)$ has the parity as `0` whereas another half has the parity as `1`. To justify this, we assume that the dimension $i$, a twisted $n$-cube (that is about to go though the twisting-and-thickening process) has $k$ arrows and $(2^{n-1} - k)$ arrows that have parity $0$ and $1$, respectively. After the twisting-and-thickening process, the first copy has $(2^{n-1} - k)$ and $k$ arrows that have parity $0$ and $1$ because all of them are reversed whereas the second copy still have $k$ and $(2^{n-1} - k)$ arrows that have parity $0$ and $1$. The dimension $i$ in the twisted $n$-cube now become dimension $(i+1)$ in the twisted $(n+1)$ cube with $(2^{n-1} - k) + k$ and $k + (2^{n-1} - k)$ arrows that have parity $0$ and $1$.

With the property above, we can select the only bucket that has arrows in the same direction as the first dimension of the main twisted $(n+1)$-cube. Next we recursively find the order of dimensions of the facet $\langle 1\star\star \dots \star \rangle$ (which is indeed a twisted $n$-cube) and treat dimension $i$ in this facet as dimension $(i+1)$ in the main twisted $(n+1)$-cube. In this way, the order of dimensions in the twisted cubes can be interestingly recovered.

Chapter

# 4

# Geometric Shapes for Higher Category Theory

Geometric shapes are often used to define various higher categories and related higher structures where a higher category is intuitively visualised as a conglomerate of cells of some particular shapes, together with some properties. In other words, a higher category will usually be encoded as a presheaf on a category of certain shapes, together with certain conditions.

There are many interesting examples of geometric shapes for higher structures in the literature [GeoHS]; for example, simplices, prisms[1], standard cubes, trees, globes, opetopes. However, I decide to focus only on simplices and standard cubes. This is because our graph-theoretic framework in chapter 5 is currently working only for geometric shapes that are convex polytopes (see convention 4.5). Moreover, simplices and standard cubes themselves are ubiquitous in the HoTT community.

[1]Not to be confused with the standard prism iterator $\texttt{prism}_{\texttt{std}}$ defined in definition 5.40

If the reader would like to learn more detail on higher structures and related shapes, then I recommend this guidebook [CL04] by Eugenia Cheng and Aaron Lauda for an introduction. For further reference, I recommend the further reading of this survey [Lei01] by Tom Leinster.

◤ **Convention 4.1** ⬣ Unlike other chapters that use the type theory mentioned in convention 2.1 as the meta-theory, some part of this chapter (and later in chapter 7) may use the traditional set-theory to explain concepts related to geometry. Even though type theory also works with geometry as well as set theory (if not actually better) doing so will require more prerequisites and overcomplicate the thesis. ◢

## § 4.1 Prerequisite Definitions and Conventions

### § 4.1.1 Topological Spaces of our Geometric Shapes

This chapter assumes some basic prerequisites on *topological spaces* that exist on elementary textbooks such as [Jän84]. We also try to make the distinction between a particular set and its usual topological space, see definition 4.2 for an example.

◤ **Definition 4.2** **(Euclidean spaces)** ⬣

Let $n : \mathbb{N}$, we define $\mathbb{R}^n$ as the Cartesian product of $\mathbb{R}$ that multiplied by itself $n$ times, we also define $\mathbb{R}^n_{\mathsf{top}}$ as the $n$-dimensional Euclidean space. This makes $\mathbb{R}^n$ becomes the set of points for $\mathbb{R}^n_{\mathsf{top}}$. ◢

◤ **Definition 4.3** **(discrete topological spaces)** ⬣

Given a set $X$, we define `TopDiscrete`$(X)$ to be a topological space that consists of the set $X$ (as the carrier set) together with the discrete topology generated from the set $X$, i.e. every subset of $X$ is an open set in `TopDiscrete`$(X)$. ◢

◤ **Lemma 4.4** ⬣ Let $X$ be a set, every function between topological spaces that has `TopDiscrete`$(X)$ as it domain is guaranteed to be continuous. ◢

*Proof.* A function between topological spaces is continuous if every open set in codomain has its preimage as an open set in codomain. This condition is true in our case because every subset of $X$ is already an open set. QED

◤ **Convention 4.5** **(compatibility of geometric shapes)** ⬣

To avoid unnecessary complexities, this thesis restricts the meaning of an $n$-dimensional geometric shape that is an $n$-dimensional *convex polytope*, which is a subspace of $\mathbb{R}^n_{\mathsf{top}}$ defined as by the *convex hull* of some subset in $\mathbb{R}^n$.

Please note that, a *convex hull* is a set (determined by points called *extreme points*) to be the intersection of all *convex sets* that contain those extreme points. A *convex set* is a set that every line segment connecting any two points in this set lies entirely with in the set. ◢

### § 4.1.2 Reedy Categories

Later in this chapter, we will extract information from each geometric shape to produce a category where objects are natural numbers represent the shape at finite dimensions and morphisms are affine transformations. Each resulting category should be *Reedy category*; in addition, if the resulting category contains only non-degenerated transformations it should be *direct category* as well.

**Definition 4.6** **(subcategories)**

Let $\mathcal{C}$ and $\mathcal{D}$ be categories, we say that $\mathcal{C}$ is a *subcategory* of $\mathcal{D}$ iff there is a functor $\mathcal{F}$ from $\mathcal{C}$ to $\mathcal{D}$ with the following conditions:

- Its object-mapping function $\mathcal{F}.\mathsf{mapObj}$ is injective; so we know that $\mathsf{ob}(\mathcal{C})$ can be seen as a subset of $\mathsf{ob}(\mathcal{D})$.
- Let $X$ and $Y$ be objects in $\mathcal{C}$, the function $\mathcal{F}.\mathsf{mapHom}(X, Y)$ is injective; so we know that $\mathcal{C}(X, Y)$ can be seen as a subset of $\mathcal{D}(\mathcal{F}(X), \mathcal{F}(Y))$.

Please note that, in most cases, $\mathsf{ob}(\mathcal{C}) \subseteq \mathsf{ob}(\mathcal{D})$ and $\mathcal{C}(X, Y) \subseteq \mathcal{D}(\mathcal{F}(X), \mathcal{F}(Y))$ so both $\mathcal{F}.\mathsf{mapObj}$ and $\mathcal{F}.\mathsf{mapHom}(X, Y)$ can simply be inclusion functions.

**Definition 4.7** **(wide subcategories)**

Let a category $\mathcal{C}$ be a subcategory of another category $\mathcal{D}$, we say that $\mathcal{C}$ is a *wide-subcategory* of $\mathcal{D}$ iff $\mathsf{ob}(\mathcal{C})$ is equal to $\mathsf{ob}(\mathcal{D})$.

◤ **Definition 4.8** **(direct categories)** ⬣

A category $\mathcal{C}$ is a *direct* category iff it is equipped with a function[2] $\mathsf{deg} : \mathsf{ob}(\mathcal{C}) \to \mathbb{N}$ (called a degree function) such that, if there is a non-identity morphism in $\mathcal{C}(X, Y)$, then $\mathsf{deg}(X) < \mathsf{deg}(Y)$ for all objects $X$ and $Y$ in $\mathcal{C}$. ◢

◤ **Definition 4.9** **(Reedy categories)** ⬣

A category $\mathcal{C}$ is a *Reedy* category iff it is equipped with a function[2] $\mathsf{deg} : \mathsf{ob}(\mathcal{C}) \to \mathbb{N}$ (called a degree function) together with other two wide-subcategories $\mathcal{L}$ and $\mathcal{R}$ that satisfy the following conditions:

- If there is a non-identity morphism in $\mathcal{L}(X, Y)$, then $\mathsf{deg}(X) > \mathsf{deg}(Y)$.
- If there is a non-identity morphism in $\mathcal{R}(X, Y)$, then $\mathsf{deg}(X) < \mathsf{deg}(Y)$.
- Each morphism in $\mathcal{C}$ factors uniquely as $(l \cdot r)$ where the morphism $l$ is in $\mathcal{L}$ and the morphism $r$ is in $\mathcal{R}$. ◢

◤ **Lemma 4.10** ⬣ Every direct category is a Reedy category. ◢

*Proof.* Given $\mathcal{C}$ as a direct category, we define $\mathcal{C}$ as a Reedy category by assigning:

- The function $\mathsf{deg}$ as the degree function that come with $\mathcal{C}$ as a direct category.
- The wide-subcategory $\mathcal{L}$ as the discrete category on $\mathsf{ob}(\mathcal{C})$.
- The wide-subcategory $\mathcal{R}$ as $\mathcal{C}$ itself.
- Each morphism $f$ in $\mathcal{C}$ factors uniquely as $f$ (which is in $\mathcal{L}$) composing with an identity morphism (which is in $\mathcal{R}$). QED

[2]In the literature, the codomain of the degree function can be any ordinal but this thesis will use only $\mathbb{N}$ as the codomain so I simplify it.

## § 4.1.3 Miscellaneous Definitions

This subsection contains other definitions that don't fit well in other places.

◤ **Definition 4.11** ⬣ Let $A$ and $B$ be sets, let $f$ be a function from $A$ to $B$, let $x$ and $y$ be elements of $A$, and let $p$ to be a proof that $x$ is equal to $y$.

We define $\texttt{cong}(f, p)$ to be a proof that $f(x)$ is equal to $f(y)$. ◢

◤ **Definition 4.12** **(finite sets)** ⬣

Let $n : \mathbb{N}$, we define $\texttt{fin}(n)$ as a subset of $\mathbb{N}$ containing all numbers less than $n$.

$$\texttt{fin}(n) \quad :\equiv \quad \{0, 1, \ldots, n-1\}$$ ◢

◤ **Definition 4.13** ⬣ We use notations $\mathbb{0}_{\text{fin}}$, $\bot$, and $\emptyset$ to denote $\texttt{fin}(0)$. ◢

◤ **Definition 4.14** ⬣ We use notations $\mathbb{1}_{\text{fin}}$ and $\top$ to denote $\texttt{fin}(1)$.
We also use $\mathtt{0} : \mathbb{1}_{\text{fin}}$ and $\mathsf{trivial} : \top$ to denote $0 : \texttt{fin}(1)$. ◢

◤ **Definition 4.15** **(booleans)** ⬣

We use a notation $\mathbb{2}_{\text{fin}}$ to denote $\texttt{fin}(2)$ and call its elements *booleans*. We also use $\mathtt{0} : \mathbb{2}_{\text{fin}}$ and $\mathtt{1} : \mathbb{2}_{\text{fin}}$ to denote $0 : \texttt{fin}(2)$ and $1 : \texttt{fin}(2)$, respectively. ◢

◤ **Definition 4.16** **(binary numbers)** ⬣

Let $n : \mathbb{N}$, we define $\texttt{binary}(n)$ to be a sequence of binary numbers of length $n$ where each digit here is a boolean (a.k.a. an element of $\mathbb{2}_{\mathsf{fin}}$).

If $\vec{b} : \texttt{binary}(n)$, we may use a notation $\langle \vec{b}_0 \vec{b}_1 \dots \vec{b}_{n-1} \rangle$ or $\langle \vec{b}_0 , \vec{b}_1 , \dots , \vec{b}_{n-1} \rangle$ to represent the binary number $\vec{b}$ where $\vec{b}_i$ is the digit of $\vec{b}$ at index $i$, which we can also denote it as $\vec{b}[i]$. ◢

◤ **Definition 4.17** **(ternary numbers)** ⬣

Let $n : \mathbb{N}$, we define $\texttt{ternary}(n)$ to be a sequence of ternary numbers of length $n$ where each digit here is an element of $\{ \texttt{0} , \texttt{1} , \star \}$.

If $\vec{t} : \texttt{ternary}(n)$, we may use a notation $\langle \vec{t}_0 \vec{t}_1 \dots \vec{t}_{n-1} \rangle$ or $\langle \vec{t}_0 , \vec{t}_1 , \dots , \vec{t}_{n-1} \rangle$ to represent the ternary number $\vec{t}$ where $\vec{t}_i$ is the digit of $\vec{t}$ at index $i$, which we can also denote it as $\vec{t}[i]$.

Please note that, although $\{ \texttt{0} , \texttt{1} , \star \}$ and $\texttt{fin}(3)$ are isomorphic to each other, but they have different underlining meanings. In particular, $\star : \{ \texttt{0} , \texttt{1} , \star \}$ should not be treated as $2 : \texttt{fin}(3)$ because $(\texttt{0} \leqslant \star \leqslant \texttt{1})$ whereas $(0 < 1 < 2)$. ◢

## § 4.2 Simplices

*Simplices* are generalisations of points, lines, triangles, and tetrahedrons into higher dimensions. If the reader is not familiar with simplices, then I recommend this paper [Fri20] by Greg Friedman.

### § 4.2.1 Geometry of Simplices

Simplices are essential for the study of algebraic topology and higher categories; in particular, the definitions of $\infty$-groupoids and quasi-categories depend on *simplicial sets* which are presheaves on the category of simplices.

◤ **Definition 4.18** ⬣ A *geometrical* $n$*-simplex* is a subspace of the $\mathbb{R}^n_{\mathsf{top}}$ determined by the convex hull of $(n+1)$ extreme points. Also, it will be called *degenerated* if it is possible to remove some extreme points and the convex hull is still be the same. ◢

In any given dimension $n$, there will be many distinct geometrical non-degenerated $n$-simplices but all of them are topologically equivalent to one another; so, we can select one of them to be the canonical geometrical $n$-simplex, denoted as $\triangle^n_{\mathsf{top}}$.

◤ **Definition 4.19** ⬣ We define the *canonical* $n$*-simplex*, denoted as $\triangle^n_{\mathsf{top}}$, to be a subspace of $\mathbb{R}^n_{\mathsf{top}}$ induced by the following set $\triangle^n_{\mathsf{set}}$.

$$\triangle^n_{\mathsf{set}} \;:\equiv\; \{ \, \langle x_0 \,,\, x_1 \,,\, \dots \,,\, x_{n-1} \rangle : \mathbb{R}^n \;\mid\; 0 \leqslant x_0 \leqslant x_1 \leqslant \cdots \leqslant x_{n-1} \leqslant 1 \, \}$$ ◢

◤ **Remark 4.20** ⬣ In literature, it is common for $\triangle^n_{top}$ to be defined by

$$\acute{\triangle}^n_{set} \;:\equiv\; \{\, \langle x_0, x_1, \ldots, x_{n-1} \rangle : \mathbb{R}^n \mid (0 \leqslant x_i \leqslant 1) \wedge (x_0 + x_1 + \cdots + x_{n-1} \leqslant 1) \,\}$$

which is more intuitive and perhaps geometrically easier to work with. However, I still prefer $\triangle^n_{set}$ over $\acute{\triangle}^n_{set}$ because the former behaves better when dealing with ordering (in particular for chapter 7). ◢

◤ **Theorem 4.21** ⬣ $\acute{\triangle}^n_{set}$ is isomorphic to $\triangle^n_{set}$ in the category **Vect**. ◢

*Proof.* Let $\phi$ and $\psi$ be matrices on $\mathbb{R}$ with size $n \times n$ where:

$$\phi_{i,j} \;:\equiv\; \begin{cases} 1 & \text{if } i \geqslant j \\ 0 & \text{otherwise} \end{cases} \qquad \psi_{i,j} \;:\equiv\; \begin{cases} 1 & \text{if } i = j \\ -1 & \text{if } i = j+1 \\ 0 & \text{otherwise} \end{cases}$$

$$\begin{bmatrix} x_0 \\ x_1 \\ x_2 \\ \vdots \\ x_{n-1} \end{bmatrix} \phi \;:\equiv\; \begin{bmatrix} x_0 \\ x_0 + x_1 \\ x_0 + x_1 + x_2 \\ \vdots \\ x_0 + x_1 + \cdots + x_{n-1} \end{bmatrix} \qquad \begin{bmatrix} x_0 \\ x_1 \\ x_2 \\ \vdots \\ x_{n-1} \end{bmatrix} \psi \;:\equiv\; \begin{bmatrix} x_0 \\ x_1 - x_0 \\ x_2 - x_1 \\ \vdots \\ x_{n-1} - x_{n-2} \end{bmatrix}$$

Then, we define an isomorphic structure ($\acute{\triangle}^n_{set} \cong \triangle^n_{set}$) as a tuple of $\phi$, $\psi$, and proofs that both $\phi \cdot \psi$ and $\psi \cdot \phi$ equal to the identity matrix. QED

◤ **Definition 4.22** ⬣ We defined the extreme point of $\triangle^n_{top}$ at *index* $i$ to be $\langle x_0, x_1, \ldots, x_{n-1} \rangle$ such that $x_j :\equiv 0$ if $(i + j < n)$, otherwise $x_j :\equiv 1$.

Thus, the extreme points of $\triangle^n_{top}$ at index 0, 1, 2, $(n-2)$, and $(n-1)$ to be $\langle 0, \ldots, 0 \rangle$, $\langle 0, \ldots, 0, 1 \rangle$, $\langle 0, \ldots, 0, 1, 1 \rangle$, $\langle 0, 1, \ldots, 1, 1 \rangle$, and $\langle 1, \ldots, 1 \rangle$, respectively. ◢

## § 4.2.2 Categories of Semi-Simplices

In this subsection, we want to gather non-degenerated simplices in subsection 4.2.1 and define the binary and combinatoric versions of the *semi-simplex category*, which are isomorphic to each other.

Please note that, the combinatoric version is more common than the binary version in the literature; nevertheless, I also define the binary version here for the pedagogical purpose, i.e. it is easier to understand and can be generalised from simplices to cubes.

◤ **Definition 4.23** ⬣ We define the category $\triangle^{\mathsf{bin}}_{\mathsf{semi}}$ to be the binary version of the semi-simplex category defined as followed.

- An object is a natural number $n$, representing $\triangle^{n}_{\mathsf{top}}$.
- A morphism from $m$ to $n$ is a binary number of length $(n+1)$ with $(m+1)$ occurrences of digit $\mathtt{1}$.
- An identity morphism of $n$ is the the binary number $\langle \mathtt{11} \dots \mathtt{1} \rangle$ with $(n+1)$ digits.
- Let $\vec{f} : \triangle^{\mathsf{comb}}_{\mathsf{semi}}(\, m\, ,\, k\, )$ and $\vec{g} : \triangle^{\mathsf{comb}}_{\mathsf{semi}}(\, k\, ,\, n\, )$, the composing morphism $\vec{f} \cdot \vec{g}$ is the binary number $\vec{g}$ but each occurrence $i$ of digit $\mathtt{1}$ is replaced by $\vec{f}[i]$. This results in binary number of length $(n+1)$ but the occurrences of digit $\mathtt{1}$ is reduced from $(k+1)$ to $(m+1)$.

Each morphism $\vec{b}$ from an object $m$ to another object $n$ represents a non-degenerated $m$-simplex of $\triangle^{n}_{\mathsf{top}}$ where each $\vec{b}[i]$ determines whether extreme point at index $i$ is included in this face: $\mathtt{1}$ means included whereas $\mathtt{0}$ means excluded. ◢

◤ **Lemma 4.24** ⬣ The category $\triangle^{\mathsf{bin}}_{\mathsf{semi}}$ is a direct category. ◢

*Proof.* We use the identity function on $\mathbb{N}$ as the degree function. For any natural numbers $m$ and $n$, we want to prove that $\deg(m) < \deg(n)$ if there is a non-identity morphism in $\triangle^{\mathsf{bin}}_{\mathsf{semi}}(m, n)$. This can be done by comparing $m$ and $n$:

- If $m < n$, then $\deg(m) < \deg(n)$ holds by unfolding the degree function.
- If $m = n$, then only the binary number $\langle 11 \ldots 1 \rangle$ can be in $\triangle^{\mathsf{bin}}_{\mathsf{semi}}(m, n)$, so it doesn't have a non-identity morphism.
- If $m > n$, then $\triangle^{\mathsf{bin}}_{\mathsf{semi}}(m, n)$ contains no morphisms because it impossible for any binary number to have occurrences of digit $1$ more than total number of digits itself. QED

◤ **Definition 4.25** ⬣ We define the category $\triangle^{\mathsf{comb}}_{\mathsf{semi}}$ to be the combinatoric version of the semi-simplex category defined as followed.

- An object is a natural number $n$, representing $\triangle^{n}_{\mathsf{top}}$.
  Alternatively, we may see the object $n$ as the set $\mathtt{fin}(n+1)$ that represents all $(n+1)$ extreme points of an $n$-simplex; therefore, we may use the set $\mathtt{fin}(n+1)$ or the natural number $n$ to represent the same object interchangeably.
- A morphism from $m$ to $n$ is a function $f : \mathtt{fin}(m+1) \to \mathtt{fin}(n+1)$ where the function $f$ is strict-monotone, i.e. $(i < j)$ implies $(f(i) < f(j))$.
- The rest of components inherits from the category **Set**. ◢

Each morphism $f : \triangle^{\text{comb}}_{\text{semi}}(m, n)$ represents the action of picking $(m+1)$ out of $(n+1)$ extreme points (to construct $\triangle^m_{\text{top}}$ in $\triangle^n_{\text{top}}$) where the extreme point at index $i : \texttt{fin}(m+1)$ of $\triangle^m_{\text{top}}$ is the extreme point at index $f(i) : \texttt{fin}(n+1)$ of $\triangle^n_{\text{top}}$.

Each function here must be strict-monotone to make sure that the picking process is in order; otherwise, there may be different picking methods that has the same result, e.g. both "picking the first then second" and "picking the second then first" have the same result containing the first two extreme points.

Actually, each function f must also be injective, i.e. $(f(i) = f(j))$ implies $(i = j)$, to make sure that there no distinct picks of the same extreme point (otherwise, the face will become degenerated). However, we don't require injectivity in this definition because strict-monotonicity implies injectivity anyway.

◤ **Lemma 4.26** ⬣ The category $\triangle^{\text{comb}}_{\text{semi}}$ is isomorphic to the category $\triangle^{\text{bin}}_{\text{semi}}$. ◢

*Proof.* Since both $\triangle^{\text{comb}}_{\text{semi}}$ and $\triangle^{\text{bin}}_{\text{semi}}$ have the same set of objects, our goal now becomes

$$\triangle^{\text{comb}}_{\text{semi}}(m, n) \quad \cong \quad \triangle^{\text{bin}}_{\text{semi}}(m, n) \qquad \text{for every} \quad m,\ n : \mathbb{N}\,.$$

In the forward direction, we define $\phi : \triangle^{\text{comb}}_{\text{semi}}(m, n) \to \triangle^{\text{bin}}_{\text{semi}}(m, n)$. Given $f : \triangle^{\text{comb}}_{\text{semi}}(m, n)$, which is a strict-monotone function from $\texttt{fin}(m+1)$ to $\texttt{fin}(n+1)$, we define $\phi(f) : \triangle^{\text{bin}}_{\text{semi}}(m, n)$ where $\phi(f)[i]$ for each index $i : \texttt{fin}(n+1)$ is the digit $\texttt{1}$ if $i$ is an image of the function $f$; otherwise, it is the digit $\texttt{0}$.
Please note that, there will be $(m+1)$ indices that have digit $\texttt{1}$ because the domain of f has $(m+1)$ elements and f is injective (derived from strict-monotonicity) so the are $(m+1)$ images from f.

In the backward direction, we define $\psi : \triangle^{\text{bin}}_{\text{semi}}(m, n) \rightarrow \triangle^{\text{comb}}_{\text{semi}}(m, n)$. Given $\vec{b} : \triangle^{\text{bin}}_{\text{semi}}(m, n)$, which is a binary number with $(n+1)$ digits such that the occurrences of digit $\mathtt{1}$ is equal to $(m+1)$, we define $\psi(\vec{b}) : \mathtt{fin}(m+1) \rightarrow \mathtt{fin}(n+1)$ as an strict-monotone function that takes number $i : \mathtt{fin}(m+1)$ then returns the index of the element in $\vec{b}$ that is the occurrence $i$ of digit $\mathtt{1}$.
Please note that, the image $\psi(\vec{b})(i)$ can be at most $n$ because the last index of $\vec{b}$ is $n$, so it fits the set $\mathtt{fin}(n+1)$.

Finally, it is obvious to see that both $\phi \cdot \psi$ and $\psi \cdot \phi$ are identity morphisms in $\triangle^{\text{comb}}_{\text{semi}}$ and $\triangle^{\text{bin}}_{\text{semi}}$, respectively. QED

◤ **Lemma 4.27** ⬣ The category $\triangle^{\text{comb}}_{\text{semi}}$ is a direct category. ◢

*Proof.* By applying lemma 4.26 to lemma 4.24 QED

## § 4.2.3 Categories of Simplices

In this subsection, we relax the semi-simplex category to the simplex category by allowing the affine transformation to be degenerated. This allow us to define the combinatoric version of the simplex category from $\triangle^{\text{comb}}_{\text{semi}}$ by relaxing the property of morphisms from strict-monotone functions to monotone functions.

◤ **Definition 4.28** ⬣ We define the category $\triangle^{\mathsf{comb}}_{\mathsf{full}}$ to be the combinatoric version of the simplex category defined as followed.

- An object is a natural number $n$, representing $\triangle^{n}_{\mathsf{top}}$.
  Alternatively, we may see the object $n$ as the set $\mathtt{fin}(n+1)$ that represents all $(n+1)$ extreme points of an $n$-simplex; therefore, we may use the set $\mathtt{fin}(n+1)$ or the natural number $n$ to represent the same object interchangeably.
- A morphism from $m$ to $n$ is a function $f : \mathtt{fin}(m+1) \to \mathtt{fin}(n+1)$ where the function $f$ is monotone, i.e. $(i \leqslant j)$ implies $(f(i) \leqslant f(j))$.
- The rest of components inherits from the category **Set**. ◢

◤ **Lemma 4.29** ⬣ The category $\triangle^{\mathsf{comb}}_{\mathsf{semi}}$ is a wide-subcategory of $\triangle^{\mathsf{comb}}_{\mathsf{full}}$. ◢

*Proof.* We can define a functor $\mathcal{F}$ from $\triangle^{\mathsf{comb}}_{\mathsf{semi}}$ to from $\triangle^{\mathsf{comb}}_{\mathsf{full}}$ such that $\mathcal{F}.\mathsf{mapObj}$ is an identity function (which is an injective function) and $\mathcal{F}.\mathsf{mapHom}(m, n)$ is an inclusion function (which is an injective function) for all $m,\ n\ :\ \mathsf{ob}(\triangle^{\mathsf{comb}}_{\mathsf{semi}})$. QED

◤ **Lemma 4.30** ⬣ The category $\triangle^{\mathsf{comb}}_{\mathsf{full}}$ is a Reedy category. ◢

*Proof.* According to definition 4.9; we define deg to be an identity function, define $\mathcal{R}$ to be the category $\triangle^{\mathsf{comb}}_{\mathsf{semi}}$, and define $\mathcal{L}$ to be a wide-subcategory of $\triangle^{\mathsf{comb}}_{\mathsf{full}}$ containing all morphisms of $\triangle^{\mathsf{comb}}_{\mathsf{full}}$ that are surjective functions.

Regarding the factorisation, each morphism $f\ :\ \triangle^{\mathsf{comb}}_{\mathsf{full}}(\,m\,,\,n\,)$, which is a monotone function from $\mathtt{fin}(m+1)$ to $\mathtt{fin}(n+1)$, we can always find a number $(k\ \leqslant\ \min(m,n))$ together with morphisms $l\ :\ \triangle^{\mathsf{comb}}_{\mathsf{full}}(\,m\,,\,k\,)$ and $r\ :\ \triangle^{\mathsf{comb}}_{\mathsf{full}}(\,k\,,\,n\,)$ such that $l \cdot r\ =\ f$. The following algorithm tells us how to find $k$, $l$, and $r$.

- Define an auxiliary set, named $\mathsf{imgs}$, as the set containing all images of $f$.
- Define $k$ to the cardinality of $\mathsf{imgs}$ minus 1.
- Define an auxiliary function $\mathsf{rank} : \mathtt{fin}(k+1) \rightarrow \mathsf{imgs}$ where $\mathsf{rank}(i)$ is the element in $\mathsf{imgs}$ at index $i$ when $\mathsf{imgs}$ is sorted in ascending order. Since the function $\mathsf{rank}$ is bijective, so we also have an inverse function $\mathsf{rank}^{-1} : \mathsf{imgs} \rightarrow \mathtt{fin}(k+1)$.
- We define $l : \mathtt{fin}(m+1) \rightarrow \mathtt{fin}(k+1)$ where $l :\equiv f \cdot \mathsf{rank}^{-1}$.
- We define $r : \mathtt{fin}(k+1) \rightarrow \mathtt{fin}(n+1)$ where $l :\equiv \mathsf{rank}$.

Now it is easy to see $l$ and $r$ are morphisms in $\mathcal{L}$ and $\mathcal{R}$, respectively. QED

◤ **Remark 4.31** **(augmented simplex category)** ⬣

There is an important variant of the simplex category called *augmented simplex category*, which is the simplex category *augmented* with an additional object called $(-1)$-simplex together with a face map from the $(-1)$-simplex to each other simplex.

Recall that an $n$-simplex can be represented by a set of $(n+1)$ extreme points; so, it is natural to algebraically extend the definition of simplices by including $(-1)$-simplex that is represented by a set of 0 extreme points.

There are some applications of the augmented simplex category but it is significantly fewer than the original simplex category; so, we can still say that there is a unique variation of the simplex category (up to the equivalence of categories). The prefix "augmented" is necessary to refer to augmented simplex category. ◢

## § 4.3 Standard Cubes

*Standard cubes* are generalisations of points, lines, squares, standard 3-cubes, and tesseracts into higher dimensions. Please note that, the meaning of cubes in topology is actually hypercubes in general context; to refer the original meaning of cubes from general context, we explicitly call them *standard* 3-*cubes* in this thesis.

### § 4.3.1 Geometry of Standard Cubes

Standard cubes are important for algebraic topology and higher categories; however simplices are usually preferred to standard cubes due to several reasons including:

- Most standard definitions in algebraic topology have tight-coupling dependency on the category of simplices. For example, a quasi-category, a popular representation of $(\infty, 1)$-category, is defined to be simplicial set satisfying a property called inner Kan conditions.
- The complexity of standard cube categories; for instance, an $n$-simplex has $(n+1)$ extreme points whereas a standard $n$-cube has $2^n$ extreme points, which is significantly larger.
- The diversity of standard cube categories [BM17] fragments the research on cubical sets: some theorem that works on one variation of standard cube category may not work in another variation. This is not the case for the simplex category.

Despite standard cubes having less popularity than simplices, there are some applications that prefer standard cubes to simplices because of their better behaviour in certain use cases; a prominent example is the categorical semantics of homotopy type theory that we have discussed in section 2.6.

◤ **Definition 4.32** ⬣ A *geometrical* $n$-*cube* is a subspace of the $\mathbb{R}^n_{\text{top}}$ determined by two points, denoted as $\langle u_0 , u_1 , \ldots , u_{n-1} \rangle$ and $\langle v_0 , v_1 , \ldots , v_{n-1} \rangle$ that generates

$$\{ \langle x_0 , x_1 , \ldots , x_{n-1} \rangle : \mathbb{R}^n \mid \min(u_i, v_i) \leqslant x_i \leqslant \max(u_i, v_i) \}$$

Also, it will be called *degenerated* if there is $i : \mathtt{fin}(n)$ such that $(u_i = v_i)$. ◢

◤ **Definition 4.33** ⬣ We define the *canonical standard* $n$-*cube*, denoted as $\square^n_{\text{top}}$, to be a subspace of $\mathbb{R}^n_{\text{top}}$ induced by points $\langle 0 , 0 , \ldots , 0 \rangle$ and $\langle 1 , 1 , \ldots , 1 \rangle$ that generate the following set

$$\square^n_{\text{set}} :\equiv \{ \langle x_0 , x_1 , \ldots , x_{n-1} \rangle : \mathbb{R}^n \mid 0 \leqslant x_i \leqslant 1 \}.$$ ◢

### § 4.3.2 Categories of Standard Semi-Cubes

◤ **Definition 4.34** ⬢ We define the category $\square^{\text{comb}}_{\text{semi}}$ to be the combinatoric version of standard semi-cube category defined as followed.

- An object is a natural number $n$, representing $\square^{n}_{\text{top}}$.
- A morphism from $m$ to $n$ is a ternary number of length $n$ with $m$ occurrences of $\star$.
- An identity morphism on object $n$ is the a ternary number $\langle\star\star\ldots\star\rangle$ with $n$ digits.
- Let $\vec{f} : \square^{\text{comb}}_{\text{semi}}(\,m\,,\,k\,)$ and $\vec{g} : \square^{\text{comb}}_{\text{semi}}(\,k\,,\,n\,)$, the composing morphism $\vec{f} \cdot \vec{g}$ is the ternary number $\vec{g}$ but each occurrence $i$ of $\star$ is replaced by $\vec{f}[i]$. This is a ternary number of length $n$ but the occurrences $\star$ is reduced from $k$ to $m$.

Each morphism $\vec{t} : \square^{\text{comb}}_{\text{semi}}(\,m\,,\,n\,)$ represents a non-degenerated $m$-face of $\square^{n}_{\text{top}}$ where:

- $\vec{t}[i] :\equiv 0$ means that the face is only at the source endpoint of dimension $i$,
- $\vec{t}[i] :\equiv 1$ means that the face is only at the target endpoint of dimension $i$,
- $\vec{t}[i] :\equiv \star$ means that the face occupy everywhere in dimension $i$. ◢

◤ **Lemma 4.35** ⬢ The category $\square^{\text{comb}}_{\text{semi}}$ is a direct category. ◢

*Proof.* We use the identity function on $\mathbb{N}$ as the degree function. For any natural numbers $m$ and $n$, we want to prove that $\deg(m) < \deg(n)$ if there is a non-identity morphism in $\square^{\text{comb}}_{\text{semi}}(\,m\,,\,n\,)$. This can be done by comparing $m$ and $n$:

- If $m < n$, then $\deg(m) < \deg(n)$ holds by unfolding the degree function.
- If $m = n$, then only the ternary number $\langle\star\star\ldots\star\rangle$ can be in $\square^{\text{comb}}_{\text{semi}}(\,m\,,\,n\,)$, so it doesn't have a non-identity morphism.
- If $m > n$, then $\square^{\text{comb}}_{\text{semi}}(\,m\,,\,n\,)$ contains no morphisms because any ternary number cannot have the occurrences of $\star$ more than total number of digits itself. QED

Chapter

# 5

# Categorifing Shapes by a Graph-Theoretic Framework

This chapter explores a novel framework to transform geometric shapes, such as simplices and cubes, into the categories of the corresponding shapes described in chapter 4, by using graphs and their morphisms as the intermediate representation.

Although the main motivation of this new framework is to rigorously define twisted cubes later in chapter 6, the framework itself deserves its own attention because it is a general framework that is applicable to any other shapes that resemble convex polytopes, that have been discussed in convention 4.5.

## § 5.1 Graph-theoretic Representation

This section defines data structures and related concepts from graph and order theories that are essential for our framework later in the rest of this chapter.

### § 5.1.1 Definitions of Graphs and Graph Morphisms

Although there are many varieties of graphs in graph theory; this thesis only considers *directed graphs* that allow loops but forbid parallel edges nor weighted edges. This restriction makes any considered graph to be equivalent to its set of nodes equipped with an *endorelation* representing its edges. For computational purposes, we further require that the each graph has finite nodes the endorelation of edges is decidable.

◤ **Definition 5.1** **(graphs)** ⬣

A record G is called a graph, if it contains these fields:

- G.Nodes — a set of things called *nodes* (a.k.a. vertices).
- G.isEdge — a function that takes two nodes s and t then returns a proposition stating whether "there exists an edge from s to t" or not.
- G.hasFinNodes — a proof ensures that the graph has finite nodes.
- G.decideEdges — a function that takes two nodes s and t then returns a proof that ensures the proposition G.isEdge$(s, t)$ is decidable.

We may also omit G.hasFinNodes and G.decideEdges if it is obvious to do so. ◢

◤ **Convention 5.2** **(alternative representation of graphs)** ⬣
Alternative to definition 5.1, we may replace the fields G.isEdge and G.decideEdges with another (pseudo) field called G.Edges, which is a decidable subset of the Cartesian product $(\mathsf{G.Nodes} \times \mathsf{G.Nodes})$ where each pair $\langle\, \mathtt{s}\, ,\, \mathtt{t}\, \rangle$ represents an *edge* from s to t.

This is because we can recreate G.isEdge as a function that takes nodes s and t then returns a proposition stating whether "the pair $\langle\, \mathtt{s}\, ,\, \mathtt{t}\, \rangle$ is in G.Edges" or not. The resulting proposition is also decidable because G.Edges is a decidable subset of a finite set (due to G.hasFinNodes); therefore we get G.decideEdges for free.

On the other hand, we can also define G.Edges by iteratively executes the function G.isEdge against every possible pair of nodes $\langle\, \mathtt{s}\, ,\, \mathtt{t}\, \rangle$; if $\mathsf{G.isEdge}(\mathtt{s}, \mathtt{t})$ is true, then $\langle\, \mathtt{s}\, ,\, \mathtt{t}\, \rangle$, which is now a legitimate edge, will be added into G.Edges.

The algorithms from two paragraphs above allow us to use definition 5.1 and the alternative definition with G.Edges interchangeably. ◢

The most basic examples of graphs are the *empty graph* (a.k.a. *null graph*), *trivial graph*, and *unit graph*, which are denoted by $\mathbb{0}_{\mathsf{grp}}$, $\epsilon_{\mathsf{grp}}$, and $\mathbb{1}_{\mathsf{grp}}$, respectively.

◤ **Example 5.3** ⬣ We define $\mathbb{0}_{\mathsf{grp}}$, $\epsilon_{\mathsf{grp}}$, and $\mathbb{1}_{\mathsf{grp}}$ to be the empty, trivial, and unit graphs, respectively, where:

| | | | | | | | | |
|---|---|---|---|---|---|---|---|---|
| $\mathbb{0}_{\mathsf{grp}}.\mathsf{Nodes}$ | $:\equiv$ | $\mathbb{0}_{\mathsf{fin}}$ | $\mathbb{0}_{\mathsf{grp}}.\mathsf{isEdge}(_,_)$ | $:\equiv$ | $\bot$ | $\mathbb{0}_{\mathsf{grp}}.\mathsf{Edges}$ | $:\equiv$ | $\emptyset$ |
| $\epsilon_{\mathsf{grp}}.\mathsf{Nodes}$ | $:\equiv$ | $\mathbb{1}_{\mathsf{fin}}$ | $\epsilon_{\mathsf{grp}}.\mathsf{isEdge}(_,_)$ | $:\equiv$ | $\bot$ | $\epsilon_{\mathsf{grp}}.\mathsf{Edges}$ | $:\equiv$ | $\emptyset$ |
| $\mathbb{1}_{\mathsf{grp}}.\mathsf{Nodes}$ | $:\equiv$ | $\mathbb{1}_{\mathsf{fin}}$ | $\mathbb{1}_{\mathsf{grp}}.\mathsf{isEdge}(_,_)$ | $:\equiv$ | $\top$ | $\mathbb{1}_{\mathsf{grp}}.\mathsf{Edges}$ | $:\equiv$ | $\{\, \langle\, 0\, ,\, 0\, \rangle\, \}$ |

Please note that, the definitions of these graphs use only the first and second columns; however, the third column shows how to alternatively define these graphs according to convention 5.2. ◢

Next, we explore the notion of notation of a *graph morphism*, which is a *homomorphism* between two graphs; that is, a function between sets of nodes, together with a requirement that such the function preserves edges.

◤ **Definition 5.4** **(graph morphisms)** ⬣

A record $\mathsf{f}$ is called a *graph homomorphism* from a graph $\mathsf{G}$ to another graph $\mathsf{G}'$, if it contains these fields:

- $\mathsf{f.mapNode}$ — a function that maps nodes from $\mathsf{G.Nodes}$ to $\mathsf{G'.Nodes}$.
  Please note that, we usually abbreviate $\mathsf{f.mapNode}(\,v\;:\;\mathsf{G.Nodes}\,)$ as $\mathsf{f}(v)$.
- $\mathsf{f.mapEdge}$ — a proof ensure that $\mathsf{f}$ preserves edges, i.e. we have

$$\mathsf{f.mapEdge}(s,t)\quad:\quad \mathsf{G.isEdge}(s,t)\;\rightarrow\;\mathsf{G'.isEdge}(\mathsf{f}(s),\mathsf{f}(t)).$$

  for each nodes $s$ and $t$ in $\mathsf{G.Nodes}$. ◢

◤ **Lemma 5.5** **(extensionality of graph morphisms)** ⬣

Let $\mathsf{f}$ and $\mathsf{g}$ be graph homomorphisms from a graph $\mathsf{G}$ to another graph $\mathsf{G}'$;
if $(\mathsf{f.mapNode}(v) \;=\; \mathsf{g.mapNode}(v))$ for every $v\;:\;\mathsf{G.Nodes}$, then $(\mathsf{f}\;=\;\mathsf{g})$. ◢

*Proof.* To prove that $(\mathsf{f} = \mathsf{g})$, we need to prove that $(\mathsf{f.mapNode} = \mathsf{g.mapNode})$ and $(\mathsf{f.mapEdge} = \mathsf{g.mapEdge})$. The former can be achieved by applying function extensionality to the assumption. The latter can also be achieved by applying function extensionality to this new auxiliary goal.

$$\mathsf{f.mapEdge}(s,t) \;=\; \mathsf{g.mapEdge}(s,t) \qquad \text{for every} \quad s,\;t\;:\;\mathsf{G.Nodes}$$

To this auxiliary goal, both terms on the equation have the same type because $\mathsf{f.mapNode}(s)$ and $\mathsf{f.mapNode}(t)$ are equal to $\mathsf{g.mapNode}(s)$ and $\mathsf{g.mapNode}(t)$, respectively.
The types belongs to `Prop`; therefore, $\mathsf{f.mapEdge}(s,t)$ is equal to $\mathsf{g.mapEdge}(s,t)$ by proposition extensionality. QED

## § 5.1.2 Graph Paths

The next entities to be defined are *paths* on graphs. We start by defining graphs that act as the "templates" of paths called *linear graphs*. Then, we define a path of any graph $G$ to be a graph homomorphism from one of these linear graphs to the graph $G$ itself.

◤ **Definition 5.6** **(linear graphs)** ⬣

Let $n : \mathbb{N}$, we define $\texttt{line}^{n}_{\mathsf{grp}}$ to be a graph where

$$\begin{aligned} \texttt{line}^{n}_{\mathsf{grp}}.\mathsf{Nodes} &:\equiv \texttt{fin}(n+1) \\ \texttt{line}^{n}_{\mathsf{grp}}.\mathsf{Edges} &:\equiv \{ \langle i, i+1 \rangle \mid i : \texttt{fin}(n) \} \end{aligned}$$ ◢

◤ **Lemma 5.7** ⬣ The empty and unit graphs can be presented as the linear graphs of lengths 0 and 1, respectively.

$$\mathbb{0}_{\mathsf{grp}} :\equiv \texttt{line}^{0}_{\mathsf{grp}} \qquad \mathbb{1}_{\mathsf{grp}} :\equiv \texttt{line}^{1}_{\mathsf{grp}}$$ ◢

◤ **Definition 5.8** **(graph paths)** ⬣

A record $p$ is called a *path* of a graph $G$, if it contains these fields:

- $p.\mathsf{length}$ — the total number of edges in the path.
- $p.\mathsf{map}$ — a graph homomorphism from $\texttt{line}^{p.\mathsf{length}}_{\mathsf{grp}}$ to $G$ itself.

Please note that, the field $p.\mathsf{map}$ is overloaded as the path $p$ itself. For example, the source and target of $p$ are $p(0)$ and $p(p.\mathsf{length} - 1)$, respectively. ◢

### § 5.1.3 Category of Graphs

Intuitively speaking, if one see a graph as an *upgraded* version of its set of nodes (similar to example 2.8), then a graph morphism is just an upgraded version of its underlining function (by equipping the condition that all edges are preserved). This allow us to define the *category of graphs* using **Set**.

◤ **Definition 5.9** ⬣ We define **Graph** to be the *category of graphs* where

- objects are graphs defined in definition 5.1 (or convention 5.2),
- morphisms are graph homomorphisms defined in definition 5.4, and
- the rest of fields can be canonically derived from nodes and node-maps which are objects and morphisms in **Set**. ◢

There are many other concepts about sets (or more precisely, *topos* [Topos; Lan92; Gol06]) that are applicable to graphs. For example, we says that a set X can be regarded as a *subset* of another set Y if there is an injective function from X to Y; this statement can be enriched for graphs as follows.

◤ **Definition 5.10** ⬣ Let $G, G' : \mathsf{ob}(\textbf{Graph})$, we says that G is a *subgraph* of $G'$ if we have $f : \textbf{Graph}(G, G')$ such that the function $f.\mathsf{mapNode}$ is injective. ◢

◤ **Lemma 5.11** ⬣ The category **Graph** has $\mathbb{0}_{\mathsf{grp}}$ and $\mathbb{1}_{\mathsf{grp}}$ as its initial object and terminal object, respectively. ◢

◤ **Lemma 5.12** ⬣ If $\mathcal{F}$ is an endofunctor on **Graph** and $G, G' : \mathsf{ob}(\textbf{Graph})$,

$$\text{then} \quad G \cong G' \quad \text{implies} \quad \mathcal{F}.\mathsf{mapObj}(G) \cong \mathcal{F}.\mathsf{mapObj}(G').$$ ◢

*Proof.* Let $\langle\ \phi : \mathbf{Graph}(G, G'),\ \ \psi : \mathbf{Graph}(G', G),\ \ \eta : \mathsf{id}(G) = \phi \cdot \psi,$ and $\varepsilon : \psi \cdot \phi = \mathsf{id}(G')\ \rangle$ to be a quadruple obtained from $(G \cong G')$. Then, we define

$$\begin{aligned}
\phi' &: \mathbf{Graph}(\mathcal{F}.\mathsf{mapObj}(G), \mathcal{F}.\mathsf{mapObj}(G')) &&:\equiv \mathcal{F}.\mathsf{mapHom}(\phi),\\
\psi' &: \mathbf{Graph}(\mathcal{F}.\mathsf{mapObj}(G'), \mathcal{F}.\mathsf{mapObj}(G)) &&:\equiv \mathcal{F}.\mathsf{mapHom}(\psi),\\
\eta' &: \mathsf{id}(\mathcal{F}.\mathsf{mapObj}(G)) = \phi' \cdot \psi' &&:\equiv \mathsf{cong}(\mathcal{F}.\mathsf{mapHom}, \eta), \quad \text{and}\\
\varepsilon' &: \psi' \cdot \phi' = \mathsf{id}(\mathcal{F}.\mathsf{mapObj}(G')) &&:\equiv \mathsf{cong}(\mathcal{F}.\mathsf{mapHom}, \varepsilon)
\end{aligned}$$

to construct $\langle \phi', \psi', \eta', \varepsilon' \rangle \ :\ \mathcal{F}.\mathsf{mapObj}(G) \cong \mathcal{F}.\mathsf{mapObj}(G')$. QED

In the homotopy type theory setting, lemma 5.12 is also true even if $\mathcal{F}$ is downgraded from an endofunctor on **Graph** to an endofunction on graphs. This is shown in lemma 5.13.

◤ **Lemma 5.13** ⬢ Given a context with homotopy type theory. If $F$ is an endofunction on graphs and $G, G' : \mathsf{ob}(\mathbf{Graph})$

$$\text{then} \quad G \cong G' \quad \text{implies} \quad F(G) \cong F(G'). \qquad ◢$$

*Proof.*

$$\begin{aligned}
(G \cong G') &\cong (G = G') && \text{[by the univalence]}\\
&\cong (\mathcal{F}.\mathsf{mapObj}(G) = \mathcal{F}.\mathsf{mapObj}(G')) && \text{[by identity on definition 5.1]}\\
(G \cong G') &\cong (\mathcal{F}.\mathsf{mapObj}(G) \cong \mathcal{F}.\mathsf{mapObj}(G')) && \text{[by the univalence]}
\end{aligned}$$

QED

### § 5.1.4 Borrowing Concepts from Order Theory

Because an endorelation is used to represents the edges of a graph; therefore, we should be able to borrow some concepts from order theory to our framework; in particular, the predicates (and closures later in subsection 5.1.5) of reflexivity and transitivity (please note that, symmetry is not presented here because we don't use it in our framework).

◤ **Definition 5.14** ⬣ We define `isRefl`, `isTran` `isIrrefl`, and `isIntran` to be predicates of reflexivity, transitivity, irreflexivity, and (strong) intransitivity, respectively. Let G be a graph and let s, t, and v be arbitrary nodes of G, then:

$$\begin{aligned}
\texttt{isRefl}(G) &:\equiv \forall(\, v : G.\mathsf{Nodes}\,) \to G.\mathsf{isEdge}(v, v) \\
\texttt{isTran}(G) &:\equiv \forall(\, s,\ v,\ t : G.\mathsf{Nodes}\,) \to \\
&\qquad G.\mathsf{isEdge}(s, v) \to G.\mathsf{isEdge}(v, t) \to G.\mathsf{isEdge}(s, t) \\
\texttt{isIrrefl}(G) &:\equiv \forall(\, v : G.\mathsf{Nodes}\,) \to \neg\ G.\mathsf{isEdge}(v, v) \\
\texttt{isIntran}(G) &:\equiv \forall(\, s,\ v,\ t : G.\mathsf{Nodes}\,) \to \\
&\qquad G.\mathsf{isEdge}(s, v) \to G.\mathsf{isEdge}(v, t) \to \neg\ G.\mathsf{isEdge}(s, t)
\end{aligned}$$ ◢

Although the predicates `isRefl` and `isIrrefl` are mutually exclusive, i.e.

$$\forall(\, G : \mathsf{ob}(\mathbf{Graph})\,) \to \neg\ (\texttt{isRefl}(G) \wedge \texttt{isIrrefl}(G)),$$

both predicates are *not* complement of each other, i.e.

$$\exists(\, G : \mathsf{ob}(\mathbf{Graph})\,) \times (\neg\ \texttt{isRefl}(G) \wedge \neg\ \texttt{isIrrefl}(G)).$$

For example, that G can be a graph such that

$$G.\mathsf{Nodes} :\equiv \mathbb{2}_{\mathsf{fin}} \quad \text{and} \quad G.\mathsf{Edges} :\equiv \{\, \langle 0 , 1 \rangle , \langle 1 , 1 \rangle \,\}.$$

The predicates `isTran` and `isIntran` also have the same relationship as `isRefl` and `isIrrefl` above. In addition, the predicate `isIntran` also implies the predicate `isIrrefl`. Please note that, the reflexivity is discussed in more detail than transitivity here because the former is used more often than the latter.

### § 5.1.5 Reflexive and Transitive Closures

This subsection defines reflexive and transitive closures. We first define these closures as functions that take a graph then return another graph as shown in definitions 5.15 and 5.16.

◤ **Definition 5.15** ⬣ For any graph G, we define a *reflexive closure* of G, denoted as $\mathtt{refl}_{\mathtt{cl}}(G)$, to be another graph where:

$$\mathtt{refl}_{\mathtt{cl}}(G).\mathsf{Nodes} \quad :\equiv \quad G.\mathsf{Nodes}$$

$$\mathtt{refl}_{\mathtt{cl}}(G).\mathsf{Edges} \quad :\equiv \quad G.\mathsf{Edges} \ \cup \ \{ \langle v , v \rangle \mid v : G.\mathsf{Nodes} \}$$ ◢

◤ **Definition 5.16** ⬣ For any graph G, we define a *transitive closure* of G, denoted as $\mathtt{tran}_{\mathtt{cl}}(G)$, to be another graph where:

$$\mathtt{tran}_{\mathtt{cl}}(G).\mathsf{Nodes} \quad :\equiv \quad G.\mathsf{Nodes}$$

$$\mathtt{tran}_{\mathtt{cl}}(G).\mathsf{Edges} \quad :\equiv \quad \{ \langle p(0) , p(p.\mathsf{length} - 1) \rangle \mid \text{each path } p \text{ in } G \text{ such that } (p.\mathsf{length} \geqslant 1) \}$$ ◢

◤ **Lemma 5.17** ⬣ Given that $G : \mathrm{ob}(\textbf{Graph})$, the following propositions hold:

$$\mathtt{isRefl}(\mathtt{refl}_{\mathtt{cl}}(G)) \qquad \mathtt{refl}_{\mathtt{cl}}(\mathtt{refl}_{\mathtt{cl}}(G)) = \mathtt{refl}_{\mathtt{cl}}(G)$$

$$\mathtt{isTran}(\mathtt{tran}_{\mathtt{cl}}(G)) \qquad \mathtt{tran}_{\mathtt{cl}}(\mathtt{tran}_{\mathtt{cl}}(G)) = \mathtt{tran}_{\mathtt{cl}}(G)$$

$$\mathtt{refl}_{\mathtt{cl}}(\mathtt{tran}_{\mathtt{cl}}(G)) = \mathtt{tran}_{\mathtt{cl}}(\mathtt{refl}_{\mathtt{cl}}(G))$$ ◢

◤ **Lemma 5.18** ⬣ Let $G$ be a graph and let $s$ and $t$ be nodes on $G$ that are not equal to each other, then $\mathtt{tran}_{\mathtt{cl}}(G).\mathsf{isEdge}(s,t)$ is equivalent to $G.\mathsf{isEdge}(s,t)$ ◢

◤ **Lemma 5.19** ⬣ Let $F$ is an endofunction on graphs that preserves reflexivity (i.e. if $\mathtt{isRefl}(G')$ then $\mathtt{isRefl}(F(G'))$ for every graph $G'$) and let $G$ be a graph, then $\mathtt{refl}_{\mathtt{cl}}(F(G)) \cong F(\mathtt{refl}_{\mathtt{cl}}(G))$ in the category **Graph**. ◢

*Proof.* We know that the graphs on both sides of the goal has the same set of nodes since both $\mathtt{refl}_{\mathtt{cl}}(F(G)).\mathsf{Nodes}$ and $F(\mathtt{refl}_{\mathtt{cl}}(G)).\mathsf{Nodes}$ can be evaluated to $F(G).\mathsf{Nodes}$; therefore, this left us to prove that, for every $F(G).\mathsf{Nodes}$,

$$\mathtt{refl}_{\mathtt{cl}}(F(G)).\mathsf{isEdge}(s,t) \quad \text{if and only if} \quad F(\mathtt{refl}_{\mathtt{cl}}(G)).\mathsf{isEdge}(s,t).$$

To prove the goal above, we compare the node $s$ against the node $t$.

- If they are equal, then the edge $\langle s, t \rangle$ must in exists in both graphs because they are reflexive. The graph $\mathtt{refl}_{\mathtt{cl}}(F(G))$ is reflexive by lemma 5.17 whereas $F(\mathtt{refl}_{\mathtt{cl}}(G))$ is reflexive because $F$ preserves the reflexivity from $\mathtt{refl}_{\mathtt{cl}}(G)$.
- If they are not equal, then both propositions on each side can be evaluated to $F(G).\mathsf{isEdge}(s,t)$ due to lemma 5.18; therefore, both propositions on each side are equivalent to each other. QED

◤ **Definition 5.20** **(endofunctor of the reflexive closure)** ⬣

We define $\texttt{refl}^{\mathsf{ftr}}_{\texttt{cl}}$ to be an endofunctor analogous to the endofunction $\texttt{refl}_{\texttt{cl}}$, where:

$$\texttt{refl}^{\mathsf{ftr}}_{\texttt{cl}}.\mathsf{mapObj} \ :\equiv \ \texttt{refl}_{\texttt{cl}} \qquad \text{and} \qquad \texttt{refl}^{\mathsf{ftr}}_{\texttt{cl}}.\mathsf{mapHom} \ :\equiv \ \texttt{refl}^{\mathsf{mor}}_{\texttt{cl}}$$

such that $\texttt{refl}^{\mathsf{mor}}_{\texttt{cl}}$ is defined to be a function that takes $f : \mathbf{Graph}(G, G')$ then returns $\texttt{refl}^{\mathsf{mor}}_{\texttt{cl}}(f) : \mathbf{Graph}(\texttt{refl}_{\texttt{cl}}(G), \texttt{refl}_{\texttt{cl}}(G'))$, where the node mapper $\texttt{refl}^{\mathsf{mor}}_{\texttt{cl}}(f).\mathsf{mapNode}$ is $f.\mathsf{mapNode}$ and the edge mapper is a function that takes $e : \texttt{refl}_{\texttt{cl}}(G).\mathsf{isEdge}(s, t)$ then returns

$$\texttt{refl}^{\mathsf{mor}}_{\texttt{cl}}(f).\mathsf{mapEdge}(s,t)(e) \ : \ \texttt{refl}_{\texttt{cl}}(G').\mathsf{isEdge}(f(s), f(t)).$$

To define the output above, we need to prove that the edge $\langle f(s), f(t) \rangle$ is in $\texttt{refl}_{\texttt{cl}}(G')$ given that $\langle s, t \rangle$ is in $\texttt{refl}_{\texttt{cl}}(G)$. We compare $s$ and $t$.

- If they are equal, then we also know that $(f(s) = f(t))$, so the edge $\langle f(s), f(t) \rangle$ must exist because $\texttt{refl}_{\texttt{cl}}(G')$ is reflexive.
- If they are not equal, then the edge $\langle s, t \rangle$ is also in the graph $G$ where it can be mapped by $f.\mathsf{mapEdge}$ then become $\langle f(s), f(t) \rangle$ in the graph $G'$, it is also in $\texttt{refl}_{\texttt{cl}}(G')$ by definition 5.15. ◢

◤ **Lemma 5.21** ⬣ Let $G$ and $G'$ be graphs such that $(G \cong G')$ in the category **Graph**, then we also have $\texttt{refl}_{\texttt{cl}}(G) \cong \texttt{refl}_{\texttt{cl}}(G')$. ◢

*Proof.* By applying the functor $\texttt{refl}^{\mathsf{ftr}}_{\texttt{cl}}$ into lemma 5.12 QED

◤ **Definition 5.22** **(endofunctor of the transitive closure)** ⬣

We define $\mathtt{tran}^{\mathsf{ftr}}_{\mathtt{cl}}$ to be an endofunctor analogous to the endofunction $\mathtt{tran}_{\mathtt{cl}}$, where:

$$\mathtt{tran}^{\mathsf{ftr}}_{\mathtt{cl}}.\mathsf{mapObj} \quad :\equiv \quad \mathtt{tran}_{\mathtt{cl}} \qquad \text{and} \qquad \mathtt{tran}^{\mathsf{ftr}}_{\mathtt{cl}}.\mathsf{mapHom} \quad :\equiv \quad \mathtt{tran}^{\mathsf{mor}}_{\mathtt{cl}}$$

such that $\mathtt{tran}^{\mathsf{mor}}_{\mathtt{cl}}$ is defined to be a function that takes $\mathsf{f} \,:\, \textbf{Graph}(\,\mathsf{G}\,,\,\mathsf{G}'\,)$ then returns $\mathtt{tran}^{\mathsf{mor}}_{\mathtt{cl}}(\mathsf{f}) \,:\, \textbf{Graph}(\,\mathtt{tran}_{\mathtt{cl}}(\mathsf{G})\,,\,\mathtt{tran}_{\mathtt{cl}}(\mathsf{G}')\,)$, where the node mapper $\mathtt{tran}^{\mathsf{mor}}_{\mathtt{cl}}(\mathsf{f}).\mathsf{mapNode}$ is $\mathsf{f}.\mathsf{mapNode}$ and the edge mapper is a function that takes $e \,:\, \mathtt{tran}_{\mathtt{cl}}(\mathsf{G}).\mathsf{isEdge}(\mathsf{s},\mathsf{t})$ then returns

$$\mathtt{tran}^{\mathsf{mor}}_{\mathtt{cl}}(\mathsf{f}).\mathsf{mapEdge}(\mathsf{s},\mathsf{t})(e) \quad : \quad \mathtt{tran}_{\mathtt{cl}}(\mathsf{G}').\mathsf{isEdge}(\mathsf{f}(\mathsf{s}),\mathsf{f}(\mathsf{t})).$$

To define the output above, we need to prove that the edge $\langle\,\mathsf{f}(\mathsf{s})\,,\,\mathsf{f}(\mathsf{t})\,\rangle$ is in $\mathtt{tran}_{\mathtt{cl}}(\mathsf{G}')$ given that $\langle\,\mathsf{s}\,,\,\mathsf{t}\,\rangle$ is in $\mathtt{tran}_{\mathtt{cl}}(\mathsf{G})$. By definition 5.16, the goal now is to find a path in $\mathsf{G}'$ from $\mathsf{f}(\mathsf{s})$ to $\mathsf{f}(\mathsf{t})$ given that there is a path in $\mathsf{G}$ from $\mathsf{s}$ to $\mathsf{t}$.
Let $\mathsf{p}$ be the assuming path, we know that $\mathsf{p}.\mathsf{map}$ is in $\textbf{Graph}(\,\mathtt{line}^{\mathsf{p}.\mathsf{length}}_{\mathsf{grp}}\,,\,\mathsf{G}\,)$; we also know that $\mathsf{f}$ is in $\textbf{Graph}(\,\mathsf{G}\,,\,\mathsf{G}'\,)$, this allow us to compose $\mathsf{p}.\mathsf{map}$ with $\mathsf{f}$ that resulting in a graph homomorphism from $\mathtt{line}^{\mathsf{p}.\mathsf{length}}_{\mathsf{grp}}$ to $\mathsf{G}'$ that can be used to construct a path in $\mathsf{G}'$ from the node $\mathsf{f}(\mathsf{s})$ to the node $\mathsf{f}(\mathsf{t})$. ◢

◤ **Lemma 5.23** ⬣ Let $\mathsf{G}$ and $\mathsf{G}'$ be graphs such that $(\mathsf{G} \,\cong\, \mathsf{G}')$ in the category **Graph**, then we also have $\quad \mathtt{tran}_{\mathtt{cl}}(\mathsf{G}) \,\cong\, \mathtt{tran}_{\mathtt{cl}}(\mathsf{G}')$. ◢

*Proof.* By applying the functor $\mathtt{tran}^{\mathsf{ftr}}_{\mathtt{cl}}$ into lemma 5.12 QED

## § 5.2 Face Graphs for Non-Degenerated Shapes

Before jumping into the full versions of the simplex or standard cube categories, it is easier to first focus on the non-degenerated versions of them. Here, we construct the graph-theoretic versions of semi-simplex category and standard-semi-cube category as full sub-categories of **Graph** determined by the families of graphs called *face graphs*, which encode all points and lines of the corresponded shapes as their nodes and edges, respectively.

### § 5.2.1 Face Graphs for Semi-Simplices

For each dimension $n : \mathbb{N}$, we will define the *face graph for the canonical* $n$*-simplex*, denoted as $\mathbb{F}^n_\triangle$, that acts as a template for a non-degenerated $n$-simplex.

◤ **Definition 5.24** (face graphs for simplices) ⬣
Let $n : \mathbb{N}$, we define a graph $\mathbb{F}^n_\triangle$ to be the face graph for $\triangle^n_{\mathsf{top}}$

$$\mathbb{F}^n_\triangle.\mathsf{Nodes} \;:\equiv\; \mathtt{fin}(n+1) \qquad\qquad \mathbb{F}^n_\triangle.\mathsf{isEdge}(s,t) \;:\equiv\; (s < t)$$ ◢

◤ **Lemma 5.25** ⬣ Let $n : \mathbb{N}$, then $\mathbb{F}^n_\triangle \cong \mathtt{tran}_{\mathtt{cl}}(\mathtt{line}^n_{\mathsf{grp}})$ in **Graph**. ◢

◤ **Definition 5.26** **(graph theoretic version of semi simplex category)** ⬣

We define the category $\triangle^{\mathsf{graph}}_{\mathsf{semi}}$ to be a full-subcategory of **Graph** induced by the family of graphs $\mathbb{F}^{n}_{\triangle}$ for all $n : \mathbb{N}$. Please note that, we may use the number $n$ or the graph $\mathbb{F}^{n}_{\triangle}$ to denote each object in $\triangle^{\mathsf{graph}}_{\mathsf{semi}}$ interchangeably. ◢

Although each object $\mathbb{F}^{n}_{\triangle} : \mathsf{ob}(\triangle^{\mathsf{graph}}_{\mathsf{semi}})$ only encodes 0-faces and 1-faces of $\triangle^{n}_{\mathsf{top}}$ as its nodes and edges but the category $\triangle^{\mathsf{graph}}_{\mathsf{semi}}$ surprisingly contains all possible faces of $\triangle^{n}_{\mathsf{top}}$ by regarding them as *subgraphs* of $\mathbb{F}^{n}_{\triangle}$. To be precise, an $m$-face of $\triangle^{n}_{\mathsf{top}}$ can be represented as an *injective* graph morphism from $\mathbb{F}^{m}_{\triangle}$ to $\mathbb{F}^{n}_{\triangle}$.

For example, a 0-face of $\triangle^{n}_{\mathsf{top}}$ that is encoded as a node $v$ of $\mathbb{F}^{n}_{\triangle}$, can be regarded as a subgraph of $\mathbb{F}^{n}_{\triangle}$ by transforming $v$ into a graph morphism $\bar{v} : \triangle^{\mathsf{graph}}_{\mathsf{semi}}(0, n)$ below. Similarly a 1-face of $\triangle^{n}_{\mathsf{top}}$ that is encoded as an edge $e$ of $\mathbb{F}^{n}_{\triangle}$ can be regarded as a subgraph of $\mathbb{F}^{n}_{\triangle}$ by transforming $e$ into a graph morphism $\bar{e} : \triangle^{\mathsf{graph}}_{\mathsf{semi}}(1, n)$.

| | | | | | |
|---|---|---|---|---|---|
| $v$ | : | $\mathbb{F}^{n}_{\triangle}.\mathsf{Nodes}$ | $\bar{v}.\mathsf{mapNode}(0)$ | $:\equiv$ | $v$ |
| $\bar{v}$ | : | $\mathbf{Graph}(\mathbb{F}^{0}_{\triangle}, \mathbb{F}^{n}_{\triangle})$ | $\bar{e}.\mathsf{mapNode}(0)$ | $:\equiv$ | $s$ |
| $e$ | : | $\mathbb{F}^{n}_{\triangle}.\mathsf{isEdge}(s, t)$ | $\bar{e}.\mathsf{mapNode}(1)$ | $:\equiv$ | $t$ |
| $\bar{e}$ | : | $\mathbf{Graph}(\mathbb{F}^{1}_{\triangle}, \mathbb{F}^{n}_{\triangle})$ | $\bar{e}.\mathsf{mapEdge}(0, 1)$ | $:\equiv$ | $e$ |

In general, recall from section 4.2 that each $m$-face of $\triangle^{n}_{\mathsf{top}}$ can be uniquely represented as a morphism $f : \triangle^{\mathsf{comb}}_{\mathsf{semi}}(m, n)$. We can transform $f$ into $\bar{f} : \mathbf{Graph}(\mathbb{F}^{m}_{\triangle}, \mathbb{F}^{n}_{\triangle})$ where $\bar{f}.\mathsf{mapNode}$ is $f$ and $\bar{f}.\mathsf{mapEdge}$ comes from the strict monotonicity property of $f$.

Since the category $\triangle^{\mathsf{graph}}_{\mathsf{semi}}$ excludes all of degenerated simplices, every morphism of $\triangle^{\mathsf{graph}}_{\mathsf{semi}}(m, n)$ must represent an $m$-face of $n$-simplex, which is a subgraph of $\mathbb{F}^{n}_{\triangle}$. This requires us to assure that there are no non-injective morphisms in $\triangle^{\mathsf{graph}}_{\mathsf{semi}}$.

◤ **Remark 5.27** **(irreflexive graphs for semi shapes)** ⬣

Instead of the operator ( $\leqslant$ ), we use ( $<$ ) in the definition of $\mathbb{F}^n_\triangle$.isEdge in order to make $\mathbb{F}^n_\triangle$ irreflexive. This essentially restricts every morphism in $\triangle^{\text{graph}}_{\text{semi}}$ to be injective; this is because any two edges must have different endpoints; therefore, any non-injective node-map cannot find a suitable edge map counterpart.

In fact, every graph $G$ : ob(**Graph**) representing a semi-shape in this framework must satisfy `isIrrefl`(G). This is to make sure that all morphisms in the resulting category are monomorphisms. ◢

To show that this new $\triangle^{\text{graph}}_{\text{semi}}$ makes sense we show that it is isomorphic to the traditional $\triangle^{\text{comb}}_{\text{semi}}$ as follows.

◤ **Lemma 5.28** ⬣ The category $\triangle^{\text{graph}}_{\text{semi}}$ is isomorphic to the category $\triangle^{\text{comb}}_{\text{semi}}$. ◢

*Proof.* Since both $\triangle^{\text{graph}}_{\text{semi}}$ and $\triangle^{\text{comb}}_{\text{semi}}$ have the same set of objects, our goal now becomes

$$\triangle^{\text{graph}}_{\text{semi}}(m, n) \quad \cong \quad \triangle^{\text{comb}}_{\text{semi}}(m, n) \qquad \text{for every} \quad m,\ n : \mathbb{N}\,.$$

In the forward direction, we define $\phi : \triangle^{\text{graph}}_{\text{semi}}(m, n) \to \triangle^{\text{comb}}_{\text{semi}}(m, n)$. Given $f : \triangle^{\text{graph}}_{\text{semi}}(m, n)$, which is a graph homomorphism from $\mathbb{F}^m_\triangle$ to $\mathbb{F}^n_\triangle$, we define $\phi(f) : \triangle^{\text{comb}}_{\text{semi}}(m, n)$ where $\phi(f) :\equiv$ f.mapNode.

- We know that $\phi(f)$ is injective because $\mathbb{F}^n_\triangle$ is irreflexive so it is impossible for f.mapNode to map two distinct nodes to the same image.
- We also know that $\phi(f)$ is strict-monotone because, given $s,\ t$ : `fin`$(m+1)$, a proof of $(s < t)$ can be translated to a proof of $\mathbb{F}^m_\triangle$.isEdge$(s, t)$, which, in turn, get mapped by f.mapEdge and become a proof of $\mathbb{F}^n_\triangle$.isEdge$(f(s), f(t))$, which, in turn again get translated back to a proof of $(f(s) < f(t))$.

In the backward direction, we define $\psi : \triangle^{\mathsf{comb}}_{\mathsf{semi}}(m, n) \to \triangle^{\mathsf{graph}}_{\mathsf{semi}}(m, n)$. Given $f : \triangle^{\mathsf{comb}}_{\mathsf{semi}}(m, n)$, which is an injective strict-monotone function from $\mathtt{fin}(m+1)$ to $\mathtt{fin}(n+1)$, we define $\psi(f) : \mathbf{Graph}(\mathbb{F}^{m}_{\triangle}, \mathbb{F}^{n}_{\triangle})$.

- Regarding $\psi(f)$.mapNode, it is the function $f$ itself.
- Regarding $\psi(f)$.mapEdge, given $s, t : \mathtt{fin}(m+1)$, a proof of $\mathbb{F}^{m}_{\triangle}$.isEdge$(s,t)$ can be translated to a proof of $(s < t)$, which, in turn, get mapped by the strict-monotonicity of $f$ and become a proof of $(f(s) < f(t))$, which, in turn again, get translated back to a proof of $\mathbb{F}^{n}_{\triangle}$.isEdge$(f(s), f(t))$.

Finally, it is obvious that both $\phi \cdot \psi$ and $\psi \cdot \phi$ are identity morphisms in **Graph**. QED

◤ **Corollary 5.29** ⬣ The category $\triangle^{\mathsf{graph}}_{\mathsf{semi}}$ is a direct category. ◢

*Proof.* By applying lemma 5.28 to lemma 4.27. QED

## § 5.2.2 Graph Definitions using Recursion

Alternative to definition 5.24, we can define $\mathbb{F}^{n}_{\triangle}$ recursively on the natural number $n$ by defining an endofunction, `cone`, that takes $\mathbb{F}^{n}_{\triangle}$ and returns $\mathbb{F}^{(n+1)}_{\triangle}$.

In addition, if $\mathtt{isRefl}(G)$, then we add an edge from the new node to itself; this is to make sure that `cone` preserves reflexivity, which is a requirement of lemma 5.19 that will be used later in section 5.3.

◤ **Definition 5.30** **(cone iterator)** ⬣

Let G be a graph, we define another graph $\texttt{cone}(G)$ as a modification of the graph G by adding a new node that attaches an edge from itself to each of original nodes. If $\texttt{isRefl}(G)$, then include a looping edge on the new node.

$$\begin{aligned}
\texttt{cone}(G).\mathsf{Nodes} &:\equiv \mathbb{1}_{\mathsf{fin}} \uplus G.\mathsf{Nodes} \\
\texttt{cone}(G).\mathsf{Edges} &:\equiv \{ \langle \texttt{inr}(s), \texttt{inr}(t) \rangle \mid \langle s, t \rangle : G.\mathsf{Edges} \} \\
&\cup \{ \langle \texttt{inl}(0), \texttt{inr}(v) \rangle \mid v : G.\mathsf{Nodes} \} \\
&\cup (\text{if } \texttt{isRefl}(G) \text{ then } \{ \langle \texttt{inl}(v), \texttt{inr}(v) \rangle \} \text{ else } \emptyset )
\end{aligned}$$ ◢

Similar to $\texttt{refl}_{\texttt{cl}}$ and $\texttt{tran}_{\texttt{cl}}$, we want to upgrade the endofunction $\texttt{cone}$ to an endofunctor; unfortunately, this is not possible because of the edge $\langle \texttt{inl}(0), \texttt{inl}(0) \rangle$ that exists conditionally on the reflexivity.

The counter example is any graph morphism in **Graph**$(G, G')$ such that the graph G is reflexive but the graph $G'$ is not; this is because there will be the edge $\langle \texttt{inl}(0), \texttt{inl}(0) \rangle$ in $\texttt{cone}(G)$ but not in $\texttt{cone}(G')$ so we can't define any graph morphism from $\texttt{cone}(G)$ to $\texttt{cone}(G')$. Nevertheless, will can still get $(\texttt{cone}(G) \cong \texttt{cone}(G'))$ by applying lemma 5.13 instead of lemma 5.12.

◤ **Lemma 5.31** ⬣ Let G and $G'$ be graphs such that $(G \cong G')$ in the category **Graph**, then we also have $\texttt{cone}(G) \cong \texttt{cone}(G')$. ◢

*Proof.* By applying the function $\texttt{cone}$ into lemma 5.13 QED

◤ **Definition 5.32** **(face graphs for simplices using recursion)** ⬣

Let $n : \mathbb{N}$, we define a graph $\acute{\mathbb{F}}^n_\triangle$ to be a recursive version of graph $\mathbb{F}^n_\triangle$.

$$\acute{\mathbb{F}}^0_\triangle :\equiv \epsilon_{\mathsf{grp}} \qquad \acute{\mathbb{F}}^{(n+1)}_\triangle :\equiv \texttt{cone}(\acute{\mathbb{F}}^n_\triangle)$$ ◢

◤ **Lemma 5.33** ⬣ Let $n : \mathbb{N}$, then $\mathbb{F}_{\triangle}^{(n+1)} \cong \texttt{cone}(\mathbb{F}_{\triangle}^{n})$. ◢

*Proof.* We define $\phi$ : **Graph**$(\mathbb{F}_{\triangle}^{(n+1)}, \texttt{cone}(\mathbb{F}_{\triangle}^{n}))$ and $\psi$ : **Graph**$(\texttt{cone}(\mathbb{F}_{\triangle}^{n}), \mathbb{F}_{\triangle}^{(n+1)})$ as graph homomorphisms for forward and backward direction, respectively, where:

$$\phi.\mathsf{mapNode} \quad : \quad \texttt{fin}(n+2) \to (\mathbb{1}_{\mathsf{fin}} \uplus \texttt{fin}(n+1))$$
$$\psi.\mathsf{mapNode} \quad : \quad (\mathbb{1}_{\mathsf{fin}} \uplus \texttt{fin}(n+1)) \to \texttt{fin}(n+2)$$

$$\phi.\mathsf{mapNode}(0) :\equiv \texttt{inl}(0) \qquad \psi.\mathsf{mapNode}(\texttt{inl}(0)) :\equiv 0$$
$$\phi.\mathsf{mapNode}(n+1) :\equiv \texttt{inr}(n) \qquad \psi.\mathsf{mapNode}(\texttt{inr}(n)) :\equiv n+1$$

Regarding $\phi$.mapEdge and $\psi$.mapEdge, it is obvious that $\phi$ and $\psi$ preserve edges by inspecting $\phi$.mapNode and $\psi$.mapNode, respectively. Finally, it is obvious that both $\phi \cdot \psi$ and $\psi \cdot \phi$ are identity morphisms in **Graph**. QED

◤ **Theorem 5.34** ⬣ Let $n : \mathbb{N}$, then $\acute{\mathbb{F}}_{\triangle}^{n} \cong \mathbb{F}_{\triangle}^{n}$ in **Graph**. ◢

*Proof.* Using induction on the natural number $n$: The base ($\acute{\mathbb{F}}_{\triangle}^{0} \cong \mathbb{F}_{\triangle}^{0}$) case is obvious because both $\mathbb{F}_{\triangle}^{0}$ and $\acute{\mathbb{F}}_{\triangle}^{0}$ have a single node with no edges. Regarding the inductive step, assuming that ($\acute{\mathbb{F}}_{\triangle}^{n} \cong \mathbb{F}_{\triangle}^{n}$), we need to construct ($\acute{\mathbb{F}}_{\triangle}^{(n+1)} \cong \mathbb{F}_{\triangle}^{(n+1)}$).

$$\acute{\mathbb{F}}_{\triangle}^{(n+1)} \cong \texttt{cone}(\acute{\mathbb{F}}_{\triangle}^{n}) \quad \text{[by definition 5.32]}$$
$$\cong \texttt{cone}(\mathbb{F}_{\triangle}^{n}) \quad \text{[by the induction hypothesis and lemma 5.21]}$$
$$\acute{\mathbb{F}}_{\triangle}^{(n+1)} \cong \mathbb{F}_{\triangle}^{(n+1)} \quad \text{[by lemma 5.33]}$$

QED

◤ **Convention 5.35** **(simplices coercion from their recursive graphs)** ⬢

We define a coercion from $\acute{\mathbb{F}}^{n}_{\triangle}$ to $\mathbb{F}^{n}_{\triangle}$ using the isomorphic structure of theorem 5.34. This convention allows us to use both definitions 5.24 and 5.32 as the definition for face graph of $n$-simplex without duplication of related components. For instance, the category $\triangle^{\mathsf{graph}}_{\mathsf{semi}}$ doesn't need its recursive version because any graph morphism of type $\mathbf{Graph}(\acute{\mathbb{F}}^{m}_{\triangle}, \acute{\mathbb{F}}^{n}_{\triangle})$ will automatically be a morphism in $\triangle^{\mathsf{graph}}_{\mathsf{semi}}(m, n)$.

Recursive and non-recursive definitions are not superior to each other. On one hand, the recursive definition makes proof-by-induction become easier (and more natural). On the other hand, the non-recursive definition provide a compact and intuitive notation to mention about nodes and edges.

The duality of recursive and non-recursive definitions are not specific for simplices but persists through other shapes, such as standard cubes and even twisted cubes. Therefore, this convention of coercion will also apply to those shapes as well (provide that the shapes have isomorphic structure between recursive and non-recursive definitions, which is also the case in this thesis anyway). ◢

### § 5.2.3 Face Graphs for Standard Semi-Cubes

Similar to $\mathbb{F}^{n}_{\triangle}$, we will define the *face graph for the canonical standard* $n$*-cube*, denoted as $\mathbb{F}^{n}_{\square}$, that acts as a template for a non-degenerated standard $n$-cube where its nodes are binary numbers and a pair of these binary numbers is an edge iff their digits are pairwisely equal to one another except for one index that the digit of former must be less than the latter.

◤ **Definition 5.36** **(face graphs for standard cubes)** ⬣

Let $n : \mathbb{N}$, we define a graph $\mathbb{F}^n_\square$ to be the face graph for $\square^n_{\text{top}}$

$$\mathbb{F}^n_\square.\text{Nodes} \quad :\equiv \quad \texttt{binary}(n)$$

$$\mathbb{F}^n_\square.\text{isEdge}(\vec{s}, \vec{t}) \quad :\equiv \quad \exists(i : \texttt{fin}(n)) \times \forall(j : \texttt{fin}(n)) \to \text{if } (i = j) \text{ then } \vec{s}[j] < \vec{t}[j] \text{ else } \vec{s}[j] = \vec{t}[j]$$ ◢

◤ **Definition 5.37** **(graph theoretic version of standard semi cube category)** ⬣

We define the category $\square^{\text{graph}}_{\text{semi}}$ to be a full-subcategory of **Graph** induced by the family of graphs $\mathbb{F}^n_\square$ for all $n : \mathbb{N}$. Please note that, we may use the number $n$ or the graph $\mathbb{F}^n_\square$ to denote each object in $\square^{\text{graph}}_{\text{semi}}$ interchangeably. ◢

◤ **Lemma 5.38** ⬣ The category $\square^{\text{graph}}_{\text{semi}}$ is isomorphic to the category $\square^{\text{comb}}_{\text{semi}}$. ◢

*Proof.* Since both $\square^{\text{graph}}_{\text{semi}}$ and $\square^{\text{comb}}_{\text{semi}}$ have the same set of objects, our goal now becomes

$$\square^{\text{graph}}_{\text{semi}}(m, n) \quad \cong \quad \square^{\text{comb}}_{\text{semi}}(m, n) \qquad \text{for every} \quad m, n : \mathbb{N}.$$

In the forward direction, we define $\phi : \square^{\text{graph}}_{\text{semi}}(m, n) \to \square^{\text{comb}}_{\text{semi}}(m, n)$. Given $f : \square^{\text{graph}}_{\text{semi}}(m, n)$, which is a graph homomorphism from $\mathbb{F}^m_\square$ to $\mathbb{F}^n_\square$, we compute $\psi(f) : \square^{\text{comb}}_{\text{semi}}(m, n)$ using the following algorithm:

- Create a new variable named $\vec{t}$ that contains a ternary number of length $n$.
- Let $\vec{a}$ be a binary number of length $m$ where every digit is $\texttt{0}$, i.e. $\vec{a} :\equiv \langle \texttt{00} \ldots \texttt{0} \rangle$.
- Assign $\vec{t}$ to be $f(\vec{a})$. Please note that, $f(\vec{a})$ is a binary number of length $n$ but it can be casted as a ternary number because a binary number is just a ternary number that is impossible to contain any occurrences of $\star$.
- Iterate a local variable $i : \texttt{fin}(m)$ from 0 to $(m-1)$. In each iteration, we compare $f(\vec{a})$ against $f(\vec{b})$ where $\vec{b}$ is $\vec{a}$ but the digit at index $i$ is replaced from $\texttt{0}$ to $\texttt{1}$. If the comparison mismatched, replace the digit of $\vec{t}$ at index $j$ to be $\star$ where $j : \texttt{fin}(n)$ is the index of the different digit in the comparison.

- Return the variable $\vec{t}$ as $\psi(f) : \square^{\text{comb}}_{\text{semi}}(m, n)$.

In the backward direction, we define $\psi : \square^{\text{comb}}_{\text{semi}}(m, n) \to \square^{\text{graph}}_{\text{semi}}(m, n)$.
Given $\vec{t} : \square^{\text{comb}}_{\text{semi}}(m, n)$, which is a ternary number of length $n$ with $m$ occurrences of $\star$, we define $\psi(\vec{t}) : \textbf{Graph}(\mathbb{F}^m_\square, \mathbb{F}^n_\square)$ where

- Regarding $\psi(\vec{t})$.mapNode, it is a function that takes $\vec{b} : \texttt{binary}(m)$ then return $\vec{t}$ except that each occurrence $i$ of $\star$ is replaced by $\vec{b}[i]$.
- Regarding $\psi(\vec{t})$.mapEdge, it is easy to see that $\psi(\vec{t})$ preserves edges.

Finally, it is obvious that both $\phi \cdot \psi$ and $\psi \cdot \phi$ are identity morphisms in **Graph**. QED

◤ **Corollary 5.39** ⬣ The category $\square^{\text{graph}}_{\text{semi}}$ is a direct category. ◢

*Proof.* By applying lemma 5.38 to lemma 4.35. QED

Similar to $\acute{\mathbb{F}}^n_\triangle$, there is a recursive version of $\mathbb{F}^n_\square$, denoted as $\acute{\mathbb{F}}^n_\square$, that use an endofunction $\texttt{prism}_{\texttt{std}}(G)$ to iterate though $\epsilon_{\text{grp}}$ in total $n$ times.

◤ **Definition 5.40** **(standard prism iterator)** ⬣
Let $G$ be a graph, we define another graph $\texttt{prism}_{\texttt{std}}(G)$ as a modification of the graph $G$ by duplicating it into two copies then adding edges from nodes on the first copy to the second copy counterpart pairwisely.

$$
\begin{aligned}
\texttt{prism}_{\texttt{std}}(G).\text{Nodes} \quad &:\equiv \quad \mathbb{2}_{\text{fin}} \times G.\text{Nodes} \\
\texttt{prism}_{\texttt{std}}(G).\text{Edges} \quad &:\equiv \quad \{ \langle \langle b, \vec{s} \rangle, \langle b, \vec{t} \rangle \rangle \mid b : \mathbb{2}_{\text{fin}}, \ \langle \vec{s}, \vec{t} \rangle : G.\text{Edges} \} \\
&\cup \quad \{ \langle \langle 0, \vec{v} \rangle, \langle 1, \vec{v} \rangle \rangle \mid \vec{v} : G.\text{Nodes} \}
\end{aligned}
$$
◢

**Definition 5.41** Let G and G′ be graphs, and $f : \textbf{Graph}(G, G')$ we define $\mathtt{prism}^{\mathsf{mor}}_{\mathtt{std}}(f)$ to be a graph homomorphism from $\mathtt{prism}_{\mathtt{std}}(G)$ to $\mathtt{prism}_{\mathtt{std}}(G')$.

$$\mathtt{prism}^{\mathsf{mor}}_{\mathtt{std}}(f).\mathsf{mapNode} \; : \; (\mathbb{2}_{\mathsf{fin}} \times G.\mathsf{Nodes}) \to (\mathbb{2}_{\mathsf{fin}} \times G'.\mathsf{Nodes})$$

$$\mathtt{prism}^{\mathsf{mor}}_{\mathtt{std}}(f).\mathsf{mapNode} \; (\langle b, \vec{v} \rangle) :\equiv \langle b, f.\mathsf{mapNode}(\vec{v}) \rangle$$

$$\mathtt{prism}^{\mathsf{mor}}_{\mathtt{std}}(f).\mathsf{mapEdge}(s,t) \; : \; \mathtt{prism}_{\mathtt{std}}(G).\mathsf{isEdge}(s,t) \to \mathtt{prism}_{\mathtt{std}}(G').\mathsf{isEdge}(f(s), f(t))$$

$$\mathtt{prism}^{\mathsf{mor}}_{\mathtt{std}}(f).\mathsf{mapEdge} \; (\langle b, \vec{s} \rangle, \langle b, \vec{t} \rangle) \; (e) :\equiv f.\mathsf{mapEdge}(\vec{s}, \vec{t})(e)$$

$$\mathtt{prism}^{\mathsf{mor}}_{\mathtt{std}}(f).\mathsf{mapEdge} \; (\langle 0, \vec{v} \rangle, \langle 1, \vec{v} \rangle) \; (e) :\equiv \mathsf{cong}(f.\mathsf{mapNode}, e)$$

**Lemma 5.42** Given two composable graph homomorphisms f and g, then

$$\mathtt{prism}^{\mathsf{mor}}_{\mathtt{std}}(f \cdot g) = \mathtt{prism}^{\mathsf{mor}}_{\mathtt{std}}(f) \cdot \mathtt{prism}^{\mathsf{mor}}_{\mathtt{std}}(g).$$

*Proof.* By applying the following proof to lemma 5.5.

$$\begin{aligned}
& \mathtt{prism}^{\mathsf{mor}}_{\mathtt{std}}(f \cdot g).\mathsf{mapNode}(\langle b, \vec{v} \rangle) \\
= \; & \langle b, (f \cdot g).\mathsf{mapNode}(\vec{v}) \rangle \\
= \; & \langle b, g.\mathsf{mapNode}(f.\mathsf{mapNode}(\vec{v})) \rangle \\
= \; & \mathtt{prism}^{\mathsf{mor}}_{\mathtt{std}}(g).\mathsf{mapNode}(\langle b, f.\mathsf{mapNode}(\vec{v}) \rangle) \\
= \; & \mathtt{prism}^{\mathsf{mor}}_{\mathtt{std}}(g).\mathsf{mapNode}(\mathtt{prism}^{\mathsf{mor}}_{\mathtt{std}}(f).\mathsf{mapNode}(\langle b, \vec{v} \rangle)) \\
= \; & (\mathtt{prism}^{\mathsf{mor}}_{\mathtt{std}}(f).\mathsf{mapNode} \cdot \mathtt{prism}^{\mathsf{mor}}_{\mathtt{std}}(g).\mathsf{mapNode})(\langle b, \vec{v} \rangle) \\
= \; & (\mathtt{prism}^{\mathsf{mor}}_{\mathtt{std}}(f) \cdot \mathtt{prism}^{\mathsf{mor}}_{\mathtt{std}}(g)).\mathsf{mapNode}(\langle b, \vec{v} \rangle)
\end{aligned}$$

QED

**Lemma 5.43** Let G be a graph, then

$$\mathtt{prism}^{\mathsf{mor}}_{\mathtt{std}}(\mathsf{id}(G)) = \mathsf{id}(\mathtt{prism}_{\mathtt{std}}(G)).$$

*Proof.* By applying the following proof to lemma 5.5.

$$\begin{aligned}\texttt{prism}^{\mathsf{mor}}_{\texttt{std}}(\mathsf{id}(G)).\mathsf{mapNode}(\langle\, b\,,\ \vec{v}\,\rangle) &= \langle\, b\,,\ \mathsf{id}(G).\mathsf{mapNode}(\vec{v})\,\rangle \\ &= \langle\, b\,,\ \vec{v}\,\rangle \\ \texttt{prism}^{\mathsf{mor}}_{\texttt{std}}(\mathsf{id}(G)).\mathsf{mapNode}(\langle\, b\,,\ \vec{v}\,\rangle) &= \mathsf{id}(\texttt{prism}_{\texttt{std}}(G)).\mathsf{mapNode}(\langle\, b\,,\ \vec{v}\,\rangle)\ \text{QED}\end{aligned}$$

◤ **Definition 5.44** **(endofunctor of the standard prism)** ⬣

We define $\texttt{prism}^{\mathsf{ftr}}_{\texttt{std}}$ to be an endofunctor analogous to the endofunction $\texttt{prism}_{\texttt{std}}$, where:

$$\begin{aligned}\texttt{prism}^{\mathsf{ftr}}_{\texttt{std}}.\mathsf{mapObj} &:\equiv \texttt{prism}_{\texttt{std}} & \texttt{prism}^{\mathsf{ftr}}_{\texttt{std}}.\mathsf{presComp} &:\equiv \text{lemma 5.42} \\ \texttt{prism}^{\mathsf{ftr}}_{\texttt{std}}.\mathsf{mapHom} &:\equiv \texttt{prism}^{\mathsf{mor}}_{\texttt{std}} & \texttt{prism}^{\mathsf{ftr}}_{\texttt{std}}.\mathsf{presIden} &:\equiv \text{lemma 5.43}\end{aligned}$$ ◢

◤ **Lemma 5.45** ⬣ Let $G$ and $G'$ be graphs such that $(G \cong G')$ in the category **Graph**, then we also have $\texttt{prism}_{\texttt{std}}(G) \cong \texttt{prism}_{\texttt{std}}(G')$. ◢

*Proof.* By applying the functor $\texttt{prism}^{\mathsf{ftr}}_{\texttt{std}}$ into lemma 5.12 QED

◤ **Definition 5.46** **(face graphs for standard cubes using recursion)** ⬣

Let $n : \mathbb{N}$, we define a graph $\acute{\mathbb{F}}^{n}_{\square}$ to be a recursive version of graph $\mathbb{F}^{n}_{\square}$.

$$\acute{\mathbb{F}}^{0}_{\square} :\equiv \epsilon_{\mathsf{grp}} \qquad \acute{\mathbb{F}}^{(n+1)}_{\square} :\equiv \texttt{prism}_{\texttt{std}}(\acute{\mathbb{F}}^{n}_{\square})$$ ◢

◤ **Lemma 5.47** ⬣ Let $n : \mathbb{N}$, then $\mathbb{F}_{\square}^{(n+1)} \cong \texttt{prism}_{\texttt{std}}(\mathbb{F}_{\square}^{n})$. ◢

*Proof.* We define $\phi$ : **Graph**( $\mathbb{F}_{\square}^{(n+1)}$ , $\texttt{prism}_{\texttt{std}}(\mathbb{F}_{\square}^{n})$ ) and $\psi$ : **Graph**( $\texttt{prism}_{\texttt{std}}(\mathbb{F}_{\square}^{n})$ , $\mathbb{F}_{\square}^{(n+1)}$ ) as graph homomorphisms for forward and backward direction, respectively, where:

$$\phi.\mathsf{mapNode} \quad : \quad \texttt{binary}(n+1) \rightarrow (\mathbb{2}_{\mathsf{fin}} \times \texttt{binary}(n))$$

$$\psi.\mathsf{mapNode} \quad : \quad (\mathbb{2}_{\mathsf{fin}} \times \texttt{binary}(n)) \rightarrow \texttt{binary}(n+1)$$

$$\phi.\mathsf{mapNode}(\langle b_0 , b_1 , \dots , b_n \rangle) \quad :\equiv \quad \langle b_0 , \langle b_1 , b_2 , \dots , b_n \rangle \rangle$$

$$\psi.\mathsf{mapNode}(\langle b_0 , \langle b_1 , b_2 , \dots , b_n \rangle \rangle) \quad :\equiv \quad \langle b_0 , b_1 , \dots , b_n \rangle$$

Regarding $\phi.\mathsf{mapEdge}$ and $\psi.\mathsf{mapEdge}$, it is obvious that $\phi$ and $\psi$ preserve edges by inspecting $\phi.\mathsf{mapNode}$ and $\psi.\mathsf{mapNode}$, respectively. Finally, it is obvious that both $\phi \cdot \psi$ and $\psi \cdot \phi$ are identity morphisms in **Graph**. QED

◤ **Theorem 5.48** ⬣ Let $n : \mathbb{N}$, then $\acute{\mathbb{F}}_{\square}^{n} \cong \mathbb{F}_{\square}^{n}$ in **Graph**. ◢

*Proof.* Using induction on the natural number $n$: The base ($\acute{\mathbb{F}}_{\square}^{0} \cong \mathbb{F}_{\square}^{0}$) case is obvious because both $\mathbb{F}_{\square}^{0}$ and $\acute{\mathbb{F}}_{\square}^{0}$ have a single node with no edges. Regarding the inductive step, assuming that ($\acute{\mathbb{F}}_{\square}^{n} \cong \mathbb{F}_{\square}^{n}$), we need to construct ($\acute{\mathbb{F}}_{\square}^{(n+1)} \cong \mathbb{F}_{\square}^{(n+1)}$).

$$\begin{aligned} \acute{\mathbb{F}}_{\square}^{(n+1)} &\cong \texttt{prism}_{\texttt{std}}(\acute{\mathbb{F}}_{\square}^{n}) && \text{[by definition 5.46]} \\ &\cong \texttt{prism}_{\texttt{std}}(\mathbb{F}_{\square}^{n}) && \text{[by the induction hypothesis and lemma 5.21]} \\ \acute{\mathbb{F}}_{\square}^{(n+1)} &\cong \mathbb{F}_{\square}^{(n+1)} && \text{[by lemma 5.47]} \end{aligned}$$

QED

◤ **Convention 5.49** **(standard cubes coercion from their recursive graphs)** ⬣
We define a coercion from $\acute{\mathbb{F}}_{\square}^{n}$ to $\mathbb{F}_{\square}^{n}$, which is analogous to convention 5.35. ◢

## § 5.3 Reflexive Graphs for (Possibly-Degenerated) Shapes

We generalise categories of non-degenerated shapes to include degenerated shapes. The main concept is to apply the reflexive closure to a face graph to become a *reflexive graph*.

### § 5.3.1 Reflexive Graphs for Simplices

Although $\mathbb{G}^n_\triangle$ can simply be defined as $\texttt{refl}_{\texttt{cl}}(\mathbb{F}^n_\triangle)$; however, there is an elegant (but equivalent) way to define $\mathbb{G}^n_\triangle$ as $\mathbb{F}^n_\triangle$ such that $( < )$ is replaced with $( \leqslant )$.

◤ **Definition 5.50** **(reflexive graphs for simplices)** ⬣

Let $n : \mathbb{N}$, we define a graph $\mathbb{G}^n_\triangle$ to be the reflexive graph for $\square^n_{\mathsf{top}}$.

$$\mathbb{G}^n_\triangle.\mathsf{Nodes} \quad :\equiv \quad \texttt{fin}\ (n+1)$$
$$\mathbb{G}^n_\triangle.\mathsf{isEdge}\ s, t \quad :\equiv \quad s \leqslant t$$ ◢

◤ **Lemma 5.51** ⬣ Let $n : \mathbb{N}$, then $\mathbb{G}^n_\triangle \cong \texttt{refl}_{\texttt{cl}}(\mathbb{F}^n_\triangle)$ in **Graph**. ◢

◤ **Corollary 5.52** ⬣ Let $n : \mathbb{N}$, then $\mathbb{G}^n_\triangle \cong \texttt{refl}_{\texttt{cl}}(\texttt{tran}_{\texttt{cl}}(\texttt{line}^n_{\mathsf{grp}}))$. ◢

◤ **Definition 5.53** **(graph theoretic version of simplex category)** ⬣

We define the category $\triangle^{\mathsf{graph}}_{\mathsf{full}}$ to be a full-subcategory of **Graph** induced by the family of graphs $\mathbb{G}^n_\triangle$ for all $n : \mathbb{N}$. Please note that, we may use the number $n$ or the graph $\mathbb{G}^n_\triangle$ to denote each object in $\triangle^{\mathsf{graph}}_{\mathsf{full}}$ interchangeably. ◢

**Lemma 5.54** The category $\triangle_{\text{full}}^{\text{graph}}$ is isomorphic to the category $\triangle_{\text{full}}^{\text{comb}}$.

*Proof.* This proof is similar to the proof of lemma 5.28, which shows that the category $\triangle_{\text{semi}}^{\text{graph}}$ is isomorphic to the category $\triangle_{\text{semi}}^{\text{comb}}$. Since both $\triangle_{\text{full}}^{\text{graph}}$ and $\triangle_{\text{full}}^{\text{comb}}$ have the same set of objects, our goal now becomes

$$\triangle_{\text{full}}^{\text{graph}}(m, n) \quad \cong \quad \triangle_{\text{full}}^{\text{comb}}(m, n) \qquad \text{for every} \quad m, n : \mathbb{N}.$$

In the forward direction, we define $\phi : \triangle_{\text{full}}^{\text{graph}}(m, n) \to \triangle_{\text{full}}^{\text{comb}}(m, n)$. Given $f : \triangle_{\text{full}}^{\text{graph}}(m, n)$, which is a graph homomorphism from $\mathbb{G}_{\triangle}^{m}$ to $\mathbb{G}_{\triangle}^{n}$, we define $\phi(f) : \triangle_{\text{full}}^{\text{comb}}(m, n)$ where $\phi(f) :\equiv f.\text{mapNode}$. We also know that $\phi(f)$ is monotone because, given $s, t : \mathtt{fin}(m+1)$, a proof of $(s \leqslant t)$ can be translated to a proof of $\mathbb{G}_{\triangle}^{m}.\text{isEdge}(s,t)$, which, in turn, get mapped by $f.\text{mapEdge}$ and become a proof of $\mathbb{G}_{\triangle}^{n}.\text{isEdge}(f(s), f(t))$, which, in turn again get translated back to a proof of $(f(s) \leqslant f(t))$.

In the backward direction, we define $\psi : \triangle_{\text{full}}^{\text{comb}}(m, n) \to \triangle_{\text{full}}^{\text{graph}}(m, n)$. Given $f : \triangle_{\text{full}}^{\text{comb}}(m, n)$, which is a monotone function from $\mathtt{fin}(m+1)$ to $\mathtt{fin}(n+1)$, we define $\psi(f) : \mathbf{Graph}(\mathbb{G}_{\triangle}^{m}, \mathbb{G}_{\triangle}^{n})$.

- Regarding $\psi(f).\text{mapNode}$, it is the function $f$ itself.
- Regarding $\psi(f).\text{mapEdge}$, given $s, t : \mathtt{fin}(m+1)$, a proof of $\mathbb{G}_{\triangle}^{m}.\text{isEdge}(s,t)$ can be translated to a proof of $(s \leqslant t)$, which, in turn, get mapped by the monotonicity of $f$ and become a proof of $(f(s) \leqslant f(t))$, which, in turn again, get translated back to a proof of $\mathbb{G}_{\triangle}^{n}.\text{isEdge}(f(s), f(t))$.

Finally, it is obvious that both $\phi \cdot \psi$ and $\psi \cdot \phi$ are identity morphisms in **Graph**. QED

**Corollary 5.55** The category $\triangle_{\text{full}}^{\text{graph}}$ is a Reedy category.

*Proof.* By applying lemma 5.54 to lemma 4.30. QED

To upgrade the recursive version of $\mathbb{G}^n_\triangle$ as reflexive graph, it is as easy as replacing $\epsilon_{\mathsf{grp}}$ with $\mathbb{1}_{\mathsf{grp}}$ in the base case.

◤ **Definition 5.56** **(reflexive graphs for simplices using recursion)** ⬣

Let $n : \mathbb{N}$, we define a graph $\acute{\mathbb{G}}^n_\triangle$ to be a recursive version of graph $\mathbb{G}^n_\triangle$.

$$\acute{\mathbb{G}}^0_\triangle \quad :\equiv \quad \mathbb{1}_{\mathsf{grp}} \qquad\qquad \acute{\mathbb{G}}^{(n+1)}_\triangle \quad :\equiv \quad \mathtt{cone}(\acute{\mathbb{G}}^n_\triangle)$$ ◢

◤ **Lemma 5.57** ⬣ Let $n : \mathbb{N}$, then $\acute{\mathbb{G}}^n_\triangle \cong \mathtt{refl}_{\mathtt{cl}}(\acute{\mathbb{F}}^n_\triangle)$ in **Graph**. ◢

*Proof.* Given that $G, G' : \mathsf{ob}(\mathbf{Graph})$, the following propositions hold:

- $\mathtt{refl}_{\mathtt{cl}}(\mathtt{cone}(G)) \cong \mathtt{cone}(\mathtt{refl}_{\mathtt{cl}}(G))$ [by lemma 5.19]
- if $(G \cong G')$ then $\mathtt{cone}(G) \cong \mathtt{cone}(G')$ [by lemma 5.31]
- if $(G \cong G')$ then $\mathtt{refl}_{\mathtt{cl}}(G) \cong \mathtt{refl}_{\mathtt{cl}}(G')$ [by lemma 5.21]

$$\begin{aligned} \text{Then,} \quad \acute{\mathbb{G}}^n_\triangle &\cong \mathtt{cone}(\mathtt{cone}((\ldots \ \mathtt{cone}(\mathbb{1}_{\mathsf{grp}})\ ))) \\ &\cong \mathtt{cone}(\mathtt{cone}((\ldots \ \mathtt{cone}(\mathtt{refl}_{\mathtt{cl}}(\epsilon_{\mathsf{grp}}))\ ))) \\ &\cong \mathtt{cone}(\mathtt{cone}((\ldots \ \mathtt{refl}_{\mathtt{cl}}(\mathtt{cone}(\epsilon_{\mathsf{grp}}))\ ))) \\ &\cong \mathtt{cone}(\mathtt{refl}_{\mathtt{cl}}((\ldots \ \mathtt{cone}(\mathtt{cone}(\epsilon_{\mathsf{grp}}))\ ))) \\ &\cong \mathtt{refl}_{\mathtt{cl}}(\mathtt{cone}((\ldots \ \mathtt{cone}(\mathtt{cone}(\epsilon_{\mathsf{grp}}))\ ))) \\ \text{Hence,} \quad \acute{\mathbb{G}}^n_\triangle &\cong \mathtt{refl}_{\mathtt{cl}}(\acute{\mathbb{F}}^n_\triangle) \end{aligned}$$

QED

◤ **Theorem 5.58** ⬣ Let $n : \mathbb{N}$, then $\acute{\mathbb{G}}^n_\triangle \cong \mathbb{G}^n_\triangle$ in **Graph**. ◢

*Proof.*

$$\begin{aligned} \acute{\mathbb{G}}^n_\triangle &\cong \mathtt{refl}_{\mathtt{cl}}(\acute{\mathbb{F}}^n_\triangle) && \text{[by lemma 5.57]} \\ &\cong \mathtt{refl}_{\mathtt{cl}}(\mathbb{F}^n_\triangle) && \text{[by lemma 5.21 and theorem 5.34]} \\ \acute{\mathbb{G}}^n_\triangle &\cong \mathbb{G}^n_\triangle && \text{[by lemma 5.51]} \end{aligned}$$

QED

## § 5.3.2 Reflexive Graphs for Standard Cubes

Although $\mathbb{G}^n_\square$ can simply be defined as $\texttt{refl}_{\texttt{cl}}(\mathbb{F}^n_\square)$; however, there is an elegant (but equivalent) way to define $\mathbb{G}^n_\square$ as $\mathbb{F}^n_\square$ such that ( $<$ ) in the then-clause is replaced with ( $\leqslant$ ).

**Definition 5.59** **(reflexive graphs for standard cubes)**

Let $n : \mathbb{N}$, we define a graph $\mathbb{G}^n_\square$ to be the reflexive graph for $\square^n_{\mathsf{top}}$

$$\mathbb{G}^n_\square.\mathsf{Nodes} \quad :\equiv \quad \texttt{binary}(n)$$

$$\mathbb{G}^n_\square.\mathsf{isEdge}(\vec{s},\vec{t}) \quad :\equiv \quad \exists (\, i : \texttt{fin}(n)\,) \times \forall (\, j : \texttt{fin}(n)\,) \to \text{if} \ (i = j) \ \text{then} \ \vec{s}[j] \leqslant \vec{t}[j] \ \text{else} \ \vec{s}[j] = \vec{t}[j]$$

**Lemma 5.60** Let $n : \mathbb{N}$, then $\mathbb{G}^n_\square \cong \texttt{refl}_{\texttt{cl}}(\mathbb{F}^n_\square)$ in **Graph**.

*Proof.* Since $\mathbb{G}^n_\square$ and $\texttt{refl}_{\texttt{cl}}(\mathbb{F}^n_\square)$ have identical set of nodes, we only need check their edges to solve the main goal. Therefore, the remaining goal become

$$\mathbb{G}^n_\square.\mathsf{isEdge}(\vec{s},\vec{t}) \quad \text{iff} \quad \texttt{refl}_{\texttt{cl}}(\mathbb{F}^n_\square).\mathsf{isEdge}(\vec{s},\vec{t}) \qquad \text{for all} \quad \vec{s},\ \vec{t} : \texttt{binary}(n)$$

To make the proof become shorter, we say

- LHS (left hand side) as $\mathbb{G}^n_\square.\mathsf{isEdge}(\vec{s},\vec{t})$,
- RHS (right hand side) as $\texttt{refl}_{\texttt{cl}}(\mathbb{F}^n_\square).\mathsf{isEdge}(\vec{s},\vec{t})$,
- variable $i$ means the binding of the outer existential quantifier, and
- variable $j$ means the binding of the inner universal quantifier.

To prove this remaining goal, we split the scenario into three cases depending on the comparison of $\vec{s}$ against $\vec{t}$ index-wisely.

- When there are no distinguishing indices, then:
  - LHS holds because, for every variable $j$, we have $(\vec{s}[j] = \vec{t}[j])$,
    which implies both then-clause and else-clause of LHS
    (regardless of variable $i$ assignment because both clauses don't depend on it).
  - RHS holds because $(\vec{s} = \vec{t})$ holds so we unfold the definition of $\texttt{refl}_{\texttt{cl}}$.

- When there are multiple distinguishing indices, then:
  - LHS doesn't hold because, for every variable $i$, there is always some variable $j$ such that $(i \neq j)$ and $(\vec{s}[j] \neq \vec{t}[j])$, which contradicts the else-clause of LHS.
  - RHS is now reduced to $\mathbb{F}^n_\square$.isEdge$(\vec{s}, \vec{t})$ because $(\vec{s} = \vec{t})$ doesn't hold so we unfold the definition of $\texttt{refl}_{\texttt{cl}}$ in contrapositive fashion.
  - The new RHS doesn't hold because the contradiction in else-clause similar to the case of LHS above.

- When there is a unique distinguishing index, then:
  - We define $k : \texttt{fin}(n)$ to be this distinguishing index.
  - RHS is now reduced to $\mathbb{F}^n_\square$.isEdge$(\vec{s}, \vec{t})$ because $(\vec{s} = \vec{t})$ doesn't hold so we unfold the definition of $\texttt{refl}_{\texttt{cl}}$ in contrapositive fashion.
  - The only different between LHS and RHS is the different in their then-clauses.
  - Both LHS and RHS can be reduced to their respective then-clauses with variables $i$ and $j$ assigned to $k$ because:
    * Let $\texttt{ite}(i', j')$ denotes the if-then-else clause inside LHS/RHS but variables $i$ and $j$ are substituted by $i'$ and $j'$, respectively.
    * Let TERM denotes $\exists\,(i) \;\times\; \forall\,(j) \;\rightarrow\; \texttt{ite}(i, j)$.

* We need to show that TERM can be reduced to $\texttt{ite}(k,k)$.
* When the existential quantifier assigns $i :\equiv k'$ for any $k' : \texttt{fin}(n)$ not equal to $k$, then the universal quantifier can assign $j :\equiv k$ to falsify $(i = j)$ and make $\texttt{ite}(k', k)$ goes through the else-clause and has a contradiction (because $j :\equiv k$ which is the distinguishing index). Therefore, $\forall (j) \rightarrow \texttt{ite}(k', j)$ reduces to $\bot$.
* Since TERM now is in a form of a disjunction between $\forall (j) \rightarrow \texttt{ite}(i, j)$ for every possible $i : \texttt{fin}(n)$; however, $((p \vee \bot) = p)$; therefore TERM is reduced to $\forall (j) \rightarrow \texttt{ite}(k, j)$.
* When the universal quantifier assigns $j :\equiv k'$ for any $k' : \texttt{fin}(n)$ not equal to $k$, then the if-condition doesn't hold and make $\texttt{ite}(k, k')$ goes through the else-clause. Therefore, $\texttt{ite}(k, k')$ holds by reflexivity.
* Since TERM now is in a form of a conjunction between $\texttt{ite}(k, j)$ for every possible $i : \texttt{fin}(n)$; however, $((p \wedge \top) = p)$; therefore TERM is reduced to $\texttt{ite}(k, k)$.

- The new LHS is really just the disjunction between the the new RHS and the proposition $(\vec{s}[k] = \vec{t}[k])$.
- The assumption of index $k$ is $(\vec{s}[k] \neq \vec{t}[k])$; therefore, LHS is now identical to RHS. QED

To upgrade the recursive version of $\mathbb{G}^n_\square$ as reflexive graph, it is as easy as replacing $\epsilon_{\mathsf{grp}}$ with $\mathbb{1}_{\mathsf{grp}}$ in the base case.

◤ **Definition 5.61** **(reflexive graphs for standard cubes using recursion)** ⬣
Let $n : \mathbb{N}$, we define a graph $\acute{\mathbb{G}}^n_\square$ to be a recursive version of graph $\mathbb{G}^n_\square$.

$$\acute{\mathbb{G}}^0_\square \quad :\equiv \quad \mathbb{1}_{\mathsf{grp}} \qquad\qquad \acute{\mathbb{G}}^{(n+1)}_\square \quad :\equiv \quad \mathtt{prism}_{\mathtt{std}}(\acute{\mathbb{G}}^n_\square)$$ ◢

◤ **Lemma 5.62** ⬣ Let $n : \mathbb{N}$, then $\acute{\mathbb{G}}^n_\square \cong \mathtt{refl}_{\mathtt{cl}}(\acute{\mathbb{F}}^n_\square)$ in **Graph**. ◢

*Proof.* Given that $G,\ G' : \mathsf{ob}(\mathbf{Graph})$, the following propositions hold:

- $\mathtt{refl}_{\mathtt{cl}}(\mathtt{prism}_{\mathtt{std}}(G)) \cong \mathtt{prism}_{\mathtt{std}}(\mathtt{refl}_{\mathtt{cl}}(G))$ [by lemma 5.19]
- if $(G \cong G')$ then $\mathtt{prism}_{\mathtt{std}}(G) \cong \mathtt{prism}_{\mathtt{std}}(G')$ [by lemma 5.45]
- if $(G \cong G')$ then $\mathtt{refl}_{\mathtt{cl}}(G) \cong \mathtt{refl}_{\mathtt{cl}}(G')$ [by lemma 5.21]

$$
\begin{aligned}
\text{Then,}\quad \acute{\mathbb{G}}^n_\square &\cong \mathtt{prism}_{\mathtt{std}}(\mathtt{prism}_{\mathtt{std}}((\ldots\ \mathtt{prism}_{\mathtt{std}}(\mathbb{1}_{\mathsf{grp}})\ )))\\
&\cong \mathtt{prism}_{\mathtt{std}}(\mathtt{prism}_{\mathtt{std}}((\ldots\ \mathtt{prism}_{\mathtt{std}}(\mathtt{refl}_{\mathtt{cl}}(\epsilon_{\mathsf{grp}}))\ )))\\
&\cong \mathtt{prism}_{\mathtt{std}}(\mathtt{prism}_{\mathtt{std}}((\ldots\ \mathtt{refl}_{\mathtt{cl}}(\mathtt{prism}_{\mathtt{std}}(\epsilon_{\mathsf{grp}}))\ )))\\
&\cong \mathtt{prism}_{\mathtt{std}}(\mathtt{refl}_{\mathtt{cl}}((\ldots\ \mathtt{prism}_{\mathtt{std}}(\mathtt{prism}_{\mathtt{std}}(\epsilon_{\mathsf{grp}}))\ )))\\
&\cong \mathtt{refl}_{\mathtt{cl}}(\mathtt{prism}_{\mathtt{std}}((\ldots\ \mathtt{prism}_{\mathtt{std}}(\mathtt{prism}_{\mathtt{std}}(\epsilon_{\mathsf{grp}}))\ )))\\
\text{Hence,}\quad \acute{\mathbb{G}}^n_\square &\cong \mathtt{refl}_{\mathtt{cl}}(\acute{\mathbb{F}}^n_\square)
\end{aligned}
$$

QED

◤ **Theorem 5.63** ⬣ Let $n : \mathbb{N}$, then $\acute{\mathbb{G}}^n_\square \cong \mathbb{G}^n_\square$ in **Graph**. ◢

*Proof.*

$$
\begin{aligned}
\acute{\mathbb{G}}^n_\square &\cong \mathtt{refl}_{\mathtt{cl}}(\acute{\mathbb{F}}^n_\square) && \text{[by lemma 5.62]}\\
&\cong \mathtt{refl}_{\mathtt{cl}}(\mathbb{F}^n_\square) && \text{[by lemma 5.21 and theorem 5.48]}\\
\acute{\mathbb{G}}^n_\square &\cong \mathbb{G}^n_\square && \text{[by lemma 5.60]}
\end{aligned}
$$

QED

Chapter

6

# Graph-Theoretic Representation of Twisted Cubes

This chapter modifies the graph-theoretic and categorial definitions of standard semi-cubes (that are described in subsection 5.2.3) into the twisted cubes counterpart.

## § 6.1 Face Graphs for Twisted Semi-Cubes

The fundamental concept here is to make the graph-theoretic version of the thickening-and-twisting process defined in subsection 3.2.3.

## § 6.1.1 The Twisted Prism Iterator

Recall from subsection 3.2.3 that a twisted $(n+1)$-cube is a twisted $n$-cube after applying the thickening-and-twisting process, the thickening phase is essentially the iterator $\texttt{prism}_{\texttt{std}}$, whereas the twisting phase reverses all edges at the first copy.

◤ **Definition 6.1** **(twisted prism iterator)** ⬣
Let G be a graph, we define another graph $\texttt{prism}_{\texttt{tw}}(G)$ as a modification of the graph G in the same way as $\texttt{prism}_{\texttt{std}}(G)$ but the edges in first copy are reversed.

$$
\begin{aligned}
\texttt{prism}_{\texttt{tw}}(G).\mathsf{Nodes} \quad &:\equiv \quad \mathbb{2}_{\mathsf{fin}} \times G.\mathsf{Nodes} \\
\texttt{prism}_{\texttt{tw}}(G).\mathsf{Edges} \quad &:\equiv \quad \{ \langle \langle 0 , \vec{t} \rangle , \langle 0 , \vec{s} \rangle \rangle \mid \langle \vec{s} , \vec{t} \rangle : G.\mathsf{Edges} \} \\
&\cup \quad \{ \langle \langle 1 , \vec{s} \rangle , \langle 1 , \vec{t} \rangle \rangle \mid \langle \vec{s} , \vec{t} \rangle : G.\mathsf{Edges} \} \\
&\cup \quad \{ \langle \langle 0 , \vec{v} \rangle , \langle 1 , \vec{v} \rangle \rangle \mid \vec{v} : G.\mathsf{Nodes} \} \qquad ◢
\end{aligned}
$$

◤ **Definition 6.2** ⬣ Let G and G′ be graphs, and $f : \mathbf{Graph}(G, G')$ we define $\texttt{prism}^{\mathsf{mor}}_{\texttt{tw}}(f)$ to be a graph homomorphism from $\texttt{prism}_{\texttt{tw}}(G)$ to $\texttt{prism}_{\texttt{tw}}(G')$.

$$
\begin{aligned}
&\texttt{prism}^{\mathsf{mor}}_{\texttt{tw}}(f).\mathsf{mapNode} \ : \ (\mathbb{2}_{\mathsf{fin}} \times G.\mathsf{Nodes}) \to (\mathbb{2}_{\mathsf{fin}} \times G'.\mathsf{Nodes}) \\
&\texttt{prism}^{\mathsf{mor}}_{\texttt{tw}}(f).\mathsf{mapNode} \ (\langle b , \vec{v} \rangle) \quad :\equiv \quad \langle b , f.\mathsf{mapNode}(\vec{v}) \rangle
\end{aligned}
$$

$$
\begin{aligned}
&\texttt{prism}^{\mathsf{mor}}_{\texttt{tw}}(f).\mathsf{mapEdge}(s,t) \ : \ \texttt{prism}_{\texttt{tw}}(G).\mathsf{isEdge}(s,t) \to \texttt{prism}_{\texttt{tw}}(G').\mathsf{isEdge}(f(s),f(t)) \\
&\texttt{prism}^{\mathsf{mor}}_{\texttt{tw}}(f).\mathsf{mapEdge} \ (\langle 0 , \vec{t} \rangle , \langle 0 , \vec{s} \rangle) \ (e) \quad :\equiv \quad f.\mathsf{mapEdge}(\vec{s},\vec{t})(e) \\
&\texttt{prism}^{\mathsf{mor}}_{\texttt{tw}}(f).\mathsf{mapEdge} \ (\langle 1 , \vec{s} \rangle , \langle 1 , \vec{t} \rangle) \ (e) \quad :\equiv \quad f.\mathsf{mapEdge}(\vec{s},\vec{t})(e) \\
&\texttt{prism}^{\mathsf{mor}}_{\texttt{tw}}(f).\mathsf{mapEdge} \ (\langle 0 , \vec{v} \rangle , \langle 1 , \vec{v} \rangle) \ (e) \quad :\equiv \quad \mathsf{cong}(f.\mathsf{mapNode}, e) \qquad ◢
\end{aligned}
$$

◤ **Lemma 6.3** ⬣ Given two composable graph homomorphisms f and g, then

$$\mathtt{prism}^{\mathsf{mor}}_{\mathtt{tw}}(f \cdot g) \;=\; \mathtt{prism}^{\mathsf{mor}}_{\mathtt{tw}}(f) \cdot \mathtt{prism}^{\mathsf{mor}}_{\mathtt{tw}}(g).$$ ◢

*Proof.* By applying the following proof to lemma 5.5.

$$\begin{aligned}
& \mathtt{prism}^{\mathsf{mor}}_{\mathtt{tw}}(f \cdot g).\mathsf{mapNode}(\langle b, \vec{v} \rangle) \\
=\; & \langle b, (f \cdot g).\mathsf{mapNode}(\vec{v}) \rangle \\
=\; & \langle b, g.\mathsf{mapNode}(f.\mathsf{mapNode}(\vec{v})) \rangle \\
=\; & \mathtt{prism}^{\mathsf{mor}}_{\mathtt{tw}}(g).\mathsf{mapNode}(\langle b, f.\mathsf{mapNode}(\vec{v}) \rangle) \\
=\; & \mathtt{prism}^{\mathsf{mor}}_{\mathtt{tw}}(g).\mathsf{mapNode}(\mathtt{prism}^{\mathsf{mor}}_{\mathtt{tw}}(f).\mathsf{mapNode}(\langle b, \vec{v} \rangle)) \\
=\; & (\mathtt{prism}^{\mathsf{mor}}_{\mathtt{tw}}(f).\mathsf{mapNode} \cdot \mathtt{prism}^{\mathsf{mor}}_{\mathtt{tw}}(g).\mathsf{mapNode})(\langle b, \vec{v} \rangle) \\
=\; & (\mathtt{prism}^{\mathsf{mor}}_{\mathtt{tw}}(f) \cdot \mathtt{prism}^{\mathsf{mor}}_{\mathtt{tw}}(g)).\mathsf{mapNode}(\langle b, \vec{v} \rangle) \qquad \text{QED}
\end{aligned}$$

◤ **Lemma 6.4** ⬣ Let G be a graph, then

$$\mathtt{prism}^{\mathsf{mor}}_{\mathtt{tw}}(\mathsf{id}(G)) \;=\; \mathsf{id}(\mathtt{prism}_{\mathtt{tw}}(G)).$$ ◢

*Proof.* By applying the following proof to lemma 5.5.

$$\begin{aligned}
\mathtt{prism}^{\mathsf{mor}}_{\mathtt{tw}}(\mathsf{id}(G)).\mathsf{mapNode}(\langle b, \vec{v} \rangle) &= \langle b, \mathsf{id}(G).\mathsf{mapNode}(\vec{v}) \rangle \\
&= \langle b, \vec{v} \rangle \\
\mathtt{prism}^{\mathsf{mor}}_{\mathtt{tw}}(\mathsf{id}(G)).\mathsf{mapNode}(\langle b, \vec{v} \rangle) &= \mathsf{id}(\mathtt{prism}_{\mathtt{tw}}(G)).\mathsf{mapNode}(\langle b, \vec{v} \rangle) \text{ QED}
\end{aligned}$$

◤ **Definition 6.5** **(endofunctor of the twisted prism)** ⬣

We define $\mathtt{prism}^{\mathsf{ftr}}_{\mathtt{tw}}$ to be an endofunctor analogous to the endofunction $\mathtt{prism}_{\mathtt{tw}}$, where:

| | | | | | |
|---|---|---|---|---|---|
| $\mathtt{prism}^{\mathsf{ftr}}_{\mathtt{tw}}.\mathsf{mapObj}$ | :≡ | $\mathtt{prism}_{\mathtt{tw}}$ | $\mathtt{prism}^{\mathsf{ftr}}_{\mathtt{tw}}.\mathsf{presComp}$ | :≡ | lemma 6.3 |
| $\mathtt{prism}^{\mathsf{ftr}}_{\mathtt{tw}}.\mathsf{mapHom}$ | :≡ | $\mathtt{prism}^{\mathsf{mor}}_{\mathtt{tw}}$ | $\mathtt{prism}^{\mathsf{ftr}}_{\mathtt{tw}}.\mathsf{presIden}$ | :≡ | lemma 6.4 ◢ |

◤ **Lemma 6.6** ⬣ Let G and G′ be graphs such that $(G \cong G')$ in the category **Graph**, then we also have $\texttt{prism}_{\texttt{tw}}(G) \cong \texttt{prism}_{\texttt{tw}}(G')$. ◢

*Proof.* By applying the functor $\texttt{prism}^{\text{ftr}}_{\texttt{tw}}$ into lemma 5.12 QED

### § 6.1.2 Non-Recursive Face Graphs for Twisted Cubes

We define $\mathbb{F}^n_{\bowtie}$ as a modification of $\mathbb{F}^n_{\square}$ by changing the then-clause from $(\vec{s}[j] < \vec{t}[j])$ to $((\vec{s}[j] \oplus p) < (\vec{t}[j] \oplus p))$ where $p : 2_{\text{fin}}$ represents the parity of this edge that has been discussed discussed in subsection 3.2.5.

◤ **Definition 6.7** **(face graphs for twisted cubes)** ⬣
Let $n : \mathbb{N}$, we define $\mathbb{F}^n_{\bowtie}$ to be the face graph for a twisted $n$-cube.

$$\mathbb{F}^n_{\bowtie}.\text{Nodes} \quad :\equiv \quad \texttt{binary}(n)$$

$$\mathbb{F}^n_{\bowtie}.\text{isEdge}(\vec{s},\vec{t}) \quad :\equiv \quad \exists(i : \texttt{fin}(n)) \times \forall(j : \texttt{fin}(n)) \to$$

$$\text{if} \quad (i = j) \quad \text{then} \quad (\vec{s}[j] \oplus p) < (\vec{t}[j] \oplus p) \quad \text{else} \quad \vec{s}[j] = \vec{t}[j]$$

$$\text{where} \quad p \quad :\equiv \quad \texttt{parity}\ \langle \vec{t}[0], \vec{t}[1], \ldots, \vec{t}[i-1] \rangle$$ ◢

### § 6.1.3 Recursive Face Graphs for Twisted Cubes

◤ **Definition 6.8** **(face graphs for twisted cubes using recursion)** ⬣
Let $n : \mathbb{N}$, we define a graph $\acute{\mathbb{F}}^n_{\bowtie}$ to be a recursive version of graph $\mathbb{F}^n_{\bowtie}$.

$$\acute{\mathbb{F}}^0_{\bowtie} :\equiv \epsilon_{\text{grp}} \qquad \acute{\mathbb{F}}^{(n+1)}_{\bowtie} :\equiv \texttt{prism}_{\texttt{tw}}(\acute{\mathbb{F}}^n_{\bowtie})$$ ◢

◤ **Lemma 6.9** ⬣ Let $n : \mathbb{N}$, then $\mathbb{F}_{\bowtie}^{(n+1)} \cong \texttt{prism}_{\texttt{tw}}(\mathbb{F}_{\bowtie}^{n})$. ◢

*Proof.* We define $\phi$ : **Graph**( $\mathbb{F}_{\bowtie}^{(n+1)}$, $\texttt{prism}_{\texttt{tw}}(\mathbb{F}_{\bowtie}^{n})$ ) and $\psi$ : **Graph**( $\texttt{prism}_{\texttt{tw}}(\mathbb{F}_{\bowtie}^{n})$, $\mathbb{F}_{\bowtie}^{(n+1)}$ ) as graph homomorphisms for forward and backward direction, respectively, where:

$$\phi.\mathsf{mapNode} \quad : \quad \texttt{binary}(n+1) \longrightarrow (\mathbb{2}_{\mathsf{fin}} \times \texttt{binary}(n))$$

$$\psi.\mathsf{mapNode} \quad : \quad (\mathbb{2}_{\mathsf{fin}} \times \texttt{binary}(n)) \longrightarrow \texttt{binary}(n+1)$$

$$\phi.\mathsf{mapNode}(\langle b_0, b_1, \dots, b_n \rangle) \quad :\equiv \quad \langle b_0, \langle b_1, b_2, \dots, b_n \rangle \rangle$$

$$\psi.\mathsf{mapNode}(\langle b_0, \langle b_1, b_2, \dots, b_n \rangle \rangle) \quad :\equiv \quad \langle b_0, b_1, \dots, b_n \rangle$$

Regarding $\phi$.mapEdge and $\psi$.mapEdge, it is obvious that $\phi$ and $\psi$ preserve edges by inspecting $\phi$.mapNode and $\psi$.mapNode, respectively. Finally, it is obvious that both $\phi \cdot \psi$ and $\psi \cdot \phi$ are identity morphisms in **Graph**. QED

◤ **Theorem 6.10** ⬣ Let $n : \mathbb{N}$, then $\acute{\mathbb{F}}_{\bowtie}^{n} \cong \mathbb{F}_{\bowtie}^{n}$ in **Graph**. ◢

*Proof.* Using induction on the natural number $n$: The base ($\acute{\mathbb{F}}_{\bowtie}^{0} \cong \mathbb{F}_{\bowtie}^{0}$) case is obvious because both $\mathbb{F}_{\bowtie}^{0}$ and $\acute{\mathbb{F}}_{\bowtie}^{0}$ have a single node with no edges. Regarding the inductive step, assuming that ($\acute{\mathbb{F}}_{\bowtie}^{n} \cong \mathbb{F}_{\bowtie}^{n}$), we need to construct ($\acute{\mathbb{F}}_{\bowtie}^{(n+1)} \cong \mathbb{F}_{\bowtie}^{(n+1)}$).

| | | | |
|---|---|---|---|
| $\acute{\mathbb{F}}_{\bowtie}^{(n+1)}$ | $\cong$ | $\texttt{prism}_{\texttt{tw}}(\acute{\mathbb{F}}_{\bowtie}^{n})$ | [by definition 6.8] |
| | $\cong$ | $\texttt{prism}_{\texttt{tw}}(\mathbb{F}_{\bowtie}^{n})$ | [by the induction hypothesis and lemma 5.21] |
| $\acute{\mathbb{F}}_{\bowtie}^{(n+1)}$ | $\cong$ | $\mathbb{F}_{\bowtie}^{(n+1)}$ | [by lemma 6.9] QED |

◤ **Convention 6.11** **(twisted cubes coercion from their recursive graphs)** ⬣
We define a coercion from $\acute{\mathbb{F}}_{\bowtie}^{n}$ to $\mathbb{F}_{\bowtie}^{n}$, which is analogous to convention 5.35. ◢

## § 6.1.4 The Twisted Semi-Cubes Category

◤ **Definition 6.12** (graph theoretic version of twisted semi cube category) ⬣

We define the category $\bowtie^{\text{graph}}_{\text{semi}}$ to be a full-subcategory of **Graph** induced by the family of graphs $\mathbb{F}^{n}_{\bowtie}$ for all $n : \mathbb{N}$. Please note that, we may use the number $n$ or the graph $\mathbb{F}^{n}_{\bowtie}$ to denote each object in $\bowtie^{\text{graph}}_{\text{semi}}$ interchangeably. ◢

◤ **Lemma 6.13** ⬣ The category $\bowtie^{\text{graph}}_{\text{semi}}$ is a direct category. ◢

*Proof.* We use the identity function on $\mathbb{N}$ as the degree function. For any natural numbers $m$ and $n$, we want to prove that $\mathsf{deg}(m) < \mathsf{deg}(n)$ if there is a non-identity morphism in $\bowtie^{\text{graph}}_{\text{semi}}(m, n)$. This can be done by comparing $m$ and $n$:

- If $m < n$, then $\mathsf{deg}(m) < \mathsf{deg}(n)$ holds by unfolding the degree function.
- If $m = n$, then only the identity morphism can be in $\bowtie^{\text{graph}}_{\text{semi}}(n, n)$.
  This is because if it contains a non-identity graph homomorphism $f$, then some nodes got swapped by $f.\mathsf{mapNode}$, which violate the unique order given by the unique Hamiltonian path of $\mathbb{F}^{n}_{\bowtie}$ later in theorem 6.17.
- If $m > n$, then $\bowtie^{\text{graph}}_{\text{semi}}(m, n)$ contains no morphisms because
  if there is $f : \bowtie^{\text{graph}}_{\text{semi}}(m, n)$, then there must be distinct nodes $v$ and $v'$ in $\mathbb{F}^{m}_{\bowtie}$ that shares the same image from $f.\mathsf{mapNode}$ because of the pigeonhole principle on the fact that $\mathbb{F}^{m}_{\bowtie}$ has more nodes than $\mathbb{F}^{n}_{\bowtie}$; however, the graph $\mathbb{F}^{n}_{\bowtie}$ is irreflexive so we can't assign $f.\mathsf{mapEdge}(v, v')$ to anywhere. QED

## § 6.2 Unique Hamiltonian path for Twisted Semi-Cubes

The most important feature of $\mathbb{F}^n_{\bowtie}$ is its unique Hamiltonian path, which is useful to prove many theorems related to twisted cubes. As a sanity check, figure 6.1 shows the unique Hamiltonian path at lower dimensions.

### § 6.2.1 Existence and Uniqueness of the Hamiltonian Path

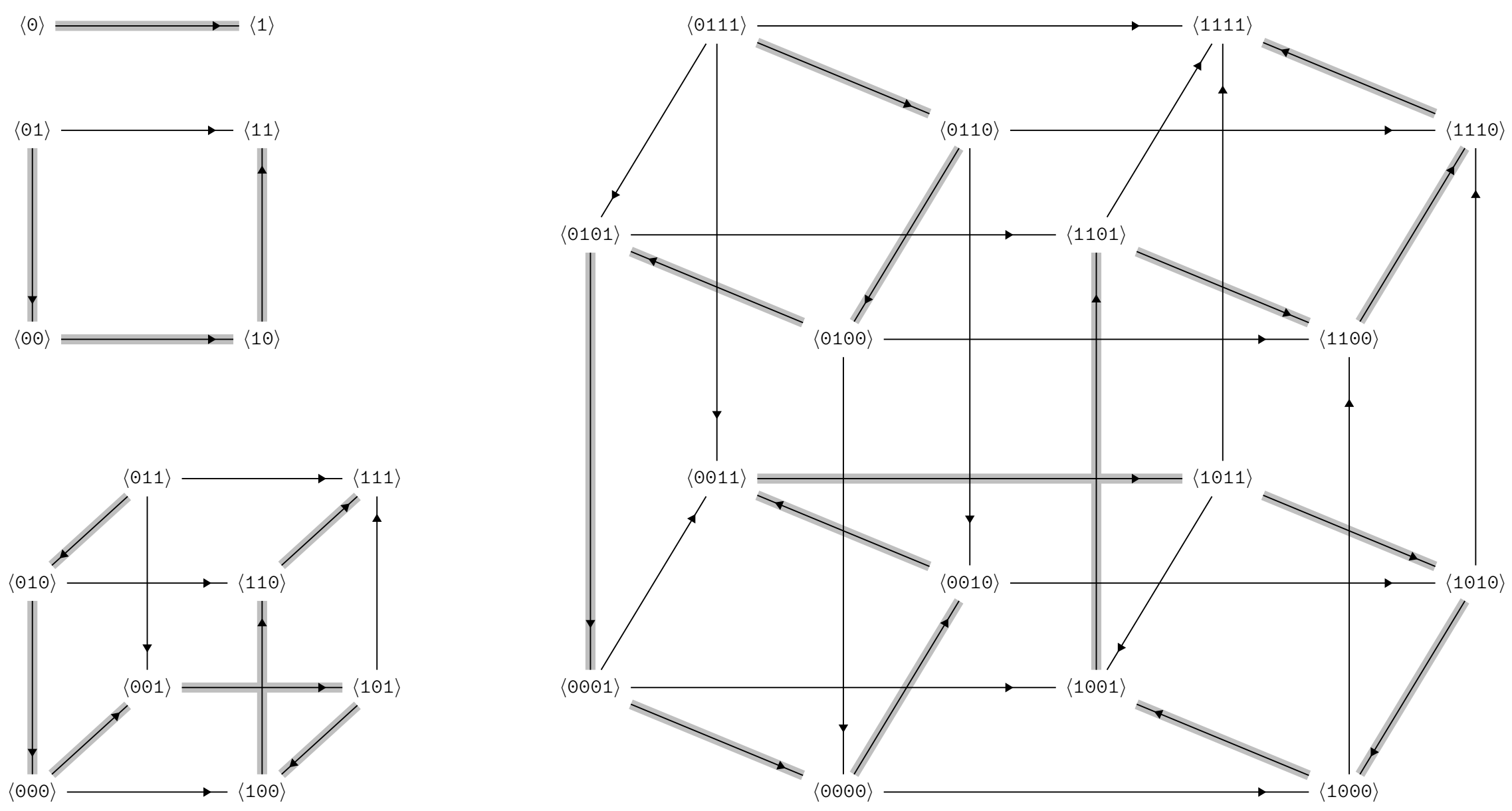

Figure 6.1: $\mathbb{F}^1_{\bowtie}$, $\mathbb{F}^2_{\bowtie}$, $\mathbb{F}^3_{\bowtie}$, and $\mathbb{F}^4_{\bowtie}$ annotated by their unique Hamiltonian path.

◤ **Definition 6.14** **(Hamiltonian paths)** ⬣

A path is a Hamiltonian path iff the path visits every node in the graph exactly once. To be more precise, a Hamiltonian path of some graph G is a path p such that p.map a is bijective function. ◢

◤ **Lemma 6.15** **(Hamiltonian path exists)** ⬣

Let $n : \mathbb{N}$, then $\mathbb{F}^n_{\bowtie}$ has a Hamiltonian path. ◢

*Proof.* We prove this by induction on $n$. For the base case, it is obvious that the Hamiltonian path of $\mathbb{F}^0_{\bowtie}$ is a path of length zero where the endpoint is $\langle\rangle$, which is the single node of $\mathbb{F}^0_{\bowtie}$. For the inductive step, the induction hypothesis is that $\mathbb{F}^n_{\bowtie}$ has a Hamiltonian path and we need to prove that it is also the case for $\mathbb{F}^{(n+1)}_{\bowtie}$.

First, we prepend digit 1 to every node in $\mathbb{F}^n_{\bowtie}$. This new graph is clearly a subgraph of $\mathbb{F}^{(n+1)}_{\bowtie}$ so the Hamiltonian path in this new graph becomes a path in $\mathbb{F}^{(n+1)}_{\bowtie}$, now denoted t, that visits every node that starts with digit 1.

Then, we prepend digit 0 to every node in $\mathbb{F}^n_{\bowtie}$ and reverse every edge in the graph. This new graph is also a $\mathbb{F}^n_{\bowtie}$ so the Hamiltonian path in this new graph, which is the same Hamiltonian path before the reversing process but in the reverse order, becomes a path in $\mathbb{F}^{(n+1)}_{\bowtie}$, now denoted s, that visits every node that starts with digit 0.

Now, let $\vec{v}$ be the first node in the Hamiltonian path of $\mathbb{F}^n_{\bowtie}$. This implies that $1\vec{v}$ is the first node of t whereas $0\vec{v}$ is the last node of s. Because there is an edge that link $0\vec{v}$ to $1\vec{v}$, so we use that edge to concatenate s with t. This results in a path that visit every node in $\mathbb{F}^{(n+1)}_{\bowtie}$; thus, it is a Hamiltonian path for $\mathbb{F}^{(n+1)}_{\bowtie}$. QED

◤ **Lemma 6.16** **(Hamiltonian path is unique)** ⬣

Let $n : \mathbb{N}$, then $\mathbb{F}^n_{\bowtie}$ has at most one Hamiltonian path. ◢

*Proof.* We prove this by contradiction. Assuming that there are distinct Hamiltonian paths for $\mathbb{F}^{n}_{\bowtie}$, so there is at least a pair of distinct nodes $\vec{v}$ and $\vec{v'}$ such that $\vec{v}$ comes before $\vec{v'}$ in the first Hamiltonian path and vice-versa in the second Hamiltonian path.

Let $p$ be a sub-path of the first Hamiltonian path that starts from $\vec{v}$ and ends at $\vec{v'}$. Let $p'$ be a sub-path of the second Hamiltonian path that starts from $\vec{v'}$ and stops at $\vec{v}$. The concatenation of $p$ and $p'$ is a cyclic path, i.e. a path of length more than zero but has the same starting point with stopping point. However, the graph $\mathbb{F}^{n}_{\bowtie}$ is acyclic; therefore, it is impossible for $\mathbb{F}^{n}_{\bowtie}$ to have multiple Hamiltonian paths. QED

◤ **Theorem 6.17** **(Hamiltonian path exists uniquely)** ⬣
Let $n : \mathbb{N}$, then $\mathbb{F}^{n}_{\bowtie}$ has a unique Hamiltonian path. ◢

*Proof.* By lemmas 6.15 and 6.16 QED

◤ **Corollary 6.18** ⬣ Let $n : \mathbb{N}$, then $\mathtt{tran}_{\mathtt{cl}}(\mathbb{F}^{n}_{\bowtie}) \cong \mathbb{F}^{(2^n)}_{\triangle}$. ◢

◤ **Theorem 6.19** ⬣ Let $n : \mathbb{N}$, then the reflexive transitive closure of face graph for twisted $n$-cube is the reflexive graph for a $2^n$-simplex. ◢

*Proof.*

$$
\begin{aligned}
& \mathtt{refl}_{\mathtt{cl}}(\mathtt{tran}_{\mathtt{cl}}(\mathbb{F}^{n}_{\bowtie})) \\
\cong\; & \mathtt{refl}_{\mathtt{cl}}(\mathbb{F}^{(2^n)}_{\triangle}) && \text{[by corollary 6.18 and lemma 5.21]} \\
\cong\; & \mathbb{G}^{(2^n)}_{\triangle} && \text{[by lemma 5.51]}
\end{aligned}
$$

QED

## § 6.2.2 Sorting the Nodes for Twisted Cubes

The existence of the unique Hamiltonian path of $\mathbb{F}^{n}_{\bowtie}$ inspires me to find an algorithm to "sort" nodes in $\mathbb{F}^{n}_{\bowtie}$. This resulting in a function called $\mathtt{sort}^{n}_{\mathsf{num}}$ that takes a node $\mathbb{F}^{n}_{\bowtie}$ in binary format then return the index of that node. If we set the codomain of the function $\mathtt{sort}^{n}_{\mathsf{num}}$ to be $\mathtt{fin}(2^{n})$, then it is easy to see that the function is indeed bijective; therefore, we can define its inverse function called $\mathtt{unsort}^{n}_{\mathsf{num}}$.

**Definition 6.20** (sorting function in Hamiltonian path order)

Let $n : \mathbb{N}$, then we define $\mathtt{sort}^{n}_{\mathsf{num}} : \mathtt{binary}(n) \to \mathtt{fin}(2^{n})$ that takes a node of $\mathbb{F}^{n}_{\bowtie}$ and returns the index that this node belongs in the Hamiltonian path.

$$\begin{aligned}
\mathtt{sort}^{0}_{\mathsf{num}} \quad & (\langle\rangle) && :\equiv \quad 0 \\
\mathtt{sort}^{(n+1)}_{\mathsf{num}} \quad & (\mathtt{0} :: \vec{b}) && :\equiv \quad 2^{n} - \mathtt{sort}^{n}_{\mathsf{num}}(\vec{b}) \\
\mathtt{sort}^{(n+1)}_{\mathsf{num}} \quad & (\mathtt{1} :: \vec{b}) && :\equiv \quad 2^{n} + \mathtt{sort}^{n}_{\mathsf{num}}(\vec{b})
\end{aligned}$$

**Definition 6.21** (unsorting function in Hamiltonian path order)

Let $n : \mathbb{N}$, then we define $\mathtt{unsort}^{n}_{\mathsf{num}} : \mathtt{fin}(2^{n}) \to \mathtt{binary}(n)$ that takes an number $i : \mathtt{fin}(2^{n})$ and return the node of $\mathbb{F}^{n}_{\bowtie}$ at index $i$ of the Hamiltonian path.

$$\begin{aligned}
\mathtt{unsort}^{0}_{\mathsf{num}} \quad & 0 && :\equiv \quad \langle\rangle \\
\mathtt{unsort}^{(n+1)}_{\mathsf{num}} \quad & \big(i : \mathtt{fin}(2^{n+1})\big) && :\equiv
\end{aligned}$$

$$\begin{cases} \mathtt{0} :: \mathtt{unsort}^{n}_{\mathsf{num}}(2^{n} - i) & \text{if } i < 2^{n} \\ \mathtt{1} :: \mathtt{unsort}^{n}_{\mathsf{num}}(2^{n} + i) & \text{if } i \geqslant 2^{n} \end{cases}$$

◤ **Lemma 6.22** ⬣ Let $n : \mathbb{N}$ and $\vec{s}, \vec{t} : \texttt{binary}(n)$,

$$\mathbb{F}^{n}_{\bowtie}.\mathsf{isEdge}(\vec{s}, \vec{t}) \quad \text{implies} \quad \texttt{sort}^{n}_{\mathsf{num}}(\vec{s}) \leqslant \texttt{sort}^{n}_{\mathsf{num}}(\vec{t})$$ ◢

◤ **Lemma 6.23** ⬣ Let $n : \mathbb{N}$ and $s, t : \texttt{fin}(2^{n})$,

$$\texttt{line}^{2^n}_{\mathsf{grp}}.\mathsf{isEdge}(s, t) \quad \text{implies} \quad \mathbb{F}^{n}_{\bowtie}.\mathsf{isEdge}(\texttt{unsort}^{n}_{\mathsf{num}}(s), \texttt{unsort}^{n}_{\mathsf{num}}(t))$$ ◢

◤ **Definition 6.24** ⬣ Let $n : \mathbb{N}$, we define $\texttt{HamPath}^{n}_{\mathsf{tw}}$ to be a path in $\mathbb{F}^{n}_{\bowtie}$.

| | | |
|---|---|---|
| $\texttt{HamPath}^{n}_{\mathsf{tw}}.\mathsf{length}$ | :≡ | $2^{n}$ |
| $\texttt{HamPath}^{n}_{\mathsf{tw}}.\mathsf{map.mapNode}$ | :≡ | $\texttt{unsort}^{n}_{\mathsf{num}}$ |
| $\texttt{HamPath}^{n}_{\mathsf{tw}}.\mathsf{map.mapEdge}$ | :≡ | lemma 6.23 ◢ |

◤ **Theorem 6.25** ⬣
Let $n : \mathbb{N}$, the path $\texttt{HamPath}^{n}_{\mathsf{tw}}$ is only the Hamiltonian-path in $\mathbb{F}^{n}_{\bowtie}$. ◢

*Proof.* We know that $\texttt{HamPath}^{n}_{\mathsf{tw}}$ is a Hamiltonian-path because $\texttt{unsort}^{n}_{\mathsf{num}}$, which defines $\texttt{HamPath}^{n}_{\mathsf{tw}}.\mathsf{map.mapNode}$, is bijective. It is also unique because of lemma 6.16. QED

In addition, we also define $\texttt{sort}^{n}_{\mathsf{bin}}$ and $\texttt{unsort}^{n}_{\mathsf{bin}}$ that do the same thing as $\texttt{sort}^{n}_{\mathsf{num}}$ and $\texttt{unsort}^{n}_{\mathsf{num}}$ but replacing $\texttt{fin}(2^{n})$ with $\texttt{binary}(n)$; so, $\texttt{sort}^{n}_{\mathsf{bin}}$ and $\texttt{unsort}^{n}_{\mathsf{bin}}$ are endofunction on $\texttt{binary}(n)$ and can be defined bit-wisely and much simpler comparing to the original counterparts.

To get a better understanding of $\texttt{sort}^{n}_{\mathsf{bin}}$ and $\texttt{unsort}^{n}_{\mathsf{bin}}$, we illustrate $\mathbb{F}^{n}_{\bowtie}$ for all $(1 \leqslant n \leqslant 3)$ in figure 6.2 where nodes of $\mathbb{F}^{n}_{\bowtie}$ are positioned linearly in order of $\texttt{HamPath}^{n}_{\mathsf{tw}}$. In addition to the original binary number that represents each node, we also label the order of each node in binary number so how each original binary number get mapped to the order using $\texttt{sort}^{n}_{\mathsf{bin}}$ and vice-versa using $\texttt{unsort}^{n}_{\mathsf{bin}}$.

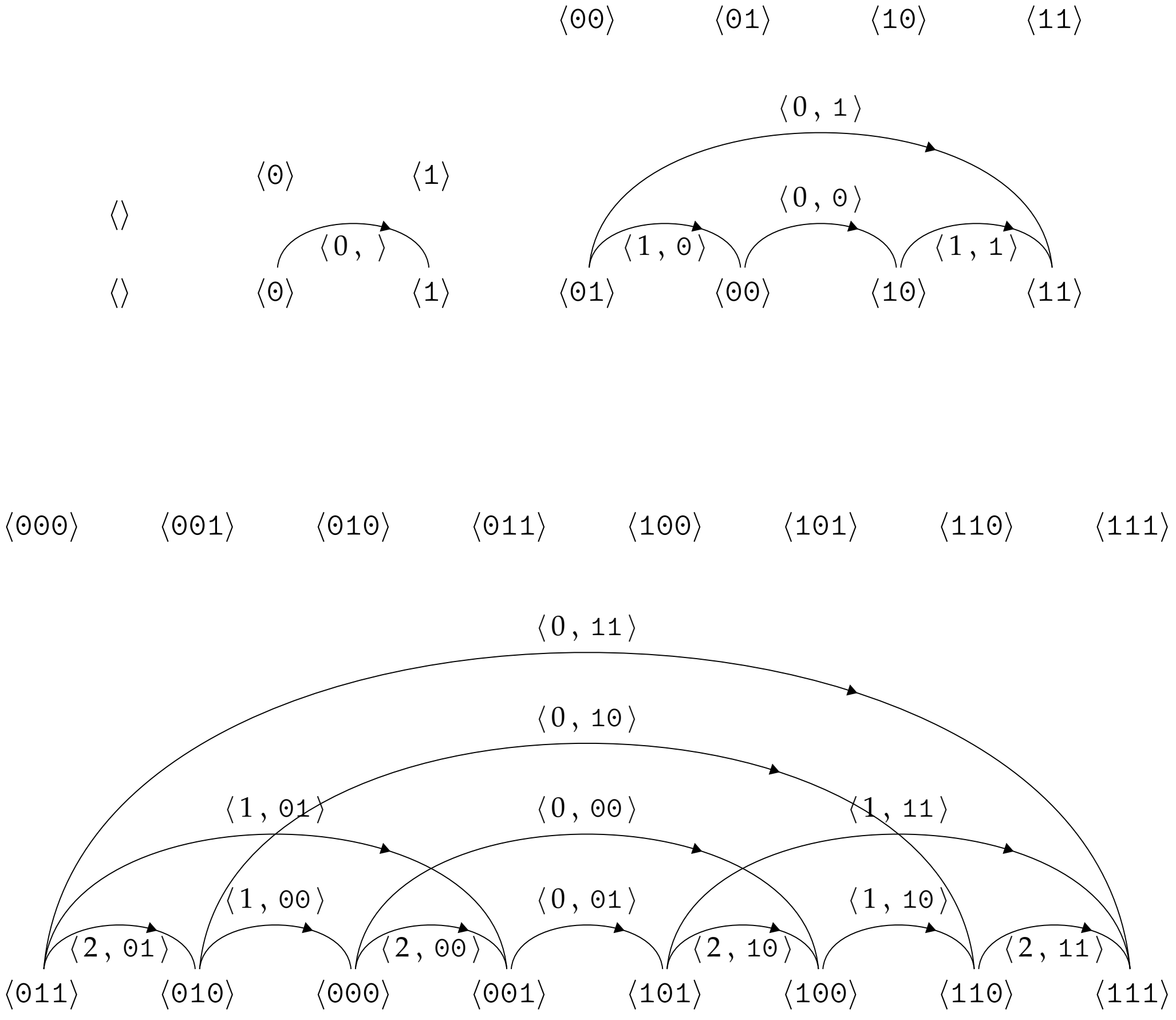

Figure 6.2: Horizontal sorting the nodes of $\mathbb{F}^1_{\bowtie}$, $\mathbb{F}^2_{\bowtie}$, and $\mathbb{F}^3_{\bowtie}$ annotated by their positional labels (lower number) and their index in the Hamiltonian path (upper number).

◤ **Definition 6.26** ⬢ Let $n : \mathbb{N}$, then we define $\texttt{sort}^n_{\textsf{bin}}$ and $\texttt{unsort}^n_{\textsf{bin}}$ to be alternative versions of $\texttt{sort}^n_{\textsf{num}}$ and $\texttt{unsort}^n_{\textsf{num}}$ but replacing $\texttt{fin}(2^n)$ with $\texttt{binary}(n)$.

$$\begin{aligned}
\texttt{sort}^n_{\textsf{bin}}\ ,\quad \texttt{unsort}^n_{\textsf{bin}} \quad &: \quad \texttt{binary}(n) \to \texttt{binary}(n) \\
\texttt{sort}^n_{\textsf{bin}}(\langle\, b_0\,,\, b_1\,,\, \dots\,,\, b_{n-1}\,\rangle) \quad &:\equiv \quad \langle\, b'_0\,,\, b'_1\,,\, \dots\,,\, b'_{n-1}\,\rangle \\
\text{where}\quad b'_i \quad &:\equiv \quad b_i \oplus \texttt{parity}(\langle b_0 b_1 \dots b_{i-1}\rangle) \\
\texttt{unsort}^n_{\textsf{bin}}(\langle\, b_0\,,\, b_1\,,\, \dots\,,\, b_{n-1}\,\rangle) \quad &:\equiv \quad \langle\, b_0\,,\, b'_1\,,\, b'_2\,,\, \dots\,,\, b'_{n-1}\,\rangle \\
\text{where}\quad b'_i \quad &:\equiv \quad b_i \oplus b_{i-1}
\end{aligned}$$
◢

### § 6.2.3 Alternative Twisting Process and the Gray Code

The definitions of $\texttt{sort}^n_{\text{bin}}$ and $\texttt{unsort}^n_{\text{bin}}$ are similar to the encoding and decoding algorithm of *Gray code*, a.k.a. reflected binary code (RBC), which has a Hamming distance 1 between consecutive words. In fact, if we tweak the thickening-and-twisting process by reversing everything at `1` instead of at `0` then the sorting and unsorting functions match exactly to the encoding and decoding algorithm of Gray code.

Alternatively, we can use the tweaked version thickening-and-twisting to reason about everything in this thesis; on one hand, this will even simplify the definition of $\texttt{parity}(\langle b_0 b_1 \ldots b_{i-1} \rangle)$ from $(1 - b_0) \oplus (1 - b_1) \oplus \ldots \oplus (1 - b_{n-1})$ to $b_0 \oplus b_1 \oplus \ldots \oplus b_{n-1}$; on the other hand, this will make the composition in section 1.3 much harder.

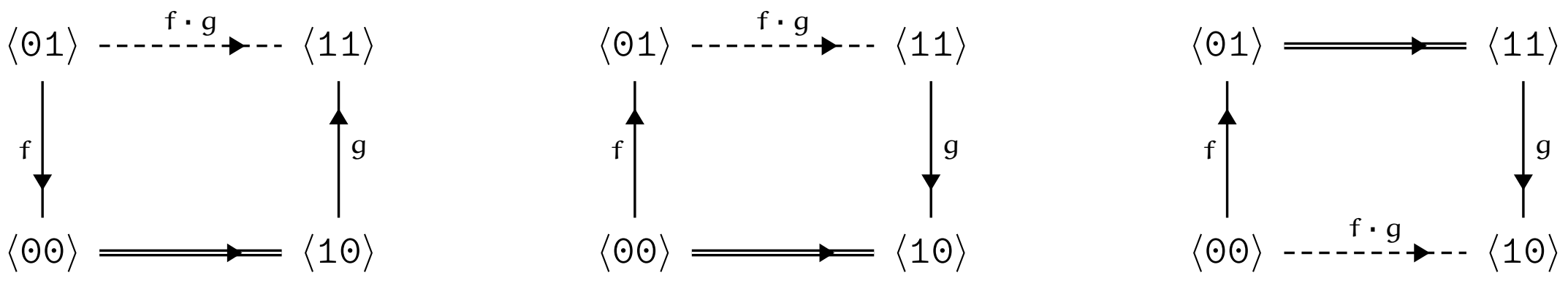

The first diagram represents a composition using the original thickening-and-twisting, which can be tweaked and become the second diagram; here, the composition is no longer makes sense so we need to swap the duty of $\langle \star 0 \rangle$ and $\langle \star 1 \rangle$ so it become the third diagram where you can see that the direction of composition is opposite to the direction of the second dimension, which is not the case for the first diagram; so it is a lot easier to be confused if we use this alternative.

Moreover, if we see an arrow as a function from its domain to its codomain, it is more sensible to reverse something in domain than codomain because pre-composition is contrapositive whereas post-composition is not. In conclusion, I think the original thickening-and-twisting is easier and it is not worth to switch to the alternative.

### § 6.2.4 Using the Order of Nodes to Redefine the Face Graphs

There is an alternative definition of $\mathbb{F}^n_{\bowtie}$, denoted as $\mathbb{F}^n_{\text{⧖}}$, that directly use the binary representation to store the order of the nodes themselves.

Surprisingly once we relabel $\mathbb{F}^n_{\bowtie}$ to $\mathbb{F}^n_{\text{⧖}}$, the definition become simpler and doesn't rely on the parity function. Moreover, $\mathbb{F}^n_{\bowtie}$ can be seen as a modification of $\mathbb{F}^n_{\square}$ where the if clause is changed from $(i = j)$ to $(i \leqslant j)$.

**Definition 6.27** **(sorted version of the face graphs for twisted cubes)**
Let $n : \mathbb{N}$, we define a graph $\mathbb{F}^n_{\text{⧖}}$ to be the sorted version of the face graph for a twisted $n$-cube where each node $\vec{b}$ is relabelled as $\mathtt{sort}^n_{\mathsf{bin}}(\vec{b})$, i.e. the binary number that label node is changed from its position to its order in the Hamiltonian path.

$$\mathbb{F}^n_{\text{⧖}}.\mathsf{Nodes} \quad :\equiv \quad \mathtt{binary}(n)$$

$$\mathbb{F}^n_{\text{⧖}}.\mathsf{isEdge}(\vec{s},\vec{t}) \quad :\equiv \quad \exists(i : \mathtt{fin}(n)) \times \forall(j : \mathtt{fin}(n)) \to \text{if } (i \leqslant j) \text{ then } \vec{s}[j] < \vec{s}[j] \text{ else } \vec{s}[j] = \vec{t}[j]$$

**Theorem 6.28** Let $n : \mathbb{N}$, then $\mathbb{F}^n_{\text{⧖}} \cong \mathbb{F}^n_{\bowtie}$ in **Graph**.

*Proof.* The isomorphism can be directly constructed using $\mathtt{sort}^n_{\mathsf{bin}}$ to map nodes forwardly and $\mathtt{unsort}^n_{\mathsf{bin}}$ to map nodes backwardly. QED

**Convention 6.29** **(twisted cubes coercion from their sorting graphs)**
We define a coercion from $\mathbb{F}^n_{\text{⧖}}$ to $\mathbb{F}^n_{\bowtie}$ using the isomorphisms in theorem 6.28.

## § 6.3 Reflexive Graphs for Twisted Cubes

The modification from $\mathbb{F}^n_\square$ to $\mathbb{F}^n_\bowtie$ and the modification from $\mathbb{F}^n_\square$ to $\mathbb{G}^n_\square$ fit well together. This results in the graph $\mathbb{G}^n_\bowtie$ that can be seen as either the former modification on $\mathbb{G}^n_\square$ or the latter modification on $\mathbb{F}^n_\bowtie$.

### § 6.3.1 Non-Recursive Reflexive Graphs for Twisted Cubes

◤ **Definition 6.30** **(reflexive graphs for twisted cubes)** ⬣

Let $n : \mathbb{N}$, we define $\mathbb{G}^n_\bowtie$ to be the reflexive graph for a twisted $n$-cube.

$$\mathbb{G}^n_\bowtie.\mathsf{Nodes} \quad :\equiv \quad \mathtt{binary}(n)$$

$$\mathbb{G}^n_\bowtie.\mathsf{isEdge}(\vec{s},\vec{t}) \quad :\equiv \quad \exists(i : \mathtt{fin}(n)) \times \forall(j : \mathtt{fin}(n)) \to$$

$$\text{if} \quad (i = j) \quad \text{then} \quad (\vec{s}[j] \oplus p) \leqslant (\vec{t}[j] \oplus p) \quad \text{else} \quad \vec{s}[j] = \vec{t}[j]$$

$$\text{where} \quad p \quad :\equiv \quad \mathtt{parity}\ \langle \vec{t}[0]\,,\, \vec{t}[1]\,,\, \dots\,,\, \vec{t}[i-1] \rangle$$ ◢

◤ **Lemma 6.31** ⬣ Let $n : \mathbb{N}$, then $\mathbb{G}^n_\bowtie \cong \mathtt{refl}_{\mathtt{cl}}(\mathbb{F}^n_\bowtie)$ in **Graph**. ◢

*Proof.* Use the proof of lemma 5.60 but replacing symbol □ with symbol ⋈. QED

### § 6.3.2 Recursive Reflexive Graphs for Twisted Cubes

To upgrade the recursive version of $\mathbb{G}^n_{\bowtie}$ as reflexive graph, it is as easy as replacing $\epsilon_{\mathsf{grp}}$ with $\mathbb{1}_{\mathsf{grp}}$ in the base case.

**Definition 6.32** (reflexive graphs for twisted cubes using recursion)

Let $n : \mathbb{N}$, we define a graph $\acute{\mathbb{G}}^n_{\bowtie}$ to be a recursive version of graph $\mathbb{G}^n_{\bowtie}$.

$$\acute{\mathbb{G}}^0_{\bowtie} :\equiv \mathbb{1}_{\mathsf{grp}} \qquad \acute{\mathbb{G}}^{(n+1)}_{\bowtie} :\equiv \texttt{prism}_{\texttt{tw}}(\acute{\mathbb{G}}^n_{\bowtie})$$

**Lemma 6.33** Let $n : \mathbb{N}$, then $\acute{\mathbb{G}}^n_{\bowtie} \cong \texttt{refl}_{\texttt{cl}}(\acute{\mathbb{F}}^n_{\bowtie})$ in **Graph**.

*Proof.* Given that $G, G' : \mathsf{ob}(\textbf{Graph})$, the following propositions hold:

- $\texttt{refl}_{\texttt{cl}}(\texttt{prism}_{\texttt{tw}}(G)) \cong \texttt{prism}_{\texttt{tw}}(\texttt{refl}_{\texttt{cl}}(G))$ [by lemma 5.19]
- if $(G \cong G')$ then $\texttt{prism}_{\texttt{tw}}(G) \cong \texttt{prism}_{\texttt{tw}}(G')$ [by lemma 6.6]
- if $(G \cong G')$ then $\texttt{refl}_{\texttt{cl}}(G) \cong \texttt{refl}_{\texttt{cl}}(G')$ [by lemma 5.21]

$$\begin{aligned}
\text{Then,}\quad \acute{\mathbb{G}}^n_{\bowtie} &\cong \texttt{prism}_{\texttt{tw}}(\texttt{prism}_{\texttt{tw}}((\ldots\ \texttt{prism}_{\texttt{tw}}(\mathbb{1}_{\mathsf{grp}})\ )))\\
&\cong \texttt{prism}_{\texttt{tw}}(\texttt{prism}_{\texttt{tw}}((\ldots\ \texttt{prism}_{\texttt{tw}}(\texttt{refl}_{\texttt{cl}}(\epsilon_{\mathsf{grp}}))\ )))\\
&\cong \texttt{prism}_{\texttt{tw}}(\texttt{prism}_{\texttt{tw}}((\ldots\ \texttt{refl}_{\texttt{cl}}(\texttt{prism}_{\texttt{tw}}(\epsilon_{\mathsf{grp}}))\ )))\\
&\cong \texttt{prism}_{\texttt{tw}}(\texttt{refl}_{\texttt{cl}}((\ldots\ \texttt{prism}_{\texttt{tw}}(\texttt{prism}_{\texttt{tw}}(\epsilon_{\mathsf{grp}}))\ )))\\
&\cong \texttt{refl}_{\texttt{cl}}(\texttt{prism}_{\texttt{tw}}((\ldots\ \texttt{prism}_{\texttt{tw}}(\texttt{prism}_{\texttt{tw}}(\epsilon_{\mathsf{grp}}))\ )))\\
\text{Hence,}\quad \acute{\mathbb{G}}^n_{\bowtie} &\cong \texttt{refl}_{\texttt{cl}}(\acute{\mathbb{F}}^n_{\bowtie})
\end{aligned}$$

QED

**Theorem 6.34** Let $n : \mathbb{N}$, then $\acute{\mathbb{G}}^n_{\bowtie} \cong \mathbb{G}^n_{\bowtie}$ in **Graph**.

*Proof.*

$$\begin{aligned}
\acute{\mathbb{G}}^n_{\bowtie} &\cong \texttt{refl}_{\texttt{cl}}(\acute{\mathbb{F}}^n_{\bowtie}) && \text{[by lemma 6.33]}\\
&\cong \texttt{refl}_{\texttt{cl}}(\mathbb{F}^n_{\bowtie}) && \text{[by lemma 5.21 and theorem 6.10]}\\
\acute{\mathbb{G}}^n_{\bowtie} &\cong \mathbb{G}^n_{\bowtie} && \text{[by lemma 6.31]}
\end{aligned}$$

QED

### § 6.3.3 Dimension-Preserving Graph Homomorphisms

Similar to the family of $\mathbb{G}^n_\triangle$ that constructs $\triangle^{\text{graph}}_{\text{full}}$, we can use the family of $\mathbb{G}^n_\bowtie$ to construct a twisted cube category (graph-theoretic version), denoted as $\bowtie^{\text{graph}}_{\text{full}}$.

◤ **Definition 6.35** **(graph theoretic version of twisted cube category)** ⬢

We define the category $\bowtie^{\text{graph}}_{\text{full}}$ to be a full-subcategory of **Graph** induced by the family of graphs $\mathbb{G}^n_\bowtie$ for all $n : \mathbb{N}$. Please note that, we may use the number $n$ or the graph $\mathbb{G}^n_\bowtie$ to denote each object in $\bowtie^{\text{graph}}_{\text{full}}$ interchangeably. ◢

Unfortunately, we still don't know whether the category $\bowtie^{\text{graph}}_{\text{full}}$ is a Reedy category or not; to work around this open problem we select only some morphisms in $\bowtie^{\text{graph}}_{\text{full}}$ that "behave well".

◤ **Definition 6.36** ⬢ We define a function

$$\texttt{dim} \;:\; \mathbb{G}^n_\bowtie.\text{isEdge}(\vec{s},\vec{t}) \;\to\; \texttt{fin}(n{+}1) \qquad \text{for every} \quad n : \mathbb{N} \quad \text{and} \quad \vec{s},\ \vec{t} \;:\; \mathbb{G}^n_\bowtie.\text{Nodes}$$

that takes an edge $\langle \vec{s}, \vec{t} \rangle$ from $\mathbb{G}^n_\bowtie$ and returns the index $i$ that $(\vec{s}[i] \neq \vec{t}[i])$, if such an index exists; otherwise, returns $n : \texttt{fin}(n+1)$.

Please note that, if a distinguishing index exists, then it is unique because the definition of $\mathbb{G}^n_\bowtie.\text{isEdge}$ prevents a pair $\langle \vec{s}, \vec{t} \rangle$ that has multiple distinguishing indices from being an edge.

Alternatively, we can think that $\texttt{dim}(\, e : \mathbb{G}^n_\bowtie.\text{isEdge}(\vec{s},\vec{t})\,)$ returns the variable $i$ in the existential quantifier in of $\mathbb{G}^n_\bowtie.\text{isEdge}(\vec{s},\vec{t})$ in definition 6.7. ◢

◤ **Definition 6.37** ⬣ We define `presDim` to be a predicate that takes $f : \textbf{Graph}(\mathbb{G}^m_{\bowtie}, \mathbb{G}^n_{\bowtie})$ for some $m, n : \mathbb{N}$ then return the following proposition.

$$\forall(\vec{s}, \vec{s}', \vec{t}, \vec{t}' : \mathbb{G}^m_{\bowtie}.\mathsf{Nodes})(e : \mathbb{G}^m_{\bowtie}.\mathsf{isEdge}(\vec{s}, \vec{t}))(e' : \mathbb{G}^m_{\bowtie}.\mathsf{isEdge}(\vec{s}', \vec{t}')) \rightarrow$$
$$(\mathtt{dim}(e) = \mathtt{dim}(e')) \rightarrow \mathtt{dim}(f.\mathsf{mapEdge}(\vec{s}, \vec{t})(e)) = \mathtt{dim}(f.\mathsf{mapEdge}(\vec{s}', \vec{t}')(e'))$$ ◢

◤ **Definition 6.38** ⬣ We define $\bowtie^{\mathsf{graph}}_{\mathsf{dim}}$ to be the wide-subcategory of $\bowtie^{\mathsf{graph}}_{\mathsf{full}}$ such that a morphism $f$ in $\bowtie^{\mathsf{graph}}_{\mathsf{full}}$ will also be in $\bowtie^{\mathsf{graph}}_{\mathsf{dim}}$ iff `presDim`$(f)$ ◢

◤ **Theorem 6.39** ⬣ Let $m, n : \mathbb{N}$, there is exactly one surjective morphism in $\bowtie^{\mathsf{graph}}_{\mathsf{dim}}$ $mn$ for $(m \geqslant n)$ (Clearly, there is none if $(m < n)$). ◢

*Proof.* See [PK20, Theorem 27] QED

◤ **Theorem 6.40** ⬣ The category $\bowtie^{\mathsf{graph}}_{\mathsf{dim}}$ is a Reedy category. ◢

*Proof.* See [PK20, Theorem 33] QED

Chapter

# 7

# Geometric Realisation of Twisted Cubes

This chapter transforms the intuition in chapter 4 from topological spaces to "direct" spaces, which we use *partially ordered spaces* to implement them. Then, differentiate the geometric realisation of twisted cubes from standard cubes, which is impossible to do so in topological spaces alone.

Please note that, the content here can't be merged into chapter 4 because the content here depends on the notion of graphs in chapter 5 and the implementation of twisted cubes in chapter 6 but these two chapters happen to depend on chapter 4; therefore, the order of these chapters must be as is.

## § 7.1 Partially Ordered Spaces

Topological spaces alone are not expressive enough to fully encode the representation of twisted cubes because paths in twisted cubes are not necessary reversible; therefore, we upgrade topological spaces to partially ordered spaces, which have sense of "direction".

◤ **Remark 7.1** ⬣ As we mention in the last paragraph of section 2.5, I would like to remind the reader that there are other definitions of *directed spaces* by Marco Grandis [Gra09] or Sanjeevi Krishnan [Kri08] that are more developed than pospaces i.e. all pospaces can be transferred to these definitions but not the other way around. However, I decide to use pospaces in this thesis because the definition is simpler yet covers all direct spaces that are important in this thesis. ◢

◤ **Definition 7.2** **(pospaces)** ⬣

A *partially ordered space*, or *pospace* for short, is a pair $\langle X, \leqslant_X \rangle$ where $X$ is a topological space and $(\leqslant_X)$ is a *closed* partial order on $X$, i.e. $(\leqslant_X)$ is a partial order on the set of points on $X$ such that $\{ \langle x, y \rangle : X^2 \mid x \leqslant_X y \}$ is a *closed set.* ◢

◤ **Definition 7.3** **(dimaps)** ⬣

A *dimap* $f$ from a pospace $\langle X, \leqslant_X \rangle$ to a pospace $\langle Y, \leqslant_Y \rangle$ is a continuous function from $X$ to $Y$ that preserves the partial order, i.e. if $(u \leqslant_X v)$ then $f(u) \leqslant_Y f(v)$. ◢

◤ **Example 7.4** ⬣ Let $n : \mathbb{N}$, then the Euclidean space $\mathbb{R}^n_{\mathsf{top}}$ can be upgraded to the *Euclidean pospace* $\mathbb{R}^n_{\mathsf{po}} :\equiv \langle \mathbb{R}^n_{\mathsf{top}}, \leqslant_{\mathbb{R}^n_{\mathsf{po}}} \rangle$, where

$$\vec{x} \leqslant_{\mathbb{R}^n_{\mathsf{po}}} \vec{y} \;:\equiv\; (\vec{x}_0 \leqslant \vec{y}_0) \wedge (\vec{x}_1 \leqslant \vec{y}_1) \wedge \cdots \wedge (\vec{x}_{n-1} \leqslant \vec{y}_{n-1})$$ ◢

## § 7.2 Generating Pospaces from Real-Valued Metric Spaces

This section introduce an algorithm that takes an arbitrary continuous function from some metric space to $\mathbb{R}_{top}$ and produces a pospace.

According to definition 7.2, the usual way to define a pospace is to; first, compare any two points in the original topological space; then, make sure that the comparison is a closed partial order; this might be counter-intuitive and tedious in some cases, e.g. the earlier attempts when I try to define twisted cubes as pospaces.

Alternatively, we can try to define a pospace by assigning a real number to each point in the original topological space, which is equivalent to defining a continuous function from the original topological space to $\mathbb{R}_{top}$, then generating a pospace from it. However, the relation arises from simply comparing real numbers between two points might not be a closed partial order; this is where I need to restrict spaces of interest from topological spaces to metric spaces so we can always get a closed partial order.

◤ **Definition 7.5** ⬣ Let $(X, d)$ be a metric space and let $f$ be a continuous function from $(X, d)$ to $\mathbb{R}_{top}$, we define $(\sqsubseteq_f)$ to be a relation on $X$ such that

$$(x \sqsubseteq_f y) \quad :\equiv \quad d(x, y) \leqslant f(y) - f(x)$$ ◢

Regarding the intuition, let $x$ and $y$ be points in $X$, if the assigned value at $x$ is less than or equal to the assigned value at $y$, i.e. $(f(x) \leqslant f(y))$, then there should be a path from $x$ to $y$. To preserve its continuity, we further require that the difference between the assigned values must be at least the distance between them. On the other hand, if $(f(y) > f(x))$, then there should be no path from $x$ to $y$.

◤ **Lemma 7.6** ⬣

Given the context as definition 7.5, the relation ( $\sqsubseteq_f$ ) is a closed partial order. ◢

*Proof.* First, we show that ( $\sqsubseteq_f$ ) is a partial order.

- Regarding reflexivity, $d(x,x) \leqslant f(x) - f(x)$ is obviously true because $d(x,x) = 0 = f(x) - f(x)$ by definiteness of the metric and algebra in $\mathbb{R}_{top}$.
- Regarding transitivity, given $d(x,y) \leqslant f(y) - f(x)$ and $d(y,z) \leqslant f(z) - f(y)$ adding these inequalities together, we get $d(x,y) + d(y,z) \leqslant f(z) - f(x)$; therefore, $d(x,z) \leqslant f(z) - f(x)$ by the triangle inequality, $d(x,z) \leqslant d(x,y) + d(y,z)$.
- Regarding antisymmetry, given $d(x,y) \leqslant f(y) - f(x)$ and $d(y,x) \leqslant f(x) - f(y)$ adding these inequalities together, we get $d(x,y) + d(y,x) \leqslant 0$.
  Then, because of the symmetry, $d(x,y) = d(y,x)$, we get $2 \cdot d(x,y) \leqslant 0$ but d can't output negative number so $d(x,y) = 0$;
  therefore, $x$ and $y$ must be the same point due to the definiteness of d.

The remaining goal is to show that is $\{ \langle x , y \rangle : X^2 \mid x \sqsubseteq_f y \}$ is a *closed set* in the product topological space $X^2$.

Since $\{ \langle x , y \rangle : X^2 \mid x \sqsubseteq_f y \}$ is the preimage of a function

$$g(x,y) \quad :\equiv \quad f(y) - f(x) - d(x,y)$$

of the closed interval $[0, \infty)$; hence, it must be closed because $g$ is continuous, by continuity of f, d, and subtraction. QED

◤ **Definition 7.7** ⬣

Given the context as definition 7.5 and let a topological space $X'$ be a subspace of X, we define $\texttt{mkPospc}(X', f)$ to be a pospace consisting of the topological space $X'$, together with the partial order ( $\sqsubseteq_f$ ) restricted to $X'$. ◢

## § 7.3 Embedding Graphs to Pospaces

We want to translate a graph into a pospace because many results in twisted cubes relay on graph theoretic representations. To do this, we treat nodes and paths in a graph G to represent points and (topological) paths of the resulting pospace $\texttt{mkGraphPospc}(G)$.

Next, we check whether each pospace (that will be defined later) is compatible its graph counterpart. In other words, let S be a shape that respects convention 4.5 and let $G_S$ be a graph that encodes the shape S, we want to ensure that the graph $\texttt{mkGraphPospc}(G_S)$ can be *embedded* into a pospace $P_S$ that we will define to realise the shape S. If this is the case then the proposition $\texttt{canEmbed}(G_S, P_S)$ holds.

**Definition 7.8**

Given an *acyclic* graph G, we define $\texttt{mkGraphPospc}(G)$ to be a pospace where:

- The topological space of $\texttt{mkGraphPospc}(G)$ is $\texttt{TopDiscrete}(G.\text{Nodes})$.
- The partial order of $\texttt{mkGraphPospc}(G)$ is $(\leqslant_G)$, where $(x \leqslant_G y)$ if and only if there is a path in the graph G from the node $x$ to the node $y$. Please note that, we require G be acyclic to make sure that $(\leqslant_G)$ has an anti-symmetric property; otherwise, $(\leqslant_G)$ will only a pre-order and not necessary partial order. ◢

**Definition 7.9**

Given an acyclic graph G and a pospace P, we define $\texttt{canEmbed}(G, P)$ to be a proposition stating that there is a dimap from $\texttt{mkGraphPospc}(G)$ to a pospace P such that the underlying function is injective and contains every extreme point of P in its image. ◢

## § 7.4 Pospaces of Standard Cubes

This section will define the pospace of a standard $n$-cube, denoted as $\square^n_{\text{po}}$. To do this, we first assign a value for each point of $\mathbb{R}^n_{\text{top}}$ according to $\texttt{rank}^n_{\text{std}}$ in definition 7.10. We also show that $\texttt{mkPospc}(\mathbb{R}^n_{\text{top}}, \texttt{rank}^n_{\text{std}})$ imitates $\mathbb{R}^n_{\text{po}}$ using lemmas 7.12 and 7.13.

Then, we follow section 7.2 to get the definition of $\square^n_{\text{po}}$ as stated in definition 7.11. Finally, to make sure that the definition of $\square^n_{\text{po}}$ makes sense, we follow section 7.3 by embedding $\mathbb{F}^n_{\square}$ into $\square^n_{\text{po}}$ using theorem 7.16.

◤ **Definition 7.10** ⬣ We define $\texttt{rank}^n_{\text{std}} : \mathbb{R}^n \to \mathbb{R}$ such that

$$\texttt{rank}^n_{\text{std}}(\langle x_0 , x_1 , \dots , x_{n-1} \rangle) \quad :\equiv \quad x_0 + x_1 + \cdots + x_{n-1}$$ ◢

◤ **Definition 7.11** ⬣ We define the *pospace of a standard $n$-cube*, denoted as $\square^n_{\text{po}}$, to be $\texttt{mkPospc}(\square^n_{\text{top}}, \texttt{rank}^n_{\text{std}})$. ◢

◤ **Lemma 7.12** ⬣
The partial order $( \leqslant_{\mathbb{R}^n_{\text{po}}} )$ is a sub-relation of $( \sqsubseteq_{\texttt{rank}^n_{\text{std}}} )$ for every $n : \mathbb{N}$. ◢

*Proof.* The goal is equivalent to

$$\begin{array}{lllcl}
\forall(\vec{x}, \vec{y} : \mathbb{R}^n) & (\vec{x} \leqslant_{\mathbb{R}^n_{\text{po}}} \vec{y}) & \to & (\vec{x} & \sqsubseteq_{\texttt{rank}^n_{\text{std}}} \quad \vec{y}) \\
\forall(\vec{x}, \vec{y} : \mathbb{R}^n) & \bigwedge_{i=0}^{n-1} (x_i \leqslant y_i) & \to & d(\vec{x}, \vec{y}) & \leqslant \quad \Sigma_{i=0}^{n-1} y_i - \Sigma_{i=0}^{n-1} x_i \\
\forall(\vec{x}, \vec{y} : \mathbb{R}^n) & \bigwedge_{i=0}^{n-1} (0 \leqslant y_i - x_i) & \to & \sqrt{\Sigma_{i=0}^{n-1} (y_i - x_i)^2} & \leqslant \quad \Sigma_{i=0}^{n-1} (y_i - x_i) \\
\forall(\vec{z} : \mathbb{R}^n) & \bigwedge_{i=0}^{n-1} (0 \leqslant z_i) & \to & \sqrt{\Sigma_{i=0}^{n-1} z_i^2} & \leqslant \quad \Sigma_{i=0}^{n-1} z_i
\end{array}$$

Now, let $\vec{z} : \mathbb{R}^n$ such that all elements of $\vec{z}$ are non-negative, the goal becomes

$$\sqrt{\Sigma_{i=0}^{n-1} z_i^2} \quad \leqslant \quad \Sigma_{i=0}^{n-1} z_i \qquad \text{for every} \quad i : \texttt{fin}(n) \,.$$

By comparing each term, we know that

$$\Sigma_{i=0}^{n-1} z_i^2 \quad \leqslant \quad (\Sigma_{i=0}^{n-1} z_i)^2 \qquad \text{for every} \quad i : \texttt{fin}(n) \,,$$

which can be used to prove the goal by the monotonicity of square root function. QED

◤ **Lemma 7.13** ⬣

The partial order ( $\leqslant_{\mathbb{R}^n_{po}}$ ) is equivalent to ( $\sqsubseteq_{\texttt{rank}^n_{\texttt{std}}}$ ) for every $n : \texttt{fin}(3)$ . ◢

*Proof.* In the forward direction, we can simply apply lemma 7.12. In the backward direction, we need to prove that

$$\sqrt{\Sigma_{i=0}^{n-1} z_i^2} \;\leqslant\; \Sigma_{i=0}^{n-1} z_i \qquad \text{implies} \qquad \bigwedge_{i=0}^{n-1} (0 \leqslant z_i)$$

for every $n : \texttt{fin}(3)$ and $\vec{z} : \mathbb{R}^n$ .

When $n :\equiv 0$, the goal is vacuously true. When $n :\equiv 1$, the assumption is $\sqrt{z_0^2} \leqslant z_0$, we also know that $0 \leqslant \sqrt{z_0^2}$ due to the square root property of real numbers, concatenate these two inequality we get, $0 \leqslant z_0$.

When $n :\equiv 2$, the assumption is $\sqrt{z_0^2 + z_1^2} \leqslant z_0 + z_1$. We can square both sides of the assumption because $\sqrt{z_0^2 + z_1^2}$ is non-negative due to the square root property of real numbers, which, in turn, makes $z_0 + z_1$ become non-negative as well. Now we have $z_0^2 + z_1^2 \leqslant z_0^2 + 2z_0z_1 + z_1^2$ therefore $0 \leqslant z_0z_1$ which means the sign of $z_0$ and $z_1$ must be the same in order to make $z_0z_1$ still be non-negative. However, both of them can't be negative because we already know that $z_0 + z_1$ is non-negative. Therefore, both $z_0$ and $z_1$ must be non-negative which satisfy the goal. QED

◤ **Definition 7.14** ⬣
We define $\mathtt{emb}^{n}_{\mathsf{std}}$ : $\mathtt{TopDiscrete}(\mathbb{F}^{n}_{\square}.\mathsf{Nodes}) \longrightarrow \square^{n}_{\mathsf{top}}$ to be a continuous function that takes a binary number $\langle b_0, b_1, \ldots, b_{n-1} \rangle$ then returns an $n$-tuple $\langle x_0, x_1, \ldots, x_{n-1} \rangle$ such that the value of $x_i : \mathbb{R}$ is either zero if $b_i :\equiv 0$ or one if $b_i :\equiv 1$ for each $i : \mathtt{fin}(n)$. Please note that, $\mathtt{emb}^{n}_{\mathsf{std}}$ is continuous because of lemma 4.4. ◢

◤ **Lemma 7.15** ⬣ Let $n : \mathbb{N}$ and $\vec{b} : \mathtt{binary}(n)$, we have

$$\mathtt{emb}^{n}_{\mathsf{std}}(\vec{b}) \quad :\equiv \quad \mathtt{emb}^{1}_{\mathsf{std}}(b_0) + \mathtt{emb}^{1}_{\mathsf{std}}(b_1) + \cdots + \mathtt{emb}^{1}_{\mathsf{std}}(b_{n-1}).$$ ◢

◤ **Theorem 7.16** ⬣ The proposition $\mathtt{canEmbed}(\mathbb{F}^{n}_{\square}, \square^{n}_{\mathsf{po}})$ holds. ◢

*Proof.* We use $\mathtt{emb}^{n}_{\mathsf{std}}$ as the underlining continuous function for the dimap of the main goal. First, it is obvious to see that the continuous function $\mathtt{emb}^{n}_{\mathsf{std}}$ is injective and contains every extreme point of $\square^{n}_{\mathsf{po}}$ in its image. Then, we ensure that the continuous function $\mathtt{emb}^{n}_{\mathsf{std}}$ preserves the relation from $\mathtt{mkGraphPospc}(\mathbb{F}^{n}_{\square})$ to $\square^{n}_{\mathsf{po}}$.

The remaining goal can be translated as

$$\mathbb{F}^{n}_{\square}.\mathsf{isEdge}(\vec{s}, \vec{t}) \qquad \text{implies} \qquad \mathtt{emb}^{n}_{\mathsf{std}}(\vec{s}) \sqsubseteq_{\mathtt{rank}^{n}_{\mathsf{std}}} \mathtt{emb}^{n}_{\mathsf{std}}(\vec{t}).$$

Please note that, we check only edges instead of all paths because each path is a collection of edges composed together anyway.

Assuming that $\langle \vec{s}, \vec{t} \rangle$ is an edge and let $j : \mathtt{fin}(n)$ is the index that $(s_i = 0)$ and $(t_i = 1)$. Please note that, such the index exists uniquely due to definition 5.36. Because of lemma 7.15, we can generalise the remaining goal to

$$\mathtt{emb}^{1}_{\mathsf{std}}(s_j) \sqsubseteq_{\mathtt{rank}^{1}_{\mathsf{std}}} \mathtt{emb}^{1}_{\mathsf{std}}(t_j) \qquad \text{for every} \quad j : \mathtt{fin}(n)$$

If $(i = j)$, then the goal holds because $(0 \sqsubseteq_{\mathtt{rank}^{1}_{\mathsf{std}}} 1)$; otherwise, we have $(s_j = t_j)$ and the goal holds because the reflexivity of $(\sqsubseteq_{\mathtt{rank}^{1}_{\mathsf{std}}})$. QED

## § 7.5 Pospaces of Twisted Cubes

This section will define the pospace of a twisted $n$-cube, denoted as $\bowtie^n_{\mathsf{po}}$. To do this, we first assign a value for each point of $\mathbb{R}^n_{\mathsf{top}}$ according to $\mathtt{rank}^n_{\mathsf{tw}}$ in definition 7.20.

Then, we follow section 7.2 to get the definition of $\bowtie^n_{\mathsf{po}}$ as stated in definition 7.21. Finally, to make sure that the definition of $\bowtie^n_{\mathsf{po}}$ makes sense, we follow section 7.3 by embedding $\mathbb{F}^n_{\bowtie}$ into $\bowtie^n_{\mathsf{po}}$ using theorem 7.27.

◤ **Definition 7.17** ⬣ We define the *canonical twisted $n$-cube*, denoted as $\bowtie^n_{\mathsf{top}}$, to be a subspace of $\mathbb{R}^n_{\mathsf{top}}$ induced by points $\langle -1\,,\ -1\,,\ \ldots\,,\ -1\rangle$ and $\langle 1\,,\ 1\,,\ \ldots\,,\ 1\rangle$ that generate the following set

$$\bowtie^n_{\mathsf{set}} :\equiv \{\ \langle x_0\,,\ x_1\,,\ \ldots\,,\ x_{n-1}\rangle\ :\ \mathbb{R}^n\ \ |\ \ -1 \leqslant x_i \leqslant 1\,\}.$$ ◢

◤ **Definition 7.18** ⬣ We define $\mathtt{mapBitToSign}^n_{\mathsf{top}}$ to be an affine transformation from $\square^n_{\mathsf{top}}$ to $\bowtie^n_{\mathsf{top}}$ such that each element of the input is multiplied by 2 then minus 1.

$$\mathtt{mapBitToSign}^n_{\mathsf{top}}(\langle x_0, x_1, \ldots, x_{n-1}\rangle) \ :\equiv\ \langle 2{\cdot}x_0{-}1,\ 2{\cdot}x_1{-}1,\ \ldots,\ 2{\cdot}x_{(n-1)}{-}1\rangle$$ ◢

◤ **Definition 7.19** ⬣ The topological space $\square^n_{\mathsf{top}}$ is *homeomorphic* to $\bowtie^n_{\mathsf{top}}$. ◢

*Proof.* It is obvious that $\mathtt{mapBitToSign}^n_{\mathsf{top}}$ is a homeomorphism from $\square^n_{\mathsf{top}}$ to $\bowtie^n_{\mathsf{top}}$. QED

◤ **Definition 7.20** ⬢ We define $\mathtt{rank}^n_{\mathsf{tw}} : \mathbb{R}^n \to \mathbb{R}$ as the following.

$$\mathtt{rank}^0_{\mathsf{tw}}\ (\langle\rangle) \quad :\equiv \quad 0$$

$$\mathtt{rank}^{n+1}_{\mathsf{tw}}(x_0 :: \vec{x}') \quad :\equiv \quad x_0 \cdot (2^n + x_0 \cdot \mathtt{rank}^n_{\mathsf{tw}}(\vec{x}'))$$

Equivalently, $\mathtt{rank}^n_{\mathsf{tw}}$ can be defined without recursion as the following.

$$\mathtt{rank}^n_{\mathsf{tw}}(\langle x_0, x_1, \dots, x_{n-1}\rangle) \quad :\equiv \quad \sum_{i=0}^{n-1}\left(x_i \cdot 2^{(n-1-i)} \cdot \prod_{j=0}^{i-1} x_j\right)$$ ◢

◤ **Definition 7.21** ⬢ We define the *pospace of a twisted $n$-cube*, denoted as $\bowtie^n_{\mathsf{po}}$, to be $\mathtt{mkPospc}(\bowtie^n_{\mathsf{top}}, \mathtt{rank}^n_{\mathsf{tw}})$. ◢

◤ **Remark 7.22** ⬢ Intuitively, $\mathtt{rank}^n_{\mathsf{tw}}$ is a modification of $\mathtt{rank}^n_{\mathsf{std}}$ where

- The term $(\prod_{j=0}^{i-1} x_j)$ is multiplied by $x_i$ because each earlier dimension $j$ could reverse the direction of dimension $i$.
- The term $2^{(n-1-i)}$ is multiplied by $x_i$ in order to ensure that the sums of later terms will not outweigh the current term i.e.

  $$\left|x_i \cdot 2^{(n-1-i)} \cdot \prod_{j=0}^{i-1} x_j\right| \quad \geqslant \quad \sum_{k=i+1}^{n-1}\left|x_k \cdot 2^{(n-1-k)} \cdot \prod_{j=0}^{k-1} x_j\right|,$$

  which is effective because $\left|x_i\right| \leqslant 1$.

Consequently, $\mathtt{rank}^n_{\mathsf{tw}}$ only works in $\bowtie^n_{\mathsf{top}}$; If we want $\mathtt{rank}^n_{\mathsf{tw}}$ to work for the entire $\mathbb{R}^n_{\mathsf{top}}$, then we needs to change the codomain of $\mathtt{rank}^n_{\mathsf{tw}}$ to *hyperreal numbers* [Rob74] and replace $2^{(n-1-i)}$ with $\omega^{(n-1-i)}$. ◢

◤ **Example 7.23** ⬣ The following diagram visualises definition 7.20 by illustrating the anatomy of the twisted $n$-cube when $1 \leqslant n \leqslant 3$.

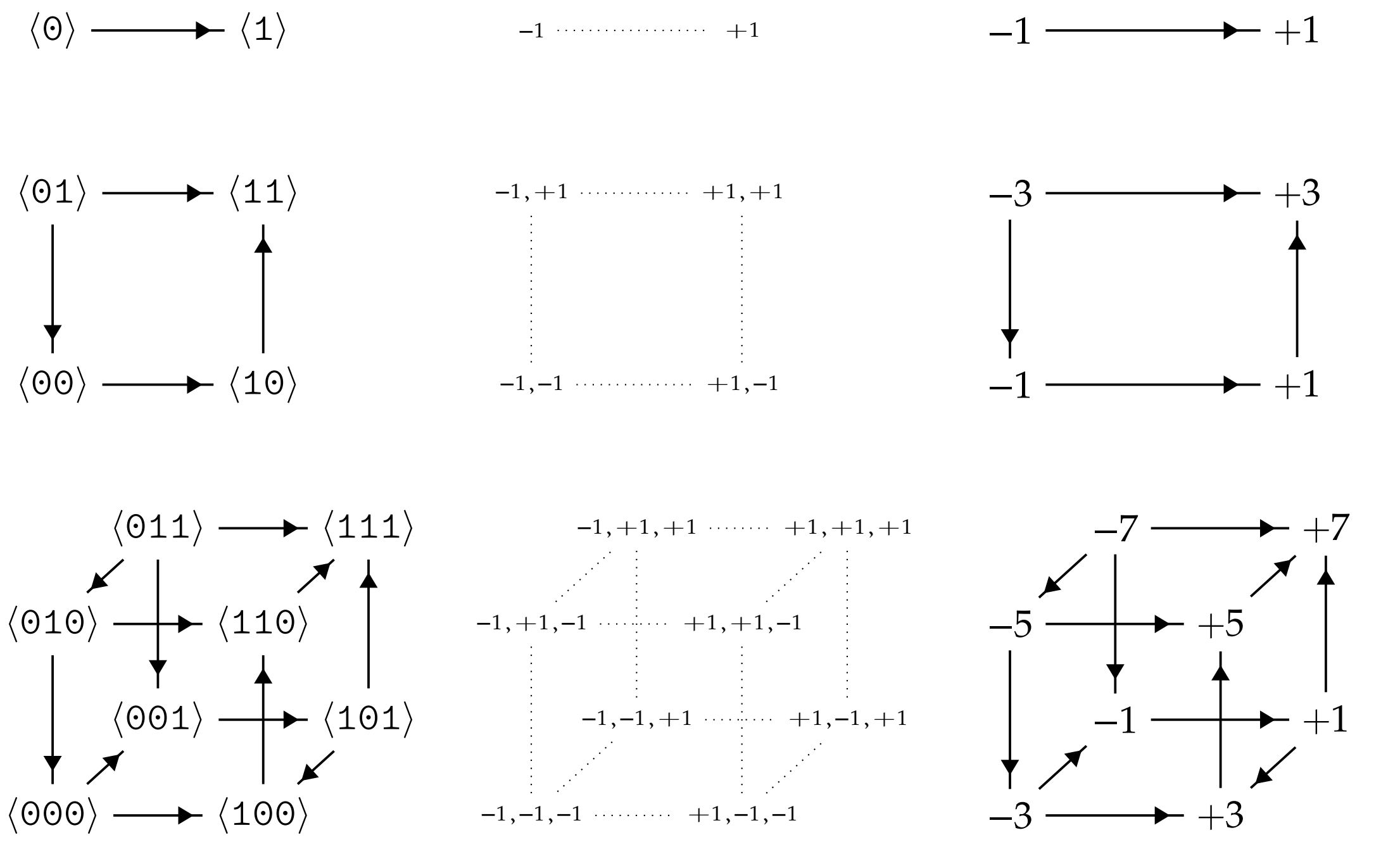

For each row where $1 \leqslant n \leqslant 3$, the left and middle columns show $\mathbb{F}^n_{\bowtie}$ and $\bowtie^n_{\mathsf{top}}$, respectively. The right column shows the value of $\mathtt{rank}^n_{\mathsf{tw}}(p)$, for each extreme point $p$ in $\bowtie^n_{\mathsf{po}}$, together with arrows in the direction that the value increases. ◢

◤ **Definition 7.24** ⬣ We define $\mathtt{emb}^n_{\mathsf{tw}}$ to be a continuous function from $\mathtt{TopDiscrete}(\mathbb{F}^n_{\bowtie}.\mathsf{Nodes})$ to $\bowtie^n_{\mathsf{top}}$ with the following definition.

$$\mathtt{emb}^n_{\mathsf{tw}} \quad :\equiv \quad \mathtt{emb}^n_{\mathsf{std}} \cdot \mathtt{mapBitToSign}^n_{\mathsf{top}}$$ ◢

◤ **Theorem 7.25** ⬣ Let $n : \mathbb{N}$ and let $\vec{b} : \mathbb{F}^n_{\bowtie}.\mathsf{Nodes}$, we have

$$\mathtt{rank}^n_{\mathsf{tw}}(\mathtt{emb}^n_{\mathsf{tw}}(\vec{b})) \quad = \quad 2 \cdot \mathtt{sort}^n_{\mathsf{num}}(\vec{b}) \quad - \quad \begin{cases} 0 & \text{if } n = 0 \\ 2^{n-1} - 1 & \text{otherwise.} \end{cases}$$ ◢

*Proof.* We prove this by induction on $n$. For the base cases:

- If $n :\equiv 0$ and $\vec{b} :\equiv \langle\rangle$, then both sides of the equation are 0.
- If $n :\equiv 1$ and $\vec{b} :\equiv \langle \mathtt{0} \rangle$, then both sides of the equation are −1.
- If $n :\equiv 1$ and $\vec{b} :\equiv \langle \mathtt{1} \rangle$, then both sides of the equation are 1.

For the induction case, we need to prove that

$$\mathtt{rank}_{\mathsf{tw}}^{n+2}(\mathtt{emb}_{\mathsf{tw}}^{n+2}\ (b_0\ ::\ \vec{b'})) \quad = \quad 2 \cdot \mathtt{sort}_{\mathsf{num}}^{n+2}\ (b_0\ ::\ \vec{b'}) \quad - \quad (2^{n+2} - 1),$$

given the induction hypothesis as

$$\mathtt{rank}_{\mathsf{tw}}^{n+1}(\mathtt{emb}_{\mathsf{tw}}^{n+1}(\vec{b'})) \quad = \quad 2 \cdot \mathtt{sort}_{\mathsf{num}}^{n+1}(\vec{b'}) \quad - \quad (2^{n+1} - 1),$$

for every $b\ :\ \mathbb{2}_{\mathsf{fin}}$ and $\vec{b'}\ :\ \mathtt{binary}(n+1)$. We now alias the left and right sides of the goal as LHS and RHS, respectively. In addition $\mathtt{0}$ and $\mathtt{1}$ will be appended to both LHS and RHS once $b_0$ is assigned to be $\mathtt{0}$ and $\mathtt{1}$, respectively.

$$\begin{array}{llll}
\mathsf{LHS} & = & \mathtt{rank}_{\mathsf{tw}}^{n+2}(\mathtt{emb}_{\mathsf{tw}}^{n+2}\ (b_0\ ::\ \vec{b'})) & [\text{unfold LHS}] \\
 & = & \mathtt{rank}_{\mathsf{tw}}^{n+2}((2 \cdot b_0 - 1)\ ::\ \mathtt{rank}_{\mathsf{tw}}^{n+1}(\mathtt{emb}_{\mathsf{tw}}^{n+1}(\vec{b'}))) & [\text{unfold } \mathtt{emb}_{\mathsf{tw}}^{n+2}] \\
 & = & (2 \cdot b_0 - 1) \cdot (2^{n+1} + \mathtt{rank}_{\mathsf{tw}}^{n+1}(\mathtt{emb}_{\mathsf{tw}}^{n+1}(\vec{b'}))) & [\text{unfold } \mathtt{rank}_{\mathsf{tw}}^{n+2}] \\
 & = & (2 \cdot b_0 - 1) \cdot (2^{n+1} + 2 \cdot \mathtt{sort}_{\mathsf{num}}^{n+1}(\vec{b'}) - (2^{n+1} - 1)) & [\text{induction hypo}] \\
\mathsf{LHS} & = & (2 \cdot b_0 - 1) \cdot (2 \cdot \mathtt{sort}_{\mathsf{num}}^{n+1}(\vec{b'}) + 1) & [\text{rearrangement}] \\
\mathsf{LHS0} & = & (-1) \cdot (2 \cdot \mathtt{sort}_{\mathsf{num}}^{n+1}(\vec{b'}) + 1) & [\text{assign } b_0 :\equiv \mathtt{0}] \\
\mathsf{LHS1} & = & (+1) \cdot (2 \cdot \mathtt{sort}_{\mathsf{num}}^{n+1}(\vec{b'}) + 1) & [\text{assign } b_1 :\equiv \mathtt{1}]
\end{array}$$

$$\begin{array}{llll}
\mathsf{RHS0} & = & 2 \cdot \mathtt{sort}_{\mathsf{num}}^{n+2}\ (\mathtt{0}\ ::\ \vec{b'}) - (2^{n+2} - 1) & [\text{unfold RHS0}] \\
 & = & 2 \cdot (2^{n+1} - \mathtt{sort}_{\mathsf{num}}^{n+1}(\vec{b'})) - (2^{n+2} - 1) & [\text{unfold } \mathtt{sort}_{\mathsf{num}}^{n+2}] \\
\mathsf{RHS0} & = & (-1) \cdot (2 \cdot \mathtt{sort}_{\mathsf{num}}^{n+1}(\vec{b'}) + 1) & [\text{rearrangement}]
\end{array}$$

$$\begin{aligned}
\mathsf{RHS1} &= 2 \cdot \mathtt{sort}^{n+2}_{\mathsf{num}}((\mathtt{1} :: \vec{b'})) - (2^{n+2} - 1) && \text{[unfold RHS1]} \\
&= 2 \cdot (2^{n+1} + \mathtt{sort}^{n+1}_{\mathsf{num}}\ \vec{b'}) - (2^{n+2} - 1) && \text{[unfold } \mathtt{sort}^{n+2}_{\mathsf{num}}\text{]} \\
\mathsf{RHS1} &= (+1) \cdot (2 \cdot \mathtt{sort}^{n+1}_{\mathsf{num}}(\vec{b'}) + 1) && \text{[rearrangement]}
\end{aligned}$$

We prove the goal by splitting $b_0 : \mathbb{2}_{\mathsf{fin}}$ into 2 cases.

- When $b_0 :\equiv \mathtt{0}$, the goal holds because ($\mathsf{LHS0} = \mathsf{RHS0}$).
- When $b_0 :\equiv \mathtt{1}$, the goal holds because ($\mathsf{LHS1} = \mathsf{RHS1}$). QED

◤ **Lemma 7.26** ⬣ Let $n : \mathbb{N}$ and $\vec{s}, \vec{t} : \mathbb{F}^n_{\bowtie}.\mathsf{Edges}$, we have

$$\mathbb{F}^n_{\bowtie}.\mathsf{isEdge}(\vec{s}, \vec{t}) \quad \text{implies} \quad (\vec{s} \sqsubseteq_{\mathtt{rank}^n_{\mathsf{tw}}} \vec{t}) \quad \text{in} \quad \bowtie^n_{\mathsf{po}}.$$ ◢

*Proof.*

$$\begin{aligned}
& \mathbb{F}^n_{\bowtie}.\mathsf{isEdge}(\vec{s}, \vec{t}) \\
\Rightarrow\ & \mathtt{sort}^n_{\mathsf{num}}(\vec{s}) \leqslant \mathtt{sort}^n_{\mathsf{num}}(\vec{t}) && \text{[By lemma 6.22]} \\
\Rightarrow\ & \mathtt{rank}^n_{\mathsf{tw}}(\mathtt{emb}^n_{\mathsf{tw}}(\vec{s})) \leqslant \mathtt{rank}^n_{\mathsf{tw}}(\mathtt{emb}^n_{\mathsf{tw}}(\vec{t})) && \text{[By theorem 7.25]} \\
\Rightarrow\ & (\vec{s} \sqsubseteq_{\mathtt{rank}^n_{\mathsf{tw}}} \vec{t}) \quad \text{in} \quad \bowtie^n_{\mathsf{po}} && \text{[By definition 7.21]}
\end{aligned}$$

QED

◤ **Theorem 7.27** ⬣ The proposition $\mathtt{canEmbed}(\mathbb{F}^n_{\bowtie}, \bowtie^n_{\mathsf{po}})$ holds. ◢

*Proof.* The proof is quite similar to the proof of theorem 7.16 but we use $\mathtt{emb}^n_{\mathsf{tw}}$ as the underlining continuous function instead of $\mathtt{emb}^n_{\mathsf{std}}$. Another difference is the remaining goal that now becomes

$$\mathbb{F}^n_{\bowtie}.\mathsf{isEdge}(\vec{s}, \vec{t}) \qquad \text{implies} \qquad \mathtt{emb}^n_{\mathsf{tw}}(\vec{s}) \sqsubseteq_{\mathtt{rank}^n_{\mathsf{tw}}} \mathtt{emb}^n_{\mathsf{tw}}(\vec{t}).$$

where $\vec{s}$ and $\vec{t}$ are node in $\mathbb{F}^n_{\bowtie}.\mathsf{Nodes}$.

This remaining goal can be proven by lemma 7.26. QED

Chapter

# 8

# Conclusions and Future Work

Although the graph-theoretic framework introduced in chapter 5 was originally developed to formalise the concept of twisted cubes, the framework itself has a value on it own right: It suggests a new perspective to reasoning about the simplices, standard cubes, and possibly other combinatorial concepts that don't necessary have geometric intuition. One possibility of related future work is to use this graph-theoretic framework to analyse other categories that have been used to construct presheaves.

Algebraic descriptions via generators and relations is also in the list of future work. Such presentations exist for many different cube categories in the literature. This, I hope, will make it possible to develop the higher categorical structures that can be encoded as presheaves on the category of twisted cubes. An ultimate goal would be to model some form of *directed cubical type theory*.

Another possible application of our twisted cube categories might be building a syntax for a parametric type theory or cubical type theory without an interval as suggested by Altenkirch and Kaposi [AK18]. A major difficulty in their development was the presence of multiple degeneracies, a problem which does not occur in the current work.

A further direction which may be worth exploring is to not consider set-valued presheaves, but type-valued presheaves instead. To facilitate this, I can consider the category of twisted semi-cubes. From there, type-valued presheaves can be encoded as Reedy-fibrant diagrams in a known style [Shu15]. We can then add a condition reminiscent of Rezk's *Segal-condition* [Rez01] by stating that the projection from twisted semi-cubical types to the sequence of types along the Hamiltonian path is an equivalence. This corresponds to saying that the partial $n$-cube with missing inner part and lid (cf. subsection 3.1.7) have a contractible type of fillers. It seems that this could be a first step towards the construction of composition and higher coherences, although further conditions seem to be necessary. The relation to the *(complete) semi-Segal types* by Capriotti and others [Ann+19; Cap16; CK17] remains to be studied.

Another possible future work is to prove conjectures 1.1 to 1.3 as a sanity check that twisted cubical sets can generalise $(\infty, n)$-*categories* well.

Finally, I would like to emphasize again that the development of twisted cubes is still in the early phase. This thesis only serves as the snapshot of the research that I have done *before* finishing this PhD. If the reader would like to see the current state-of-the-art regarding the twisted cubes, please check

`https://orcid.org/0000-0002-8483-5261`

which is my OCRID webpage, thank you for your interest.

This page intentionally left blank.

(as the back cover of the thesis)